\documentclass[aps,prl,twocolumn,floats,superscriptaddress,tighten,letterpaper,floatfix]{revtex4-1}
\usepackage{amssymb}
\usepackage{amsbsy}
\usepackage{amsmath}
\usepackage{graphicx}
\usepackage{graphics}
\usepackage{setspace}
\usepackage{array}
\usepackage{color}
\usepackage{xcolor}
\usepackage{fontenc}
\usepackage{textcomp}
\usepackage{rotating}
\usepackage{bm}
\usepackage{subfigure}
\usepackage{hyperref}
\hypersetup{pdfpagemode=UseNone}

\DeclareMathOperator{\Tr}{Tr}
\DeclareMathOperator{\sgn}{sgn}

\newcommand{\e}{\varepsilon}

\newcommand{\vex}[1]{\bm{\mathrm{#1}}}
\newcommand{\Nabla}{\bm{\nabla}}

\newcommand{\pupsf}[1]{{\scriptscriptstyle{(\mathsf{{#1}})}}}

\newcommand{\tauh}{\hat{\tau}}

\newcommand{\sigh}{\hat{\sigma}}
\newcommand{\sigb}{\hat{\bm{\sigma}}}

\newcommand{\T}{\mathsf{T}}
\newcommand{\Mp}{\hat{M}_{\mathsf{P}}}

\newcommand{\Ms}{\hat{M}_{\mathsf{S}}}
\newcommand{\hs}{\hat{h}}

\newcommand{\parr}{\partial}
\newcommand{\parb}{\bar{\partial}}

\newcommand{\Ac}{\mathsf{A}}
\newcommand{\Ab}{\bar{\mathsf{A}}}

\newcommand{\bsub}{\begin{subequations}}
\newcommand{\esub}{\end{subequations}}

\newcommand{\EE}{\mathsf{E}}
\newcommand{\sci}{S_{\scriptscriptstyle{\mathrm{CI}}}}
\newcommand{\sct}{S_{\scriptscriptstyle{\mathrm{C}}}}

\begin{document}
\title{Critical Percolation Without Fine Tuning on the Surface of a Topological Superconductor}
\author{Sayed Ali Akbar Ghorashi}
\affiliation{Texas Center for Superconductivity and Department of Physics, University of Houston, Houston, Texas 77204, USA}
\author{Yunxiang Liao}
\affiliation{Department of Physics and Astronomy, Rice University, Houston, Texas 77005, USA}
\author{Matthew S.\ Foster}
\affiliation{Department of Physics and Astronomy, Rice University, Houston, Texas 77005, USA}
\affiliation{Rice Center for Quantum Materials, Rice University, Houston, Texas 77005, USA}
\date{\today\\}

\newcommand{\be}{\begin{equation}}
\newcommand{\ee}{\end{equation}}
\newcommand{\bea}{\begin{eqnarray}}
\newcommand{\eea}{\end{eqnarray}}
\newcommand{\h}{\hspace{0.30 cm}}
\newcommand{\vs}{\vspace{0.30 cm}}
\newcommand{\n}{\nonumber}

\begin{abstract}	
We present numerical evidence that most two-dimensional surface states of a bulk 
topological superconductor (TSC) sit at an integer quantum Hall plateau transition.  
We study TSC surface states in class CI with quenched disorder. 
Low-energy (finite-energy) surface states were expected to be critically delocalized 
(Anderson localized). We confirm the low-energy picture, but find instead that finite-energy 
states are also delocalized, with universal statistics that are independent of the TSC winding number, 
and consistent with the spin quantum Hall plateau transition (percolation).
\end{abstract}

\maketitle

When fluid floods a landscape, the percolation threshold is the stage at which 
travel by land or by sea becomes equally difficult.
In the quantum Hall effect (QHE),
the sea level corresponds to the Fermi energy 
and the landscape is the electrostatic impurity potential \cite{Trugman1983}.
The critical wave function that sits exactly at the plateau transition corresponds 
to the percolation threshold, while closed contours that encircle isolated lakes or 
islands correspond to Anderson localized states within a plateau. 
Although the critical statistics of the usual plateau transition differ \cite{Huckestein1995,Evers2008},
the plateau transition in the \emph{spin quantum Hall effect} (SQHE) 
\cite{Kagalovsky1999,Gruzberg1999,Senthil1999,Cardy2000,Beamond2002,Evers2003,Mirlin2003,Cardy2010}
can be mapped exactly to classical percolation \cite{Gruzberg1999}. 
The SQHE was introduced in the context of spin singlet, two-dimensional (2D)
superconductivity. 
Gapless quasiparticles can conduct a spin
current, and under the right conditions (broken time-reversal symmetry but negligible 
Zeeman coupling, as in a $d$+$i$$d$ superconductor \cite{Senthil1999}), 
the spin Hall conductance within the spin Hall plateau is precisely quantized. 
The SQHE belongs to the Altland-Zirnbauer class C \cite{Evers2008}. 
 
For both classical percolation and the plateau transition in the SQHE, the ``sea level'' has to be fine-tuned to the 
percolation threshold; in the SQHE, this means that almost all states are Anderson localized,
except those at the transition. 
In this Letter, we uncover a new realization of 2D critical percolation that requires
no fine-tuning. In particular, we provide numerical evidence for an \emph{energy band} of  
states, where each state exhibits statistics consistent with critical percolation.
We show that this band 
of 
states appears at the surface of a three-dimensional (3D)
topological superconductor 
(TSC)
in class CI, 
subject to 
quenched surface disorder that 
preserves spin SU(2) symmetry and time-reversal invariance. 

Our results 
suggest an unexpected, direct link between
3D time-reversal invariant 
TSCs and 2D QHEs. 
We 
use
a generalized 
surface model that 
works for any bulk winding number,
but we find the same ``percolative'' states at finite energy in all cases.  
Together with previous results for 
class AIII 
\cite{Ludwig1994,Ostrovsky2007,Chou2014}, 
it is natural to conjecture that the three classes of 3D TSCs (CI, AIII, DIII \cite{SRFL2008,Kitaev09}) possess
surface states that at finite energy and 
\emph{any winding number} 
are equivalent to the corresponding plateau transitions of the SQHE (class C), 
integer quantum Hall effect (class A), and thermal quantum Hall effect (class D). 
The critical surface state band
found here 
will dominate finite-temperature
response
(a ``multifractal spin metal'').
 
Effective field theories for TSC surface states \cite{SRFL2008,Foster2012,Foster2014,Ghorashi2017,Roy2017} 
were 
originally
studied 
\cite{Ludwig1994,Nersesyan1994,Mudry1996,Caux1996,Bhaseen2001}
as examples of exactly solvable, critical delocalization in 2D.
Only recently was it understood that these must be attached to a higher-dimensional bulk, 
owing to certain anomalies \cite{SRFL2008,Ryu2012,Stone2012}. 
TSC surface states can appear as multiple species of 2D Dirac or Majorana fermions \cite{SRFL2008}. 
In class CI these are Dirac
owing to the conservation of 
spin, 
the $z$-component of which plays the role of a U(1) ``electric'' charge. 
Nonmagnetic intervalley impurity scattering takes the form of an
SU(2) vector potential, due to the anomalous version of time-reversal symmetry \cite{Foster2012,Foster2014}. 
The exact solvability (and proof of critical delocalization) holds only at
zero
energy in these theories \cite{Nersesyan1994,Mudry1996,Caux1996,Bhaseen2001}. 

Topological protection \cite{SRFL2008} requires
that 
at least one surface or edge state 
must evade Anderson localization in the presence of 
(nonmagnetic) quenched disorder \cite{Essin2015,Volovik,Essin2011}.
For a class CI TSC surface, the zero-energy wave function is critically delocalized, 
with statistics that are exactly solved by a certain conformal field theory (CFT) \cite{Ryu2009,Foster2012}. 
The standard symmetry-based argument \cite{Evers2008,Essin2015} 
is
that all finite-energy states of a class CI Hamiltonian should reside 
in the ``orthogonal'' metal class (AI), 
known to possess only
Anderson-localized states in 2D \cite{Evers2008}.

A superficial argument can be given for
why any nonstandard class (such as CI)
with a special chiral or particle-hole symmetry becomes a standard Wigner-Dyson class
(here AI) at finite energy: 
Adding the energy perturbation to the Hamiltonian matrix
$\hat{h} \rightarrow \hat{h} - \e \, \hat{1}$
breaks the special symmetry for any $\e \neq 0$.  
This logic is flawed, however, because $\e$ couples to the identity operator $\hat{1}$,
which \emph{commutes} with $\hat{h}$. 
The argument works for a random symmetry-breaking perturbation $\e \rightarrow \e(\vex{r})$ 
($\vex{r}$ is the position vector), but that is a different problem.   

A physical argument for the reduction to Wigner-Dyson in a nonstandard class describing 
Bogoliubov-de Gennes quasiparticles in superconductors \cite{SRFL2008,Evers2008} is the following.
For single-particle energies much larger than the BCS gap, 
the wave functions should resemble those
of the parent normal metal, while $\e = 0$ is the only symmetry-distinguished energy. 
However, TSC surface states can evade this argument as well, since
the bulk gap is the \emph{maximum} allowed surface state energy; above this, 2D surface states 
hybridize with the
3D bulk. All TSC surface states are (Andreev) bound states. 

We find energy stacks of delocalized class C, SQHE plateau transition states
for any bulk TSC winding number. The SQHE states are identified by their 
multifractal spectrum \cite{Evers2003,Mirlin2003,Evers2008}.
The absence of Anderson localization throughout the \emph{surface} energy spectrum 
is qualitatively similar 
to 1D edge states of quantum Hall, as well as edge and surface states of 2D and 3D topological insulators.
Our work suggests that this may be a general principle of fermionic topological matter. 

Our results generalize a previous observation for a simpler
model in class AIII. This model consists of a single 2D Dirac
fermion coupled to abelian vector potential disorder; it is critically delocalized
and exactly solvable at zero energy \cite{Ludwig1994}. 
It can also be interpreted as the surface state of TSC with winding number $\nu = 1$ \cite{SRFL2008,Foster2014}. 
It was later claimed \cite{Ostrovsky2007}
that all finite-energy states of this model should reside at the plateau transition 
of the (class A) integer quantum Hall effect, and this was verified numerically \cite{Chou2014}.


\textit{Model}.---We employ a $k$-\emph{generalized}
two-species Dirac model to capture the surface
states of a class CI TSC with even winding number $\nu = 2 k$, 
\begin{equation}\label{hsDef}
	\hs=
	\begin{bmatrix}
	0 									& (-i\parr)^k + \Ac^a \, \tauh^a + \Ac^0	\\
	(-i\parb)^k + \Ab^a \, \tauh^a + \Ab^0					& 0
	\end{bmatrix},
\end{equation}
where 
$\parr \equiv \partial_x - i \partial_y$
and
$\Ac \equiv A_x - i A_y$, with $\parb$ and $\Ab$ respective complex conjugates of these. 

For $k = 1$, this is the surface theory for the lattice model in \cite{Ryu2009}. 
Quenched disorder enters via
the abelian vector potential $\Ac^0(\vex{r})$ 
or 
the 
nonabelian SU(2) vector potential $\Ac^a(\vex{r}) \, \tauh^a$,
where $\vex{r} = \{x,y\}$ is the position vector, 
and 
$\tauh^{1,2,3}$ denotes Pauli matrices acting on the space of the two species. 
The case with $k > 1$ was inspired by higher-dispersion surface bands obtained
in spin-3/2 class DIII TSC models \cite{Fang2015,Yang2016,Ghorashi2017,Roy2017}. 
The bulk winding number can be inferred by turning on the time-reversal symmetry-breaking
mass term and calculating the surface Chern number \cite{Volovik,Ghorashi2017}. 
For $k > 1$, Eq.~(\ref{hsDef}) is not gauge invariant, but this is 
of no consequence because the vector potentials merely represent the most
relevant type of quenched disorder allowed by symmetry. 
Class CI has $P^2 = -1$ particle-hole symmetry \cite{SRFL2008}. 
In order to realize $P$, we take $\Ac^0 = 0$ for odd $k$, 
while we take
$\Ac^3 = 0$ for even $k$ \cite{Supp}. 
Time-reversal invariance is equivalent to the block
off-diagonal form of $\hs$ \cite{Foster2012,Foster2014}. 

We analyze $\hs$ in Eq.~(\ref{hsDef}) numerically via exact diagonalization.
Calculations are performed in momentum space $\vex{q} = \{q_x,q_y\}$ to avoid 
doubling 
the 
surface theory 
\cite{Chou2014}. 
The Fourier components of any nonzero vector potential
$A_{x,y}^{0,1,2,3}(\vex{q})$ are parameterized via
$
	A(\vex{q}) 
	= 
	\left(\sqrt{\lambda} / L \right) 
	\exp\left[
	i \theta(\vex{q}) - q^2 \xi^2 / 4
	\right],
$
where $\theta(\vex{q}) = - \theta(-\vex{q})$, but these are otherwise independent, uniformly distributed random phases.
Here $L$, $\xi$ and $\lambda$ denote the system length, 
correlation length, and disorder strength respectively;
the latter is dimensionless for $k = 1$.  
We choose periodic boundary conditions so that $q_i = (2 \pi / L) n_i$, with $-N \leq n_i \leq N$, for $i \in \{x,y\}$.
Here $N$ determines the size of the vector space, which is $4(2 N + 1)^2$. 
The correlation length $\xi = 0.25 (L / N)$ for all calculations.

Except for states deep in the high-energy ``Lifshitz tails'' 
(see Fig.~\ref{Fig--DirtS}), 
we find no evidence of localization
in the surface eigenstate spectrum, although we cannot rule it out for much larger system sizes.  
Localization at high energies 
would not be unexpected, because the model is not terminated in a physical way 
(which would instead involve hybridizing the 2D surface with the 3D bulk). 

All of the states that we find in the bulk of the surface energy spectrum look ``critically delocalized,''
i.e.\ $|\psi|^2(\vex{r})$ is small over most of the surface, but is sporadically punctuated by 
probability peaks of variable height. 
We analyze these states via multifractal analysis \cite{Huckestein1995,Evers2008}.
One breaks the system up into boxes of size $b$, and defines the box probability $\mu_n$ 
and inverse participation ratio (IPR) $\mathcal{P}_q$ via
$
		\mu_n
		=
		\int_{\mathcal{A}_n}d^2\vex{r} \, |\psi(\vex{r})|^2,
$
$
		\mathcal{P}_q\equiv\sum_n\mu^q_n,
$
where $\mathcal{A}_n$ denotes the $n$th box. 
The multifractal spectrum $\tau(q)$ governs the scaling of the IPR,
$
		\mathcal{P}_q
		\sim
		(b/L)^{\tau(q)}.
$

For critically delocalized states, the form of $\tau(q)$ is expected to be self-averaging in the
infinite system size limit \cite{Chamon1996}. 
For class CI surface states at zero energy 
(class C SQHE plateau transition states),
the spectrum is exactly (to a good approximation) parabolic, 
and is given by 
$\tau(q) =  (q-1)(2 - \theta\, q)$ for $q < |q_c|$, 
and 
$\tau(q) =  [\sqrt{2}- \sgn(q) \sqrt{\theta}]^2 q$ for $q > |q_c|$. 
Here $q_c \equiv \sqrt{2 / \theta}$.
The parameter $\theta$ determines the degree of critical rarification:
$\theta = 0$ ($\theta > 0$) for a plane wave (multifractal) state. 
In the above, $q_c$ denotes the \emph{termination threshold} \cite{Chamon1996,Foster2009};
the spectrum is linear for $|q| > q_c$, and the slopes govern the 
scaling of the peaks and valleys of $|\psi|^2(\vex{r})$. 
Note that an accurate calculation of $\tau(q)$ for negative $q$ requires significant coarse-graining,
since it entails taking negative powers of a function that is small almost everywhere. 
For this reason negative-$q$ results are always worse than positive $q$ (and are
often not reported). 

For class CI, the $\text{Sp}(2n)_k$ CFT predicts that $\theta_k = 1/2(k+1)$ \cite{Foster2012}.  
Analytical and numerical results on the SQH plateau transition instead give $\theta \simeq 1/8$ \cite{Evers2003,Mirlin2003}.


\begin{figure*}[t!]
\centering
\includegraphics[width=.834\textwidth]{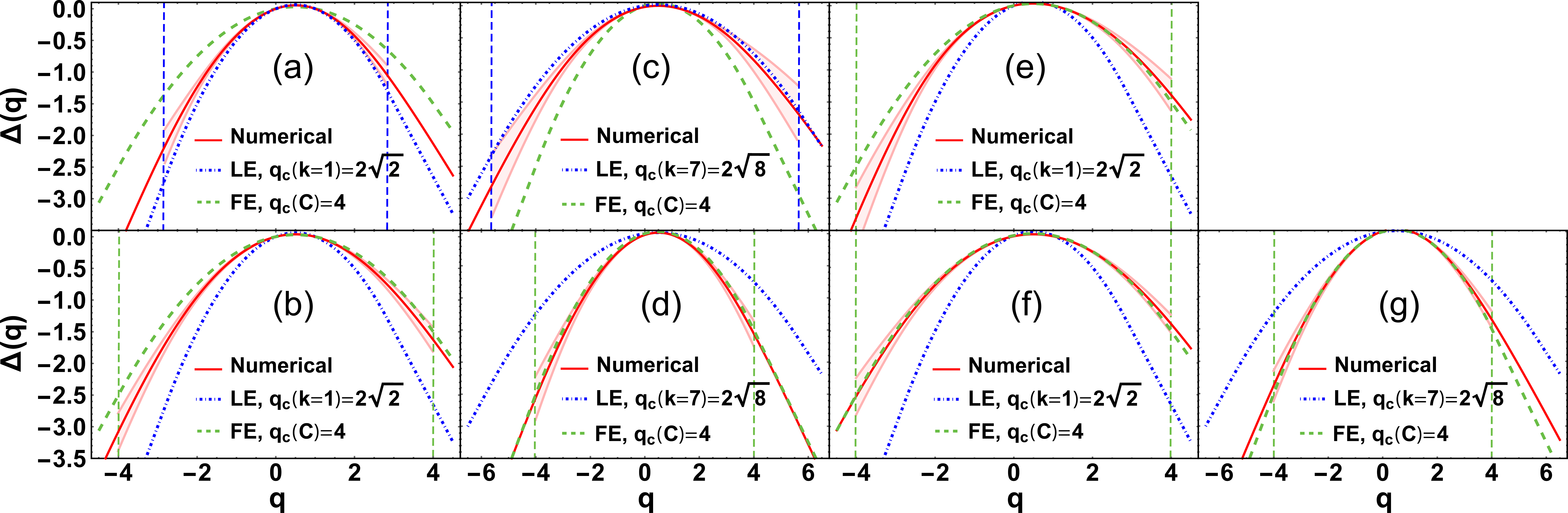}
\caption{Anomalous part of the multifractal spectrum 
$\Delta(q) \equiv \tau(q) - 2 (q-1)$ for low-energy (a,c) 
and 
finite-energy (b,d) states, for the 
TSC surface state model
in Eq.~(\ref{hsDef}) with $k = 1$ (a,b) and $k = 7$ (c,d).
The solid red curves obtain from momentum-space exact diagonalization
\cite{Chou2014}. 
The blue dot-dashed curve (green dashed curve) is the $k$-dependent (independent) 
class CI (class C SQH plateau transition) prediction. 
The solid red curve in each panel is obtained by averaging over states within 
a narrow energy bin (see text); the shaded red region 
indicates the standard deviation within the bin. 
Panel (e) shows the same for low-energy states when time-reversal symmetry is broken
explicitly, while spin SU(2) (particle-hole) symmetry is preserved. 
The disorder strength $\lambda = 1.6\pi$ ($16 \pi$) for $k = 1$ ($7$). 
Parameters for (e) are specified in \cite{Supp}.  
The system is a $(2N+1)\times(2N+1)$ momentum grid;
$N = 40$ for (a)--(e). Panels (f,g) are the same as (b,d), but
for a larger system size ($N = 46$). 
} 
\label{Fig--RepMFCs}
\end{figure*}

\begin{figure}
\includegraphics[width=0.41\textwidth]{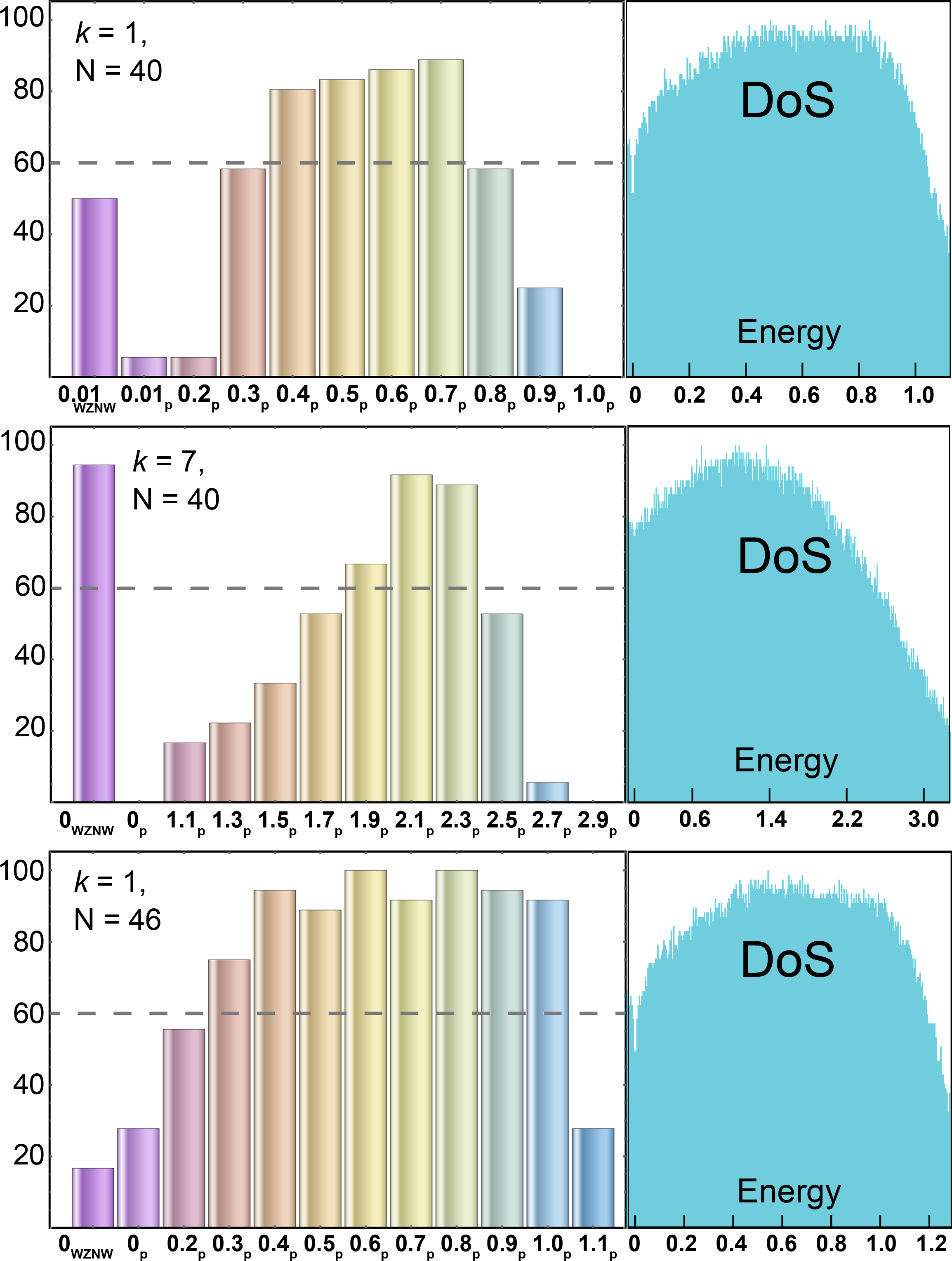}
\caption{
Population statistics for critically delocalized eigenstates. The bars in each graph give the percentage
of states with consecutive energy eigenvalues lying within a narrow energy bin that match a certain fitness criterion.
The bar labeled ``$\EE_{\mathsf{WZNW}}$'' (``$\EE_\mathsf{P}$'') denotes the percentage of 
eigenstates in the bin beginning with energy $\EE$ that match the class CI (class C) prediction for the multifractal spectum (see text for details). 
The bar energy labels should be compared to the corresponding density of states (DoS).
The top two plots have $N = 40$, while the bottom has $N = 46$. 
In the latter case even the lowest energy bin has more class C than class CI states.
The disorder strength $\lambda = 1.6 \pi$ ($16\pi$) for $k = 1$ ($7$).
}
\label{Fig--Bars}
\end{figure}

\textit{Numerical Results}.---In Fig.~\ref{Fig--RepMFCs}, we plot the anomalous part of the multifractal spectrum
$\Delta(q) \equiv \tau(q) - 2(q-1)$ for $k = 1$ (a,b) and $k = 7$ (c,d).
The class CI and class C (percolation) analytical predictions are respectively depicted as blue dot-dashed and green dashed lines.
In Figs.~\ref{Fig--RepMFCs}(a,c), we plot the numerical result for the low-energy states of the spectrum,
which show good agreement with the $k$-dependent class CI prediction. 
Calculations are performed for a typical realization of the random phase disorder, without disorder-averaging,
over a square grid of momenta.   
The solid red line in each panel is obtained by averaging
over a narrow energy bin of 36 consecutive low-energy states. For $k = 7$ these correspond to the lowest
positive energies in the spectrum, while for $k = 1$ we neglect states very close to zero energy, keeping
those in the energy bin (0.01-0.0141) 
(see Fig.~\ref{Fig--Bars} for the numerical density of states versus energy). 
The average plus or minus the standard deviation is indicated by the light red shaded region in each panel.
We plot the deviation only for $|q| < q_c$, where $q_c$ is the termination threshold for the low-energy
class CI prediction (a,c) or finite-energy SQHE class C prediction (b,d). 
Since the spectrum becomes linear outside of this range, the error in $\Delta(q)$ 
also grows linearly for $|q| > q_c$, but 
only the slope discrepancy near $q = \pm q_c$ is meaningful.  

Figs.~\ref{Fig--RepMFCs}(b,d) show the 
results for finite-energy states. 
For both $k = 1$ (b) and $k = 7$ (d),
the solid red curve in each panel agrees well with
the class C prediction (dashed green curve). 
The finite-energy bin for each $k$ is selected as the one with the highest percentage of states matching the
spin quantum Hall prediction, as indicated by a certain fitness criterion described below.
Figs.~\ref{Fig--RepMFCs}(f,g) are the same as Figs.~\ref{Fig--RepMFCs}(b,d), but for a larger system size.

Fig.~\ref{Fig--RepMFCs}(e) shows the low-energy spectrum of the $k = 1$ model, but now
with time-reversal symmetry broken explicitly. This is obtained by turning on random mass and 
nonabelian potential terms with vanishing average value, but nonzero variance \cite{Kagalovsky1999,Supp}.
This preserves spin SU(2) symmetry.
A nonzero average mass corresponds to a ``spin Hall Chern insulator''; tuning this to zero
while retaining a nonzero variance was expected to give the SQHE plateau transition \cite{Ryu2009,Ryu2012}.
Fig.~\ref{Fig--RepMFCs}(e) matches the states in Figs.~\ref{Fig--RepMFCs}(b,d,f,g).

In Fig.~\ref{Fig--Bars}, 
we compare the computed $\tau(q)$ spectrum for every state in regularly spaced
energy bins to the class CI and C predictions, for $k = \{1,7\}$. 
We introduce a ``fitness'' criteria, 
defined as follows. For each eigenstate $\psi(\vex{r})$, we compute the error between the numerical spectrum 
[$\equiv \tau_N(q)$] and the appropriate analytical prediction [$\equiv \tau_A(q)$], 
error$(q) \equiv |\tau_N(q)-\tau_A(q)|/\tau_A(q)$. 
If the error is less than or equal to 6\% for 75\% of the evaluated $q$-points in the interval $0 \leq q \leq q_c$, we keep the state. 
We consider bins of 36 states each; the states within each bin have consecutive eigenenergies.  
The height of each bar marked ``$\EE_{\mathsf{WZNW}}$'' (``$\EE_\mathsf{P}$'') denotes the percentage of 
eigenstates in the bin starting with energy $\EE$ that match the class CI 
(class C) prediction. The energies in each left panel of Fig.~\ref{Fig--Bars} should 
be compared to the numerical density of states shown in the corresponding right panel.

\begin{figure}[b!]
\centering
\includegraphics[width=.35\textwidth]{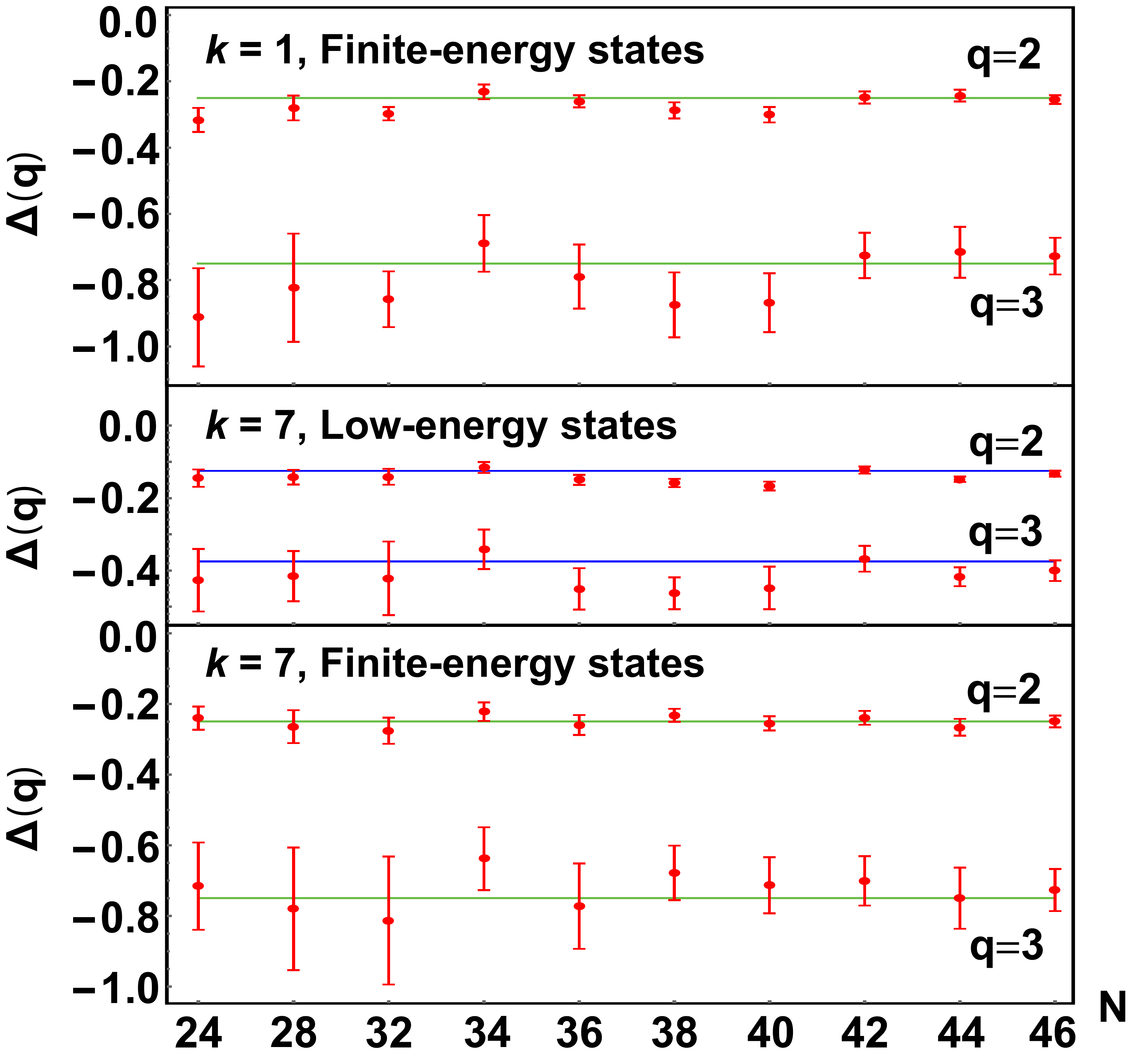}
\caption{Finite-energy and low-energy $\Delta(q)$ as in Fig.~\ref{Fig--RepMFCs}, 
but for fixed $q = 2,3$ and for varying system sizes $N$.  
The intrinsic disorder correlation length and strength are kept fixed \cite{Supp}. 
The blue and green lines are the exact analytical predictions for $\Delta(2,3)$ \cite{Foster2012,Mirlin2003}.
The solid points show the average, while error bars indicate the standard deviation within the energy bin. 
The main effect of increasing $N$ is to reduce the fluctuations, although the reduction is slower for $q$ closer to
$q_c$ ($=4$ for class C). 
See \cite{Supp} for full $\Delta(q)$ and population statistics.
} 
\label{Fig--FiniteSize}
\end{figure}

Fig.~\ref{Fig--Bars} indicates that
for $k = 1$ the finite energy states match well the class C SQH prediction 
(bins $\EE_{\mathsf{P}}$ with $\EE \in \{0.4,\ldots,0.7\}$ for $N = 40$).
The plot for $k = 7$ shows a narrower band of finite energy states that match class C
for the chosen disorder strength;
results for $k = 8$ are similar \cite{Supp}. 
Results for a larger $k = 1$ system ($N = 46$) appear at the bottom of Fig.~\ref{Fig--Bars},
showing that the 
statistics at finite energy
\emph{improve} with
increasing system size.

\begin{figure}[t!]
\centering
\includegraphics[width=.48\textwidth]{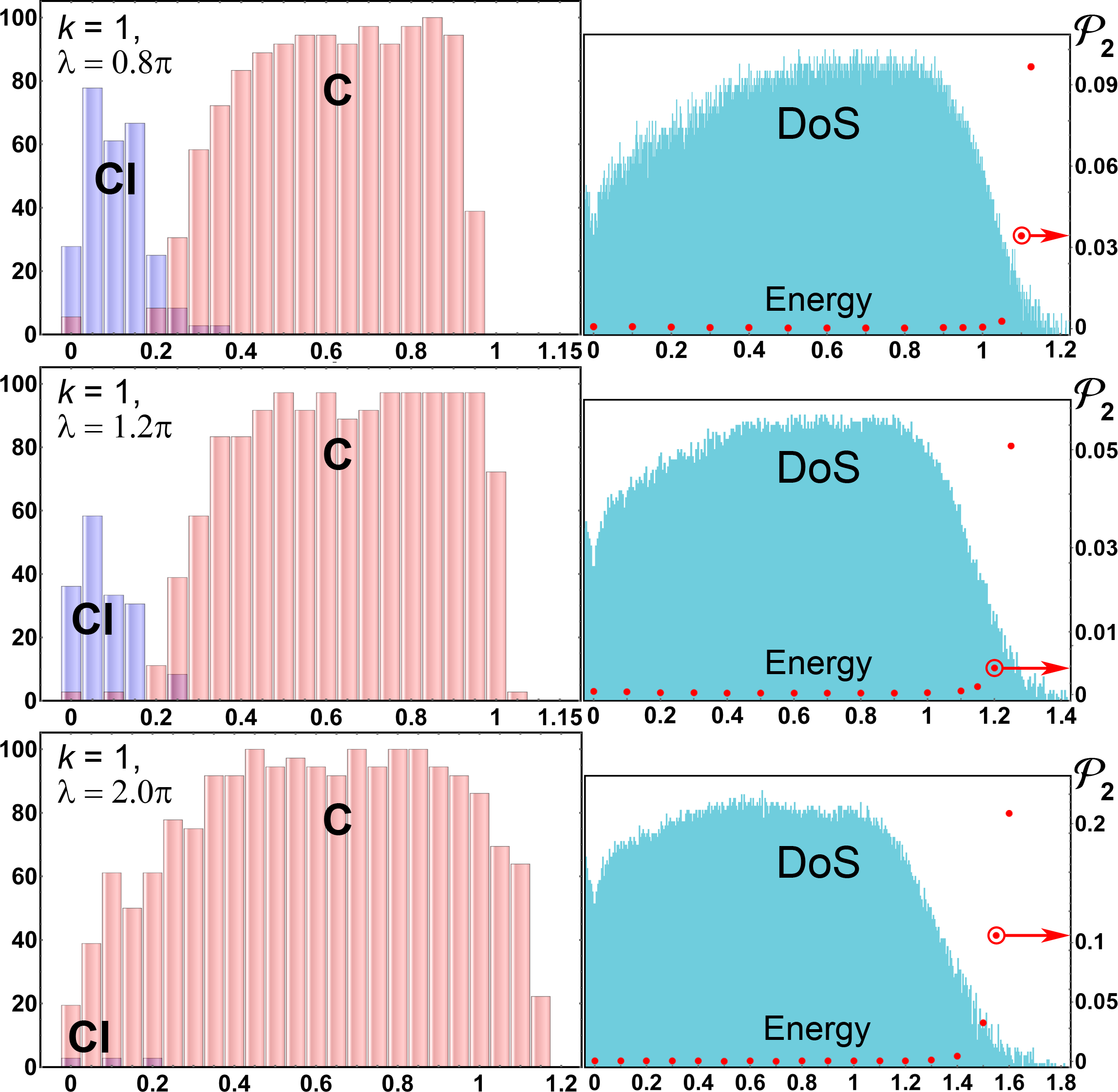}
\caption{Disorder strength ($\lambda$)-dependence of the population statistics for $k = 1$. 
Blue (red) bars marked CI (C) describe the percentage of states that match the class CI (C)
multifractal spectrum; results are shown for $\lambda \in \{0.8,1.2,2.0\}\pi$ and $N = 46$. 
$\lambda = 2 \pi$ corresponds to strong disorder 
(the freezing transition for the abelian AIII model \cite{Ludwig1994,Chou2014,Chamon1996}).
The main effect of increasing $\lambda$ is to decrease the percentage of CI-like states. 
The second IPR $\mathcal{P}_2$ (red dots) is superimposed over the 
associated 
DoS in the 
right-hand plots. Even for strong disorder, this becomes appreciable only 
in the high-energy ``Lifshitz tail,'' possibly signaling localization there \cite{LocNote}.} 
\label{Fig--DirtS}
\end{figure}

The finite-energy results for $\tau(q)$ shown in Figs.~\ref{Fig--RepMFCs}(b,d,f,g)
were obtained from the bin with the highest percentage of states matching the class C SQH prediction
for each value of $k$. I.e., for $k = 7$, the red solid curve in Fig.~\ref{Fig--RepMFCs}(d) was
obtained by averaging over states in the bin starting with energy $\EE = 2.1$ (Fig.~\ref{Fig--Bars}). 
We emphasize that while the fitness criterion introduced above is arbitrary, the trends are not \cite{Supp}. 

We show finite-size trends in Fig.~\ref{Fig--FiniteSize}. Here we plot $\Delta(q)$ for $q = 2,3$, which are well-distinguished for 
the class CI (low-energy) and class C (finite-energy) analytical predictions \cite{Foster2012,Mirlin2003}. 
The trends for increasing $N$ suggest convergence towards the analytical results.

Population statistics (computed as in Fig.~\ref{Fig--Bars}) for variable disorder are shown 
in Fig.~\ref{Fig--DirtS}. 
Stronger disorder converts more of the low-energy spectrum from class CI to class C in a fixed system size. 
We also plot the second inverse participation ratio $\mathcal{P}_2$, which becomes
appreciable only in the high-energy tail. 
Since $\mathcal{P}_2 \sim \zeta^{-2}$ for a state with localization length $\zeta$
\cite{Huckestein1995,Evers2008}, localization 
(if it occurs \cite{LocNote}) is restricted to high energies, and 
does not encroach upon the wide ``multifractal spin metal'' class C region.


\textit{Conclusion}.---Our results
are completely different from
a 2D system with delocalization at only one energy, as
in the quantum Hall plateau transition \cite{Huckestein1995}. 
Although localized 
states arbitrarily 
close to that transition 
appear critical on large scales, the criticality reflects
the transition itself. By contrast, here we find
(1) the expected class CI criticality near zero energy, 
and 
(2) a robust, \emph{completely different, non-standard} class C criticality over
a wide swath of finite energy states. 
This swath does not shrink as the disorder strength or system size is increased. 

Although we observe ``localization'' \cite{LocNote} 
in the high-energy tail (Fig.~\ref{Fig--DirtS}), we believe
that this is an artifact of the unphysical UV truncation of the TSC surface
Hamiltonian. A real 3D TSC would hybridize 2D surface states with the bulk
when the surface energy crosses the bulk gap. We expect that \emph{all} states
in this case are delocalized, and that all of the 2D surface states below
the gap show class C criticality (except the one at zero energy).

The class CI zero energy state is described by a  
Wess-Zumino-Novikov-Witten (WZNW) nonlinear sigma model \cite{Foster2012,Foster2014}.
The energy perturbation breaks the symmetry from $G \times G$ down to $G$, where $G = \text{Sp}(4n)$ (using replicas,
with $n\rightarrow0$). 
This relevant perturbation presumably
induces a renormalization group (RG) flow
to another sigma model with lower symmetry. This argument is insufficient to 
choose between 
the orthogonal metal 
class AI
[manifold $\text{Sp}(4n)/\text{Sp}(2n)\times\text{Sp}(2n)$] 
and 
class C 
[manifold $\text{Sp}(4n)/\text{U}(2n)$] 
\cite{Evers2008}. 
If the class CI model is deformed to class C ``by hand,'' the WZNW term becomes a theta term 
(with theta proportional to $k$) \cite{Supp}. 

We thank Ilya Gruzberg, Victor Gurarie, Lucas Wagner, Robert Konik, Alexei Tsvelik,
Antonello Scardicchio, Vadim Oganesyan, and Sarang Gopalakrishnan
for useful discussions. 
This research was supported by the Welch Foundation grants No.~E-1146 (S.A.A.G.) and No.~C-1809 (Y.L. and M.S.F.), 
as well as NSF CAREER Grant No.~DMR-1552327 (Y.L. and M.S.F.), and by the Texas center for the superconductivity (S.A.A.G.).
M.S.F. thanks the Aspen Center for Physics, which is supported
by the NSF Grant No.~PHY-1066293, for its hospitality while part of this work was performed.


\newpage \clearpage

\onecolumngrid

\begin{center}
	{\large
	Critical Percolation Without Fine Tuning on the Surface of a Topological Superconductor
	\vspace{4pt}
	\\
	SUPPLEMENTAL MATERIAL
	}
\end{center}
\makeatletter
\setcounter{equation}{0}
\setcounter{figure}{0}
\setcounter{table}{0}
\setcounter{page}{1}
\renewcommand{\theequation}{S\arabic{equation}}
\renewcommand{\thefigure}{S\arabic{figure}}
\renewcommand{\bibnumfmt}[1]{[S#1]}
\renewcommand{\citenumfont}[1]{S#1}

\begingroup
\hypersetup{linkbordercolor=white}
\tableofcontents
\endgroup

\section{{\bf I.\ Particle-hole symmetry for odd and even $k$}}

The Hamiltonian $\hs$ can be succinctly encoded by introducing
an additional species of Pauli matrices $\sigh^{1,2,3}$ as a basis 
for the decomposition in Eq.~(1). Then
\begin{align}\label{S--hs}
	\hs 
	= 
	\frac{1}{2}
	\left[
	\sigh^+
	\left(-i \parr\right)^k
	+
	\sigh^-
	\left(-i \parb\right)^k
	\right]
	+ 
	\sigb \cdot \vex{A}^a(\vex{r}) \, \tauh^a
	+
	\sigb \cdot \vex{A}^0(\vex{r}),
\end{align}
where $\sigh^{\pm} \equiv \sigh^1 \pm i \sigh^2$
and $\sigb \cdot \vex{A} = \sigh^1 A_x + \sigh^2 A_y$. 
Physical time-reversal invariance is encoded in the chiral condition \cite{S--SRFL2008,S--Foster2012,S--Foster2014}
\begin{align}\label{S--ChDef}
	- \Ms \, \hs \, \Ms = \hs, \qquad \Ms = \sigh^3.
\end{align}

For class CI, particle-hole symmetry $P$ must involve an antisymmetric matrix $\Mp = - \Mp^\T$
(``$P^2 = -1$'') \cite{S--SRFL2008}. The particle-hole condition is  
\begin{align}\label{S--PDef}
	- \Mp \, \hs^\T \, \Mp = \hs,
\end{align}
where $\T$ denotes the transpose operation and it is understood that derivative operators are odd
under transposition: $(\partial_{x,y})^\T = - \partial_{x,y}$. 

For $k \in \{1,3,5,\ldots\}$ (odd), we take \cite{S--Foster2012,S--Foster2014}
\begin{align}\label{S--Podd}
	\Mp^{\pupsf{odd}} = \sigh^1 \tauh^2.
\end{align}
Eq.~(\ref{S--hs}) is invariant under Eq.~(\ref{S--PDef}) with this choice provided we set the abelian
vector potential equal to zero, $\vex{A}^0 = 0$.
For $k \in \{2,4,6,\ldots\}$ (even), we take 
\begin{align}\label{S--Peven}
	\Mp^{\pupsf{even}} = \sigh^2 \tauh^1.
\end{align}
Eq.~(\ref{S--hs}) is invariant under Eq.~(\ref{S--PDef}) with this choice provided we set the 
third component of the SU(2) vector potential equal to zero, $\vex{A}^3 = 0$.


\section{{\bf II.\ Class C model with broken TRI}}

In Fig.~1(e), we exhibit the low-energy anomalous multifractal spectrum $\Delta(q) = \tau(q) - 2(q-1)$ for 
the $k = 1$ class CI model in Eq.~(\ref{S--hs}), except that we have now explicitly broken time-reversal symmetry
(in every fixed realization of disorder, but not on average). 
The Hamiltonian in this case resides in class C \cite{S--Kagalovsky1999},
\begin{align}\label{S--hCDef}
	\hat{h}_{\scriptscriptstyle{\mathrm{C}}}
	\equiv
	\hs
	+
	m(\vex{r}) \, \sigh^3
	+
	v^a(\vex{r}) \, \tauh^a,
\end{align}
where $a \in \{1,2,3\}$. 
The mass $m(\vex{r})$ and nonabelian valley potentials $v^{1,2,3}(\vex{r})$ break physical
time-reversal symmetry [Eq.~(\ref{S--ChDef})], but preserve $P^2 = -1$ particle-hole 
[Eqs.~(\ref{S--PDef}) and (\ref{S--Podd}) for $k = 1$]. The latter is tantamount to spin SU(2) invariance.


\section{{\bf III.\ Parameter specification for the numerics}}

The numerical results presented in Figs.~1--4 were obtained via the exact diagonalization of $\hs$ in Eq.~(1).
Calculations are performed in momentum space. 
All disorder potentials are parameterized as described in the text,
i.e.\ 
\begin{align}
	A_i^\mu(\vex{q}) = \frac{\sqrt{\lambda}}{L} \exp\left[i \, \theta_i^\mu(\vex{q}) - \frac{q^2 \xi^2}{4}\right],
\end{align} 
where $i \in \{x,y\}$ and $\mu \in\{0,1,2,3\}$. 
The phases $\theta_i^\mu(\vex{q}) = - \theta_i^\mu(-\vex{q})$, but are otherwise identical, independent
random variables uniformly distributed over $[0,2\pi)$. 
This approach is equivalent to 
disorder-averaging, up to finite-size corrections \cite{S--Chou2014}.
We assign the same disorder strength $\lambda$ to all nonzero components of the abelian $\Ac^0$
and nonabelian $\Ac^{1,2,3}$ vector potentials. 
We choose $\lambda = 1.6 \pi$ ($16 \pi$) for the $k = 1$ ($k = 7$) calculations with $N = 40$
[except Fig.~1(e), described below]. 
Different disorder strengths are studied in Fig.~4 (with $N = 46$), as described in the caption to that figure. 

The arbitrary dimensionful system length $L$ determines the ultraviolet cutoff $\Lambda = 2 \pi N/L$.
The correlation length of the impurity potentials is chosen to be $\xi = 0.25 (L/N)$ \cite{S--Chou2014}.  
For $N \neq 40 \equiv N_0$ in Figs.~1--3, we rescale the disorder parameter $\lambda \rightarrow (N / N_0)^{2(k - 1)} \lambda$, 
which corresponds to fixing the intrinsic disorder strength relative to the appropriate power of the momentum cutoff. 
There is no rescaling in the $k = 1$ (Dirac dispersion) case, where $\lambda$ is dimensionless. 

The multifractal spectra exhibited in 
Figs.~1(a,b,e) and (c,d) 
are extracted using box sizes 
$b = 2$ and $b = 8$  ($k = 1$)
and
$b = 3$ and $b = 10$ ($k = 7$),
respectively. 
Box sizes for Figs.~1(f,g) are 
$b = 2$ and $b = 13$  ($k = 1$)
and
$b = 2$ and $b = 31$ ($k = 7$).

For the low-energy states in class C with $k = 1$ [Fig.~1(e)], 
in addition to the nonabelian vector potentials $A_{x,y}^{1,2,3}$, 
we include random mass $m(\vex{r})$ and 
(valley-graded) scalar potential $v^3(\vex{r})$ disorders, see Eq.~(\ref{S--hCDef}). 
The disorder strengths are $\lambda_A = \lambda_{v^3} = 0.8 \pi$, while the mass variance
is $\lambda_m = 16 \pi$. We neglect the off-diagonal potentials $v^{1,2}(\vex{r}) = 0$ 
[Eq.~(\ref{S--hCDef})].


\section{{\bf IV.\ Sigma models for class CI TSC surface states and the class C SQHE}}

The nonlinear sigma model representations 
for 
(a) 
the class CI conformal field theory
describing zero-energy TSC surface state wave functions 
and 
(b)
the class C spin quantum Hall effect are captured by the following 
actions \cite{S--Senthil1998,S--Foster2014,S--Liao2017}, 
\begin{subequations}
\begin{align}
	\sci 
	=&\, 
	\frac{k}{8 \pi} 
	\int d^2\vex{r} \Tr\left[\Nabla \hat{q}^\dagger \cdot \Nabla \hat{q}\right]
	+
	i 
	\,
	\frac{\omega}{2}
	\int d^2\vex{r} \Tr\left[\hat{\Lambda} \left(\hat{q} + \hat{q}^\dagger\right)\right]
	-
	i 
	\frac{k}{12 \pi} 
	\int d^2\vex{r}
	\,
	d R 
	\,
	\epsilon^{a b c}
	\Tr\left[
		\left(\hat{q}^\dagger \partial_a \hat{q} \right)
		\left(\hat{q}^\dagger \partial_b \hat{q} \right)
		\left(\hat{q}^\dagger \partial_c \hat{q} \right)
	\right],
\label{S--CI}
\\
	\sct 
	=&\, 
	\frac{\sigma_{11}}{8} 
	\int d^2\vex{r} \Tr\left[\Nabla \hat{q} \cdot \Nabla \hat{q}\right]
	+
	i \, \omega
	\int d^2\vex{r} \Tr\left[\hat{\Lambda} \, \hat{q} \right]
	- 
	\frac{\sigma_{12}}{8} 
	\int d^2\vex{r}
	\,
	\epsilon^{i j}
	\Tr\left[\hat{q} \, \partial_i \, \hat{q} \, \partial_j \, \hat{q}\right].
\label{S--C}
\end{align}
\end{subequations}
In Eq.~(\ref{S--CI}), $\hat{q}(\vex{r})$ is a $4n\times4n$ unitary matrix field that is also 
a $\text{Sp}(4 n)$ group element; $\vex{r} = \{x,y\}$ is the position vector 
that spans over the 2D TSC surface.  
We assume that disorder-averaging has been accomplished with the replica trick;
$n \rightarrow 0$ counts the number of replicas \cite{S--Altland}. 
Here and in Eq.~(\ref{S--C}), $\omega$ is a real parameter that denotes the ac 
frequency of the spin conductivity that the sigma model is designed to compute \cite{S--Liao2017,S--Altland}.
The matrix $\hat{\Lambda}$ is diagonal and equal to 
$\hat{\Lambda} = \mathrm{diag}(1,1,\ldots,1,-1,-1,\ldots,-1)$ [$2n$ (+1)s and $2n$ (-1)s]. 
The last term in Eq.~(\ref{S--CI}) is the Wess-Zumino-Novikov-Witten (WZNW) term. This term
is defined over a three-dimensional ball (coordinates $\{\vex{r},R\}$, $0 \leq R \leq 1$), such that 
$\hat{q}(\vex{r},R = 1) = \hat{q}(\vex{r})$ is the field on the 2D surface, 
while $\hat{q}(\vex{r},R)$ for $R < 1$ is a smooth deformation of this into the ball interior. 
The WZNW term ensures that the action in Eq.~(\ref{S--CI}) is conformally invariant for $\omega = 0$ 
\cite{S--Foster2014,S--BYB}. In the context of a 3D topological superconductor,
the ball can be identified with the bulk if the surface has genus zero \cite{S--Konig2012}. 

The structure of the first two terms of the class C sigma model can be obtained from the corresponding ones 
in class CI by imposing the additional constraint on $\hat{q}(\vex{r})$ by hand, 
\begin{align}\label{S--Constraint}
	\hat{q}^\dagger(\vex{r}) = \hat{q}(\vex{r}).
\end{align}
The matrix $\hat{q}(\vex{r})$ now belongs to the space $\text{Sp}(4n)/\text{U}(2n)$ 
\cite{S--Senthil1998,S--Evers2008}.
The last term in Eq.~(\ref{S--C}) is the Pruisken or theta term, which assigns
winding numbers to different topological sectors of the $\hat{q}$-field and evaluates
to a pure imaginary phase \cite{S--Altland}. 
The coefficients of the first and third terms in Eq.~(\ref{S--C})
are respectively proportional to the longitudinal conductivity $\sigma_{11}$ and Hall 
conductivity $\sigma_{12}$; the class CI WZNW model has $\sigma_{11} = k/\pi$
and $\sigma_{12} = 0$ (in units of the spin conductance quantum) \cite{S--SRFL2008,S--Foster2014}.  

When the constraint in Eq.~(\ref{S--Constraint}) is imposed on $\hat{q}(\vex{r})$ in 
Eq.~(\ref{S--CI}), we can show that the WZNW term in the latter becomes the theta
term in Eq.~(\ref{S--C}), with 
\begin{align}\label{S--sigma12}
	\sigma_{12} = k/2.
\end{align}
However, we stress that in the context of the finite-energy TSC surface states
discussed in this Letter, there is no reason to trust this assignment of $\sigma_{12}$,
because it does not follow from a physical RG flow. As emphasized at the end of the main
text, the only statement we can make is that the energy perturbation 
(the operator coupling to $\omega \neq 0$) is relevant, and breaks the symmetry of
the class CI model from $G \times G$ down to $G$, where 
$G = \text{Sp}(4n)$ \cite{S--Footnote}.

If $\omega \neq 0$ induces a flow to the class C SQHE plateau transition, 
it is must be due to the effect of the WZNW term on the RG. Without the 
WZNW term, the class CI model describes (gapless) quasiparticles in an ordinary 
(non-topological) spin-singlet superconductor, and the single-particle wave functions
must become those of the orthogonal metal class AI at large energies. 
Moreover, in this case all states (at zero and finite energy) are localized
in 2D for arbitrarily weak disorder \cite{S--Senthil1998,S--Evers2008}.

\subsection{A. From WZNW to theta}

The derivation of Eq.~(\ref{S--sigma12}) requires a little care, since the 
restriction in Eq.~(\ref{S--Constraint}) induces topologically distinct sectors
of the $\hat{q}$-matrix; this is why the theta term in Eq.~(\ref{S--C}) can produce
an effect. It means that the field $\hat{q}(\vex{r},R)$ appearing in the WZNW
term of Eq.~(\ref{S--CI}) cannot be strictly restricted in this way, because
otherwise it is not possible to deform generic $\hat{q}(\vex{r},R < 1)$ in the interior
to some particular $\hat{q}(\vex{r},R = 1)$ at the surface. 
Here we show how the WZNW term in Eq.~(\ref{S--CI})
becomes the theta term in Eq.~(\ref{S--C}), employing the method used in Ref.~\cite{S--Bocquet2000}. 

We extend $\hat{q}(\vex{r})$ on the 2D surface to $\hat{q}(\vex{r},R)$ on the three-dimensional ball through the following equation:
\begin{align}\label{S--extension}
\begin{aligned}
	\hat{q}(\vex{r},R) 
	=\, &
	-i \cos \left( \frac{\pi R}{2}\right) \hat{1}
	+
	\sin \left( \frac{\pi R}{2}\right) \hat{q}(\vex{r}),
	\qquad 
	0 \leq R \leq 1.
\end{aligned}
\end{align}
In this extension scheme, we have $\hat{q}(\vex{r},R=0) = -i \hat{1}$ (where $\hat{1}$ is the identity)
and $\hat{q}(\vex{r},R=1)=\hat{q}(\vex{r})$.
Note that, unlike $\hat{q}(\vex{r})$ on the surface, $\hat{q}(\vex{r},R)$ does not obey the restriction in Eq.~(\ref{S--Constraint}).
 
Inserting Eq.~(\ref{S--extension}) into the WZNW term of Eq.~(\ref{S--CI}), the latter reduces to
\begin{align}\label{S--Gamma}
\begin{aligned}
	&- \frac{k}{8 }
	\int d^2\vex{r}
	\,
	d R 
	\,
	\epsilon^{abc}
	\Tr 
	\left[ 
	\hat{q}(\vex{r})
	\partial_b \hat{q}(\vex{r})
	\partial_c \hat{q}(\vex{r})
	\right] 
	\sin^2 \left( \frac{ \pi R}{2}\right) 
	\partial_a R 
	\\
	=\,&
	-\frac{k}{8}
	\int d^2\vex{r}
	\,
	d R 
	\,
	\epsilon^{a b c}
	\Tr 
	\left\lbrace 
	\partial_a
	\left[ 
	\hat{q}(\vex{r})
	\partial_b \hat{q}(\vex{r})	
	\partial_c \hat{q}(\vex{r})	
	\left( \frac{R}{2}-\frac{1}{2\pi} \sin (\pi R) \right)
	\right] 
	\right\rbrace 
	\\
	=\,&
	-\frac{k}{16}
	\int d^2\vex{r}
	\,
	\epsilon^{ij}
	\Tr 
	\left[ 
	\hat{q}(\vex{r}) \partial_i \hat{q}(\vex{r}) \partial_j \hat{q}(\vex{r})
	\right],					
\end{aligned}
\end{align}
where in the second equality we have used the divergence theorem and the fact that $R=1$ on the 2D surface.
Comparing this expression with the theta term in Eq.~(\ref{S--C}), one obtains Eq.~(\ref{S--sigma12}).

\makeatother


\section{{\bf V.\ Additional numerical results}}

\begin{figure}
\includegraphics[width=0.8\textwidth]{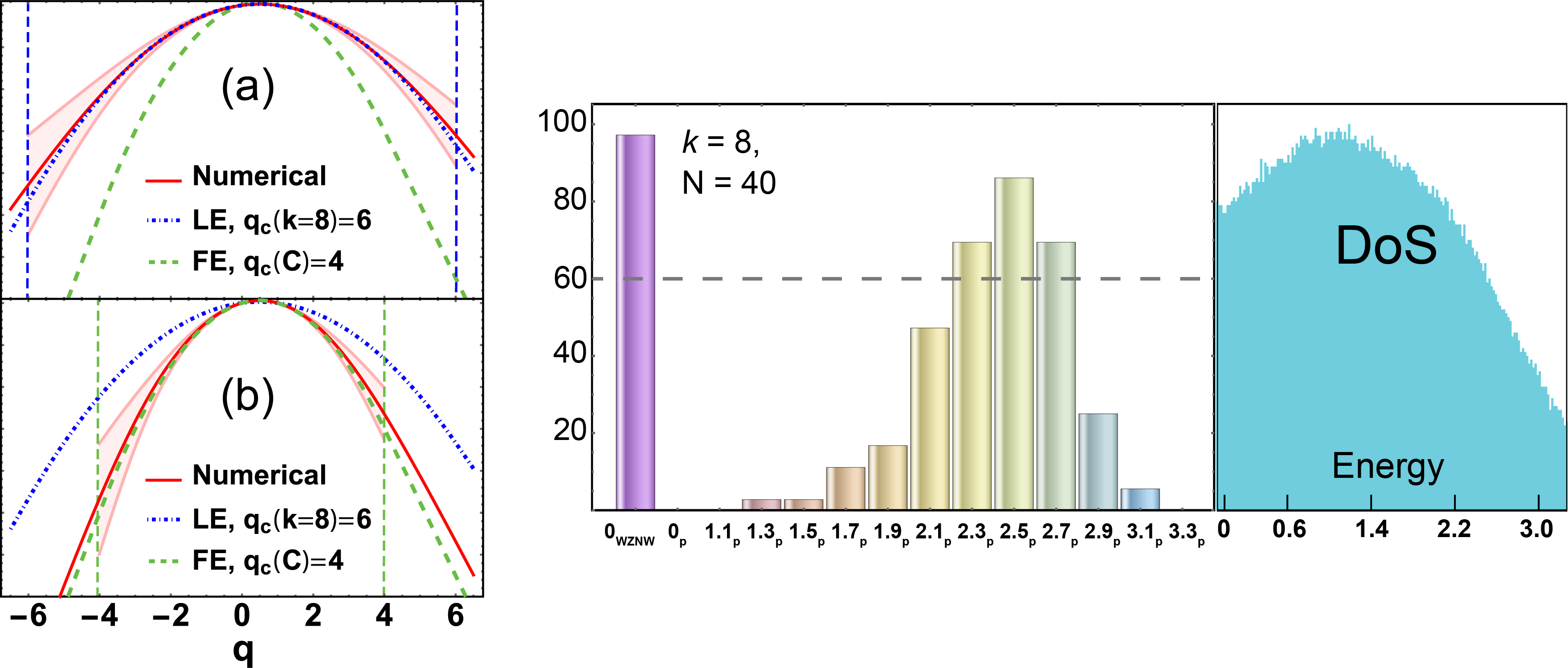}
\caption{Results for $k = 8$ and $N = 40$. Left: anomalous part of the multifractal spectrum  $\Delta(q) \equiv \tau(q) - 2 (q-1)$ as in Fig.~1 of the main text. The top panel (a) [bottom panel (b)] shows the low-energy (finite-energy) states, compared to the analytical predictions for class CI (blue dot-dashed curve) and class C (green dashed curve). 
The right panel shows the population statistics for critically delocalized eigenstates as in Fig.~2 of the main text. 
} 
\label{Fig--k=8}
\end{figure}

\subsection{A. $k = 8$ results}

Our results in this Letter generalize a previous observation for a simpler
model in class AIII. This model consists of a single 2D Dirac
fermion coupled to abelian vector potential disorder; it is critically delocalized
and exactly solvable at zero energy \cite{S--Ludwig1994}. 
It can also be interpreted as the surface state of TSC with winding number $\nu = 1$ \cite{S--SRFL2008,S--Foster2014}. 
It was claimed \cite{S--Ostrovsky2007} that all finite-energy states of this model should reside at the plateau transition 
of the (class A) integer quantum Hall effect, and this was verified numerically \cite{S--Chou2014}. 
The same logic employed by Haldane \cite{S--Haldane-EO} and Pruisken \cite{S--Pruisken-EO} 
implies that there should be an ``even-odd'' effect whereby AIII surface states
for a TSC with even winding number are localized at finite energy, while
those with odd form stacks of critical plateau-transition states \cite{S--Ostrovsky2007}.  
Here we find critical delocalization for all class CI winding numbers, with no ``even-odd'' effect. 

Results for $k = 8$ and $N = 40$ are shown in Fig.~\ref{Fig--k=8}. 
The box sizes chosen to obtain the multifractal spectrum are $b = 5$ and $b = 10$.

\subsection{B. Density of states}

\begin{figure}[b!]
\centering
\includegraphics[width=0.45\textwidth]{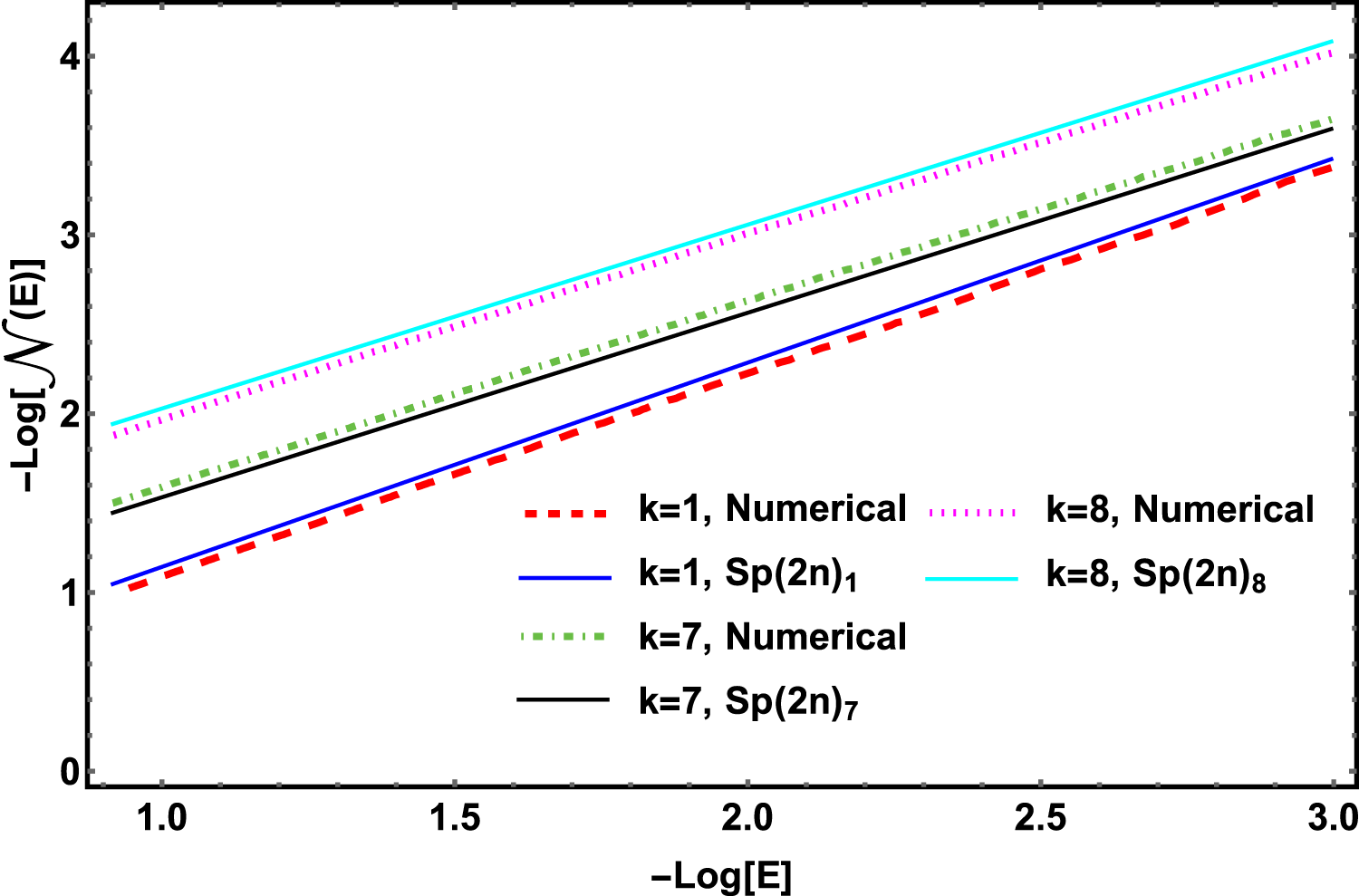}
\caption{Log-log plot of the integrated density of states $\mathcal{N}(\EE)$ obtained via exact diagonalization of $\hs$ in Eq.~(1), 
versus the exact analytical prediction of the class CI conformal field theory \cite{S--LeClair2008,S--Foster2012}.
Results are shown for $k = \{1,7,8\}$, with parameters chosen the same as in Figs.~1, 2, and \ref{Fig--k=8} ($N = 40$).} 
\label{Fig--IDOS}
\end{figure}

\begin{figure}
\includegraphics[width=0.4\textwidth]{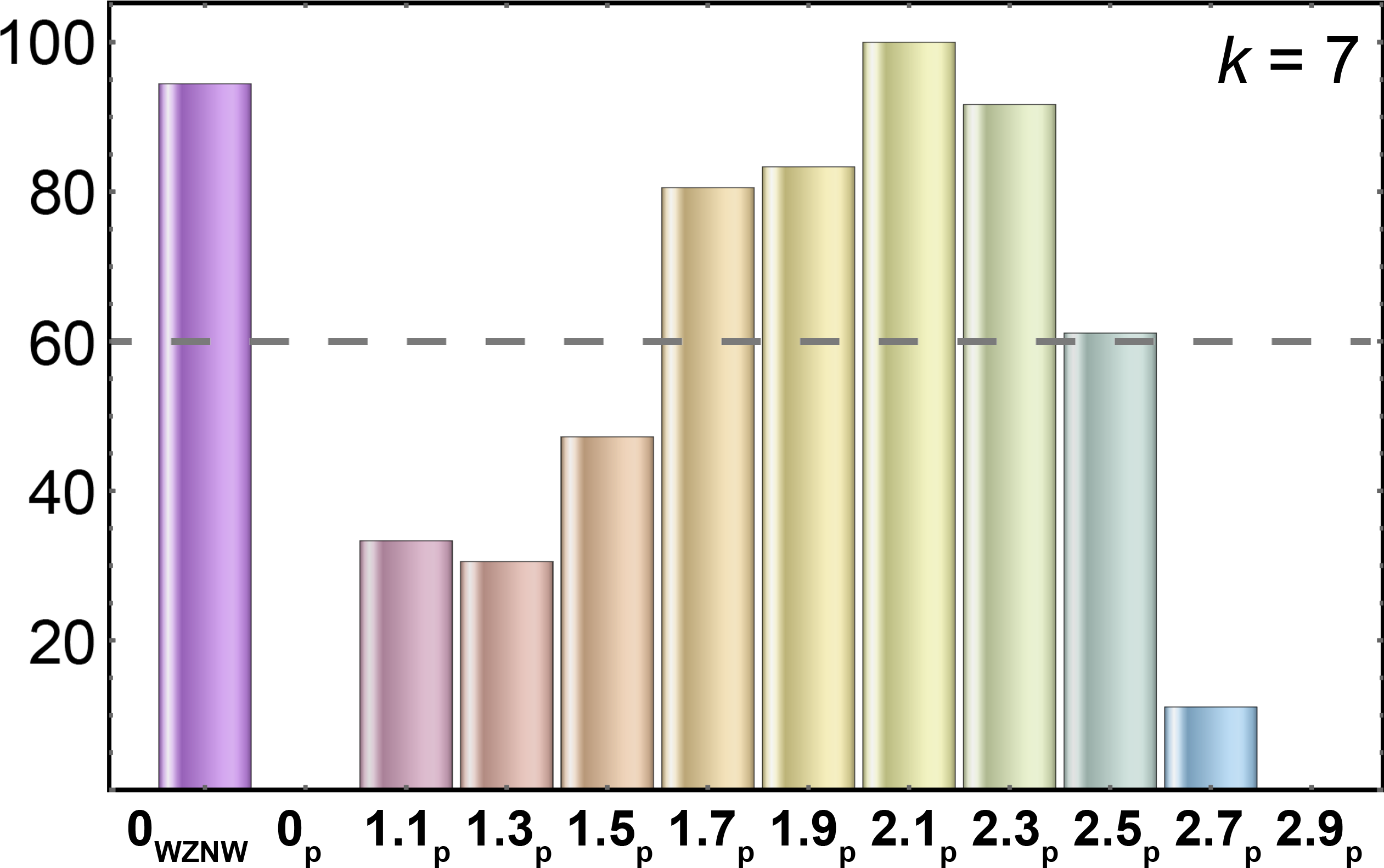}
\caption{
The same as the second left-hand panel of Fig.~2 ($k = 7$), but with a 7\% error threshold for states matching
the class CI (``WZNW'') or C (``P'') prediction for the multifractal spectrum $\tau(q)$, over the 
range $0 \leq q \leq q_c$.  
}
\label{Fig--k7_N40_7p_Bars}
\end{figure}

The numerical density of states (DoS) is depicted via the histograms in the right panels 
of Fig.~2 for $k \in \{1,7\}$ and in Fig.~\ref{Fig--k=8} for $k = 8$. 
The DoS shows a ``dip'' upon approaching zero energy $\EE = 0$. 
This is expected for the critically delocalized class CI surface states,
where the $\text{Sp}(2n)_k$ conformal field theory predicts universal scaling 
for the DoS $\nu(\EE)$ \cite{S--LeClair2008,S--Foster2012}
\begin{align}
	\nu(\EE) \sim |\EE|^{1/(4 k + 3)},
\end{align}
a result that is independent of the multifractal spectrum \cite{S--Foster2012,S--Foster2014}. 
In Fig.~\ref{Fig--IDOS}, we compare the numerical integrated density of states $\mathcal{N}(\EE) \equiv \int^\EE \nu(\EE') \, d \EE'$
to the class CI prediction, and observe good agreement for $k = \{1,7,8\}$ ($N = 40$).

\subsection{C. Alternative fitness threshold}

To obtain the population statistics exhibited in Fig.~2, we employ
an arbitrary ``fitness'' criteria, described in the main text. 
While the exact percentages of states above the fitness threshold (encoded in the bar heights in Fig.~2) 
depend somewhat sensitively on these criteria, the overall trends as a function of energy $\EE$ do not. 
In Fig.~\ref{Fig--k7_N40_7p_Bars}, we replot the same data as in $k = 7$ panel of Fig.~2, 
but for the criterion that a state is kept if it matches the appropriate analytical
prediction for $\tau(q)$ with no more than 7\% error, over 75\% or more 
of the $q$-values in the range $0 \leq q \leq q_c$.


\subsection{D. Representative class C and high-energy tail-state wave functions for $k = 1$, $N = 46$}

Here we plot representative wave function probability profiles for $k = 1$ and $N = 46$,
for two different disorder strengths $\lambda = \{1.2,2.0\}\pi$. The corresponding population statistics,
density of states, and second inverse participation ratio $\mathcal{P}_2$ are shown in Fig.~4 of the main text. 
The latter figure demonstrates that only states deep in the high-energy ``tail'' exhibit appreciable  $\mathcal{P}_2$.
We therefore expect that these states could exhibit Anderson localization. 

\begin{figure}[h!]
\subfigure[$\lambda = 1.2\pi$, Energy $0.75$]{
\includegraphics[width=0.313\textwidth]{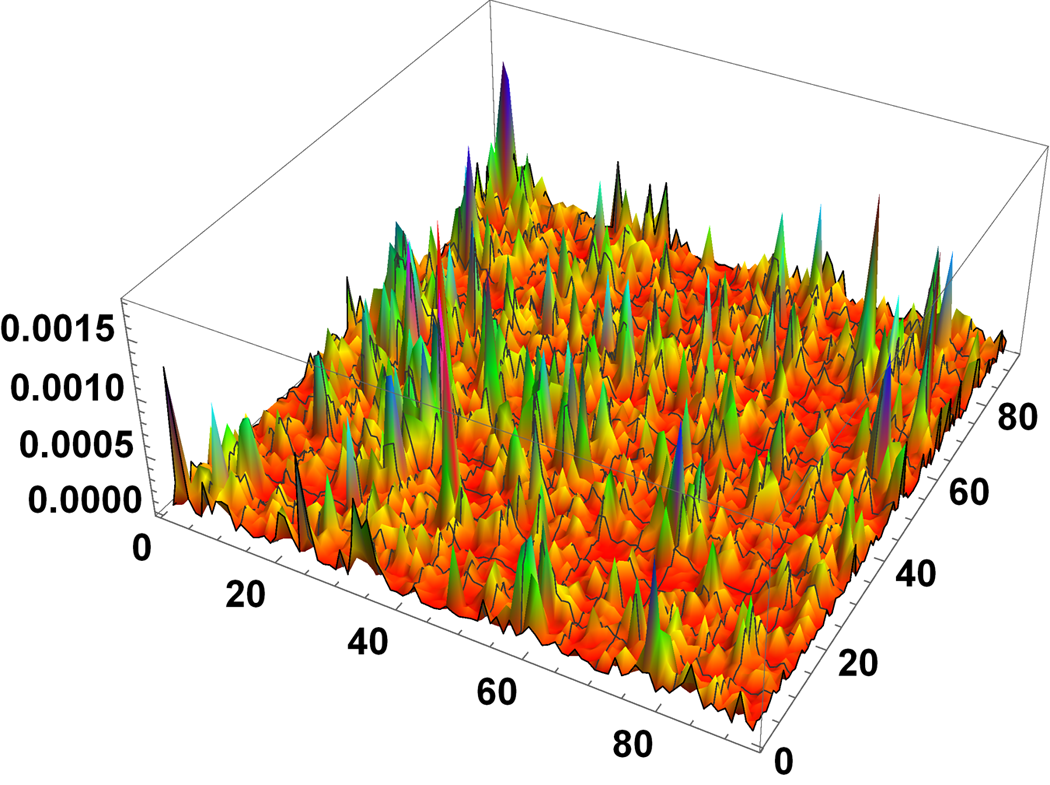}
}
\subfigure[$\lambda = 1.2\pi$, Energy $1.15$]{
\includegraphics[width=0.313\textwidth]{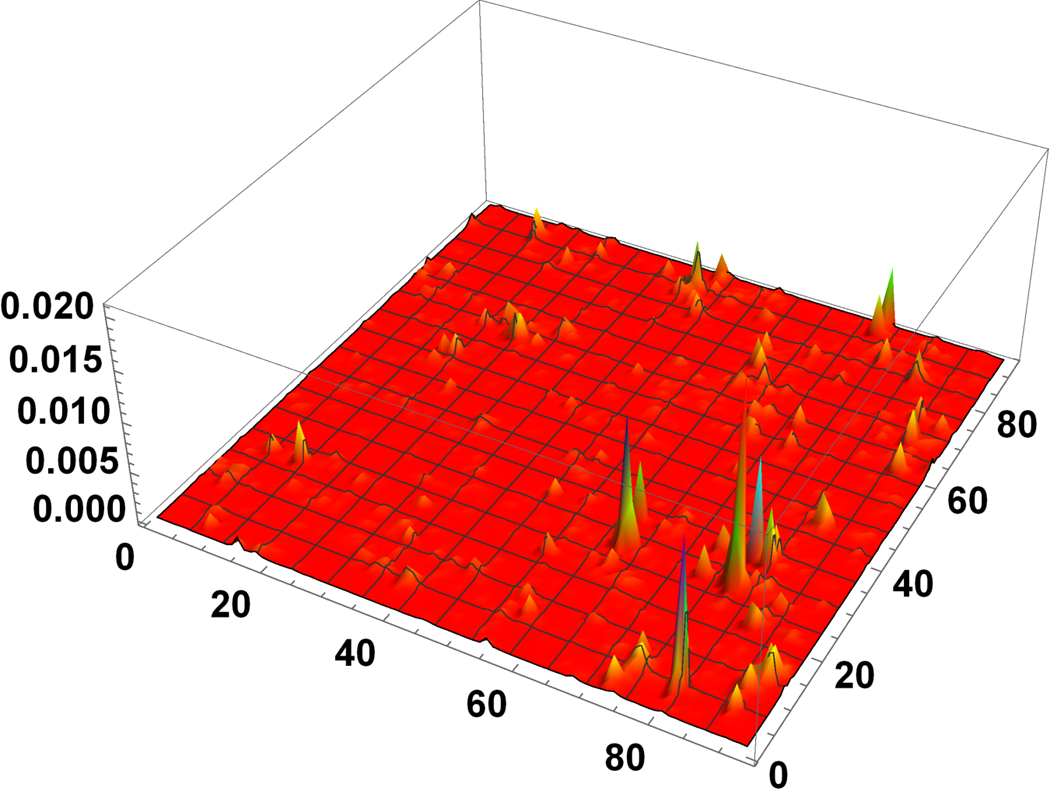}
}
\\
\subfigure[$\lambda = 1.2\pi$, Energy $1.2$]{
\includegraphics[width=0.313\textwidth]{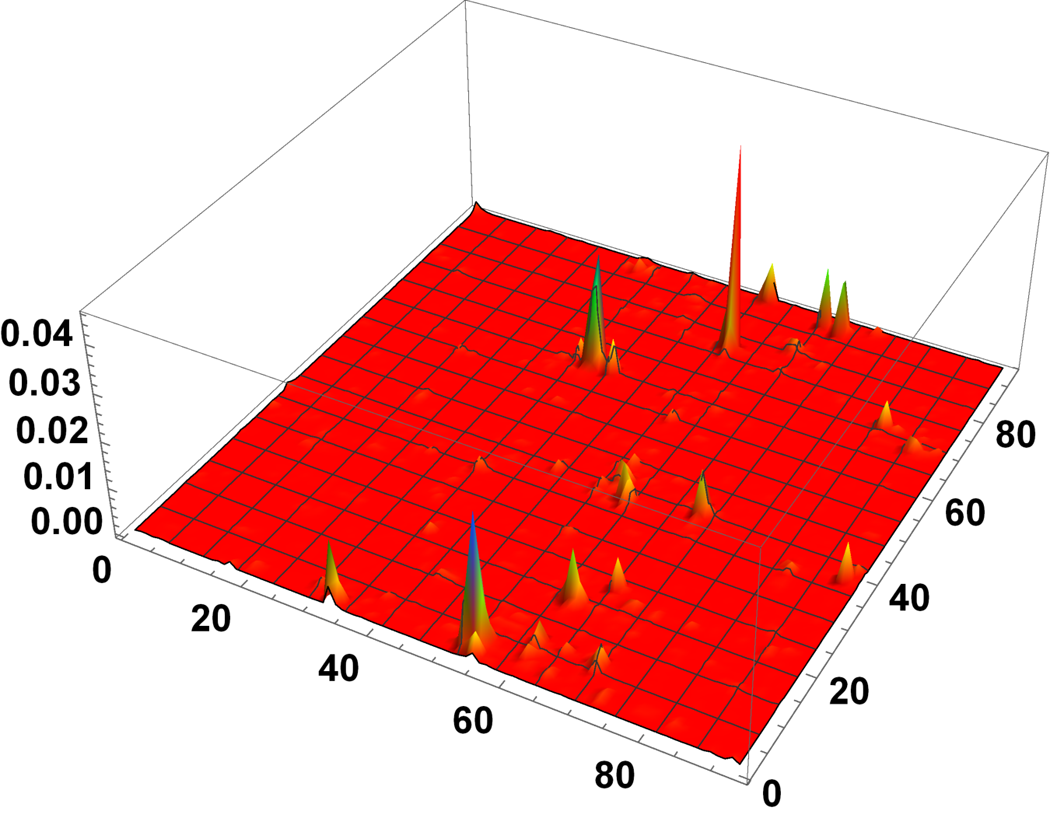}
}
\subfigure[$\lambda = 1.2\pi$, Energy $1.25$]{
\includegraphics[width=0.313\textwidth]{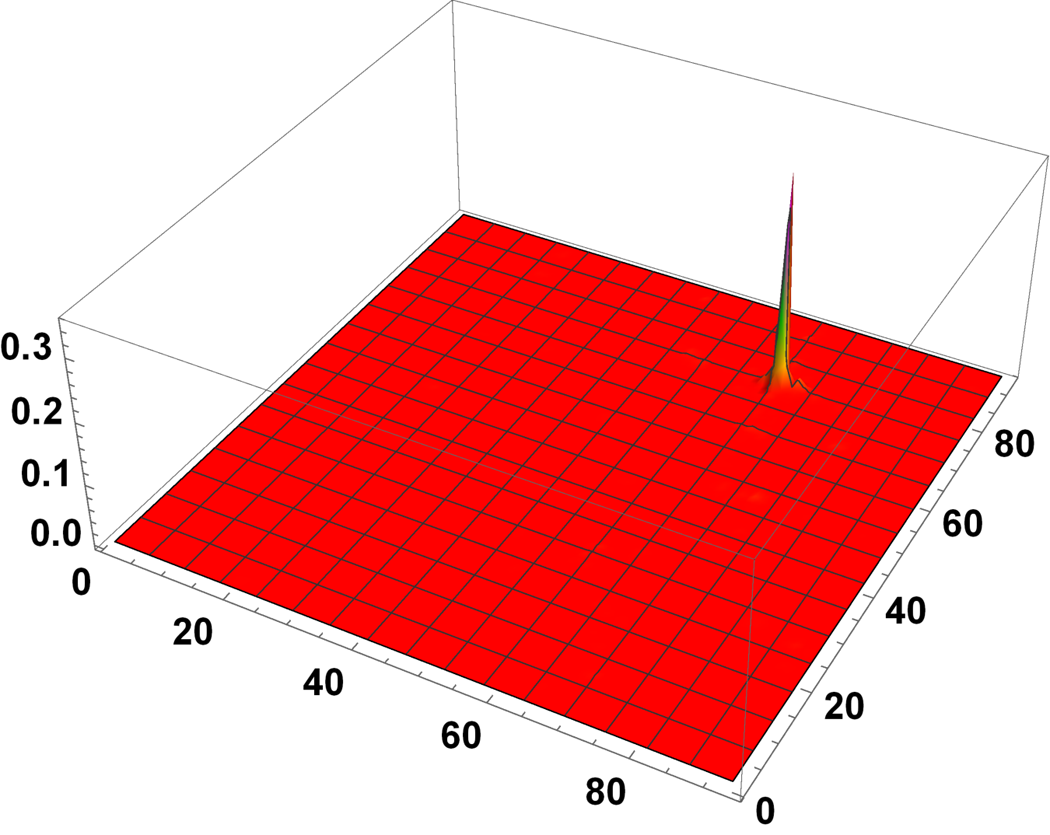}
}
\\
\subfigure[$\lambda = 2.0\pi$, Energy $0.7$]{
\includegraphics[width=0.313\textwidth]{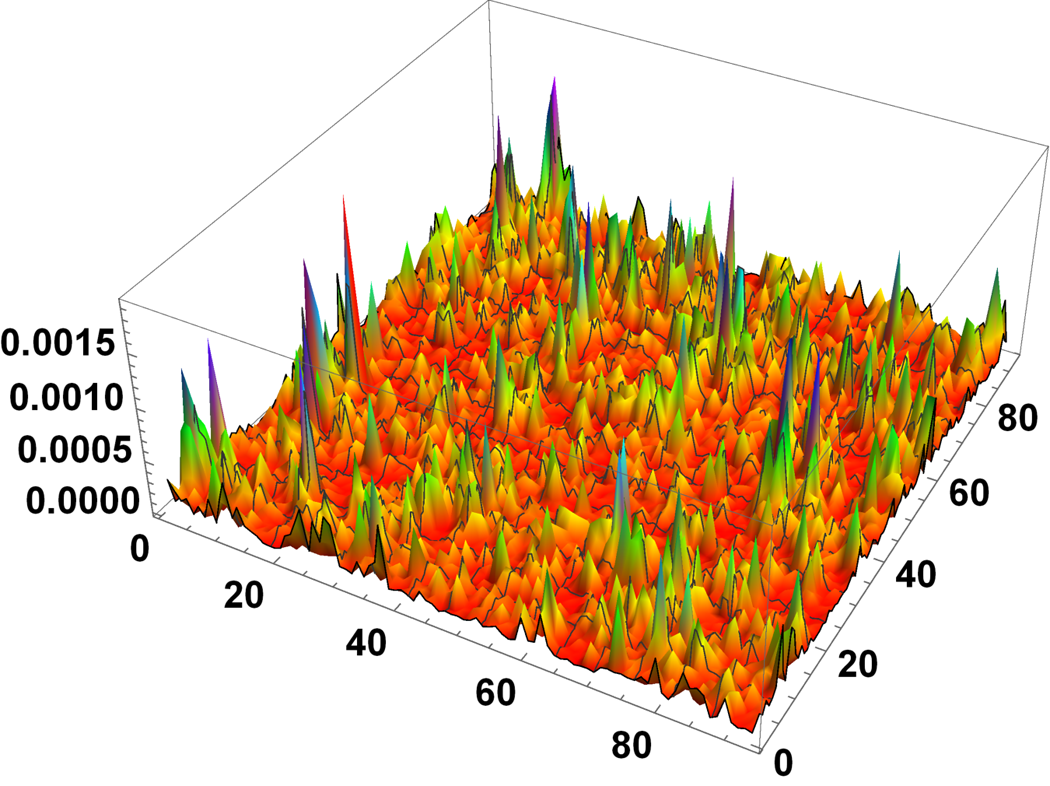}
}
\subfigure[$\lambda = 2.0\pi$, Energy $1.4$]{
\includegraphics[width=0.313\textwidth]{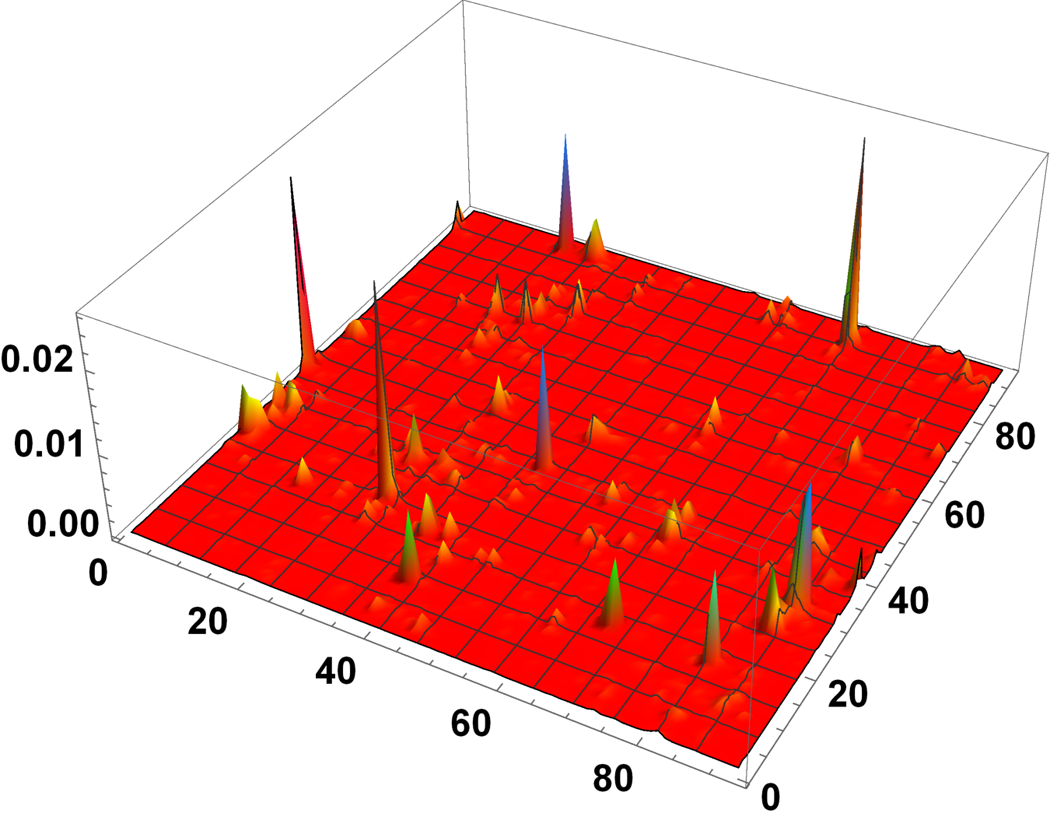}
}
\\
\subfigure[$\lambda = 2.0\pi$, Energy $1.5$]{
\includegraphics[width=0.313\textwidth]{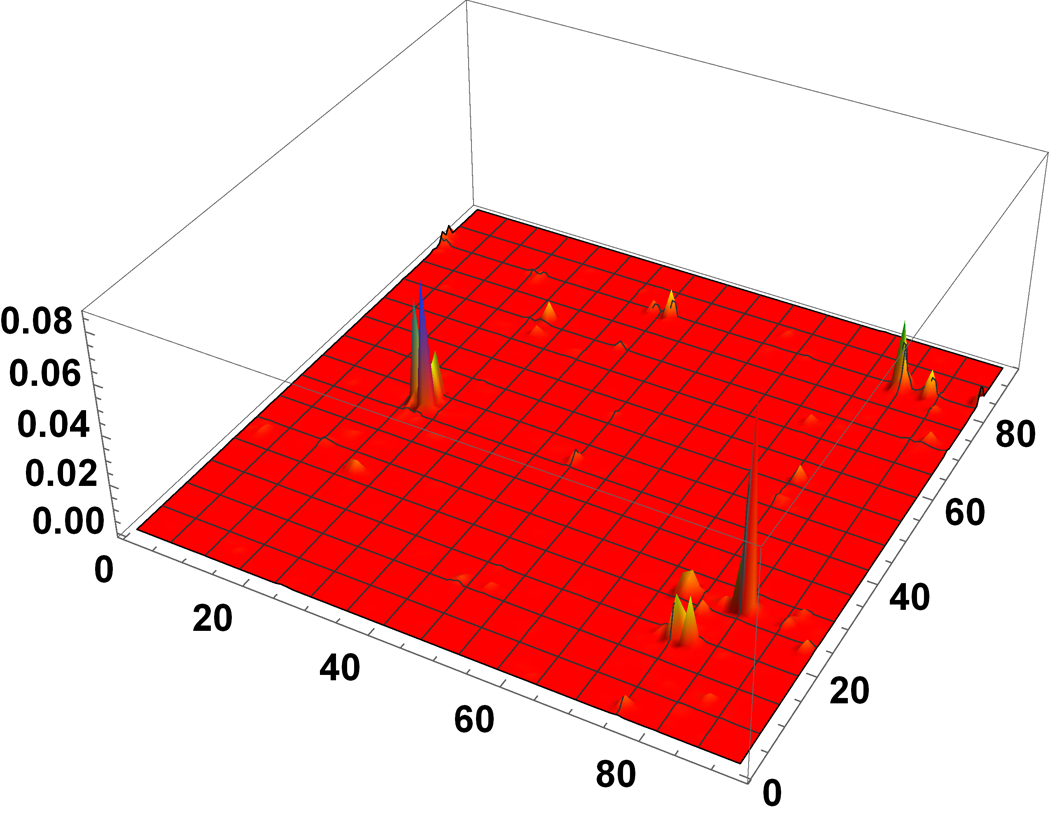}
}
\subfigure[$\lambda = 2.0\pi$, Energy $1.6$]{
\includegraphics[width=0.313\textwidth]{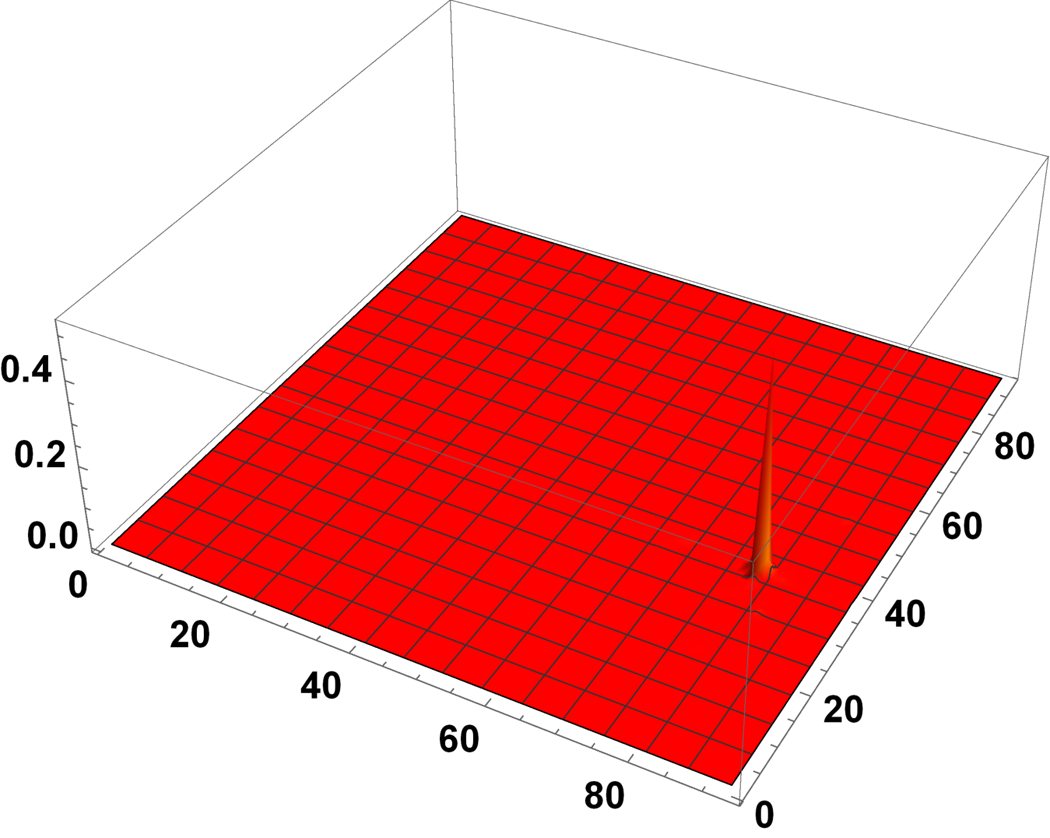}
}
\caption{
Representative wave function 
position space 
probability profiles $|\psi(\vex{r})|^2$
for $k = 1$, $N = 46$,
and disorder strengths $\lambda = 1.2\pi$ and $2.0\pi$ (see Fig.~4 in the main text).
The wave functions (a) and (e) are representative of class C
multifractal states. The other states reside in the high-energy tail of the 
DoS (see Fig.~4), which we have argued arises from the unphysical truncation
of the surface state spectrum. While the states (d) and (h) appear localized,
the intermediate states suggest that the tail states might instead exhibit \emph{multifractal
freezing} \cite{S--Chamon1996,S--Foster2009}.  
}
\label{DirtyWFs}
\end{figure}

Fig.~\ref{DirtyWFs} shows that while representative states in the class C ``bulk'' are clearly critically delocalized,
the behavior of the tail states is less clear. In fact, it is possible that the tail states are becoming \emph{frozen} instead
of localized. A single frozen state is ``quasilocalized,'' and consists of a \emph{few} peaks that are separated by arbitrarily large
distances \cite{S--Chamon1996,S--Foster2009}. These states are known to arise for class AIII surface states with winding number 
$\nu = 1$ and strong disorder ($\lambda \geq 2 \pi$) \cite{S--Chamon1996,S--Chou2014}. 

We do not pursue the nature
of the tail states further in this work, since we believe that these are an artifact of the UV cutoff. For 2D surface states attached
to a 3D topological superconducting bulk, surface states with energies approaching the gap would bleed into the bulk, ultimately 
hybridizing with bulk quasiparticles states.

\subsection{E. Finite-size trends for $k = 1$ and $k = 7$}

Here we plot the anomalous spectrum $\Delta(q) = \tau(q) - 2(q-1)$ for low- and finite-energy states
as in Fig.~1, as well as the population statistics and numerical density of states as in Fig.~2,
for the full range of $N$ used to obtain the results shown in Fig.~3. 
Box sizes for multifractal analysis are chosen to be almost commensurate with the linear system size $2 N + 1$.

\begin{figure}[h!]
\subfigure{
\includegraphics[width=0.4\textwidth]{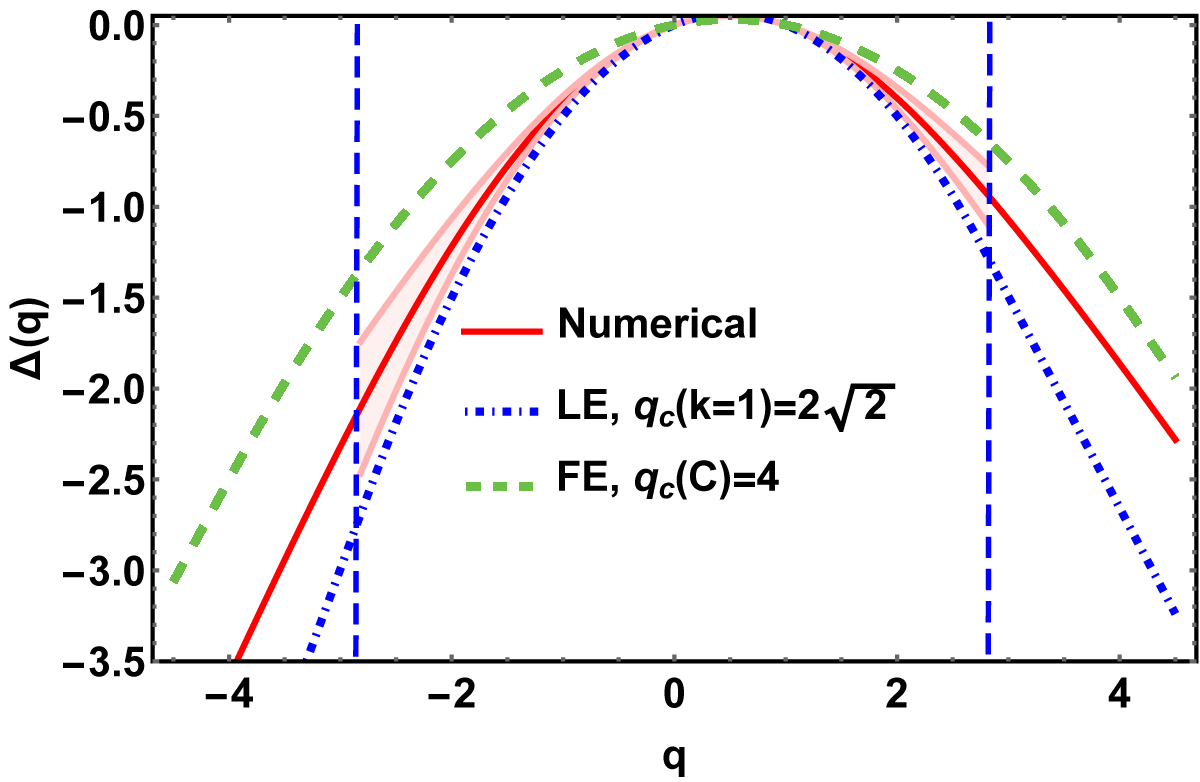}
}
\subfigure{
\includegraphics[width=0.4\textwidth]{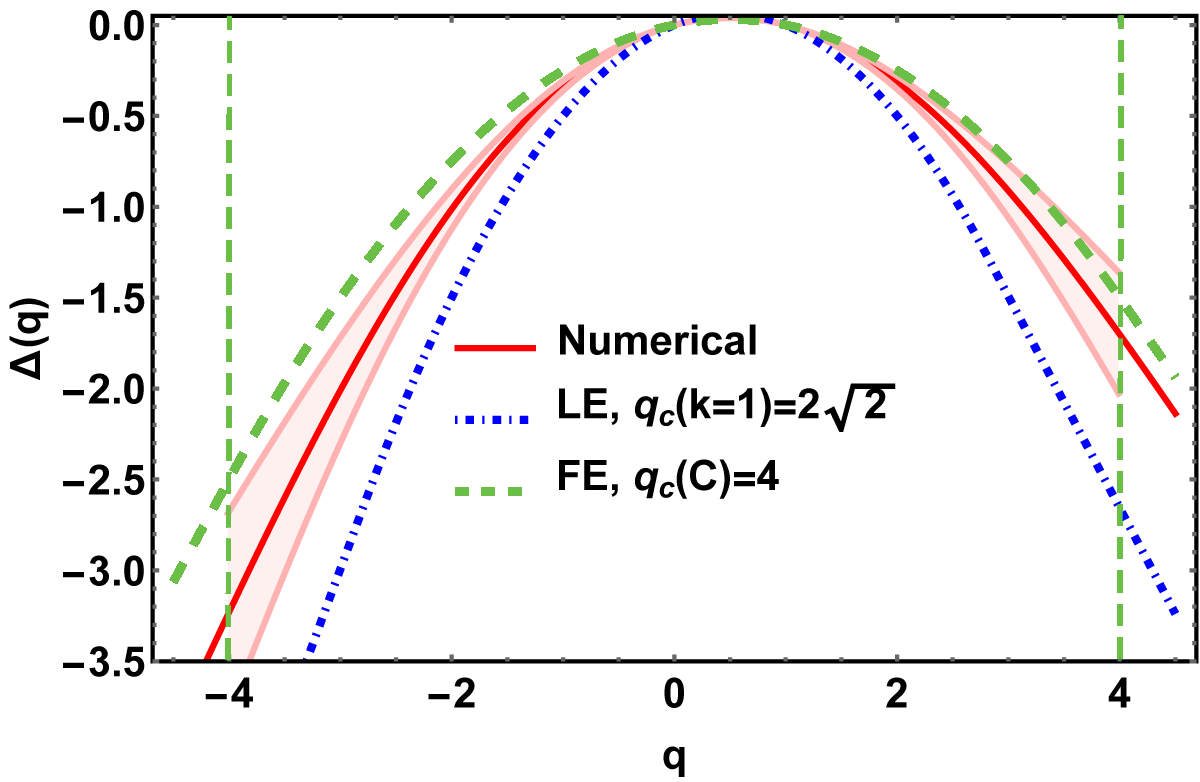}
}
\\
\subfigure{
\includegraphics[height=0.2\textheight]{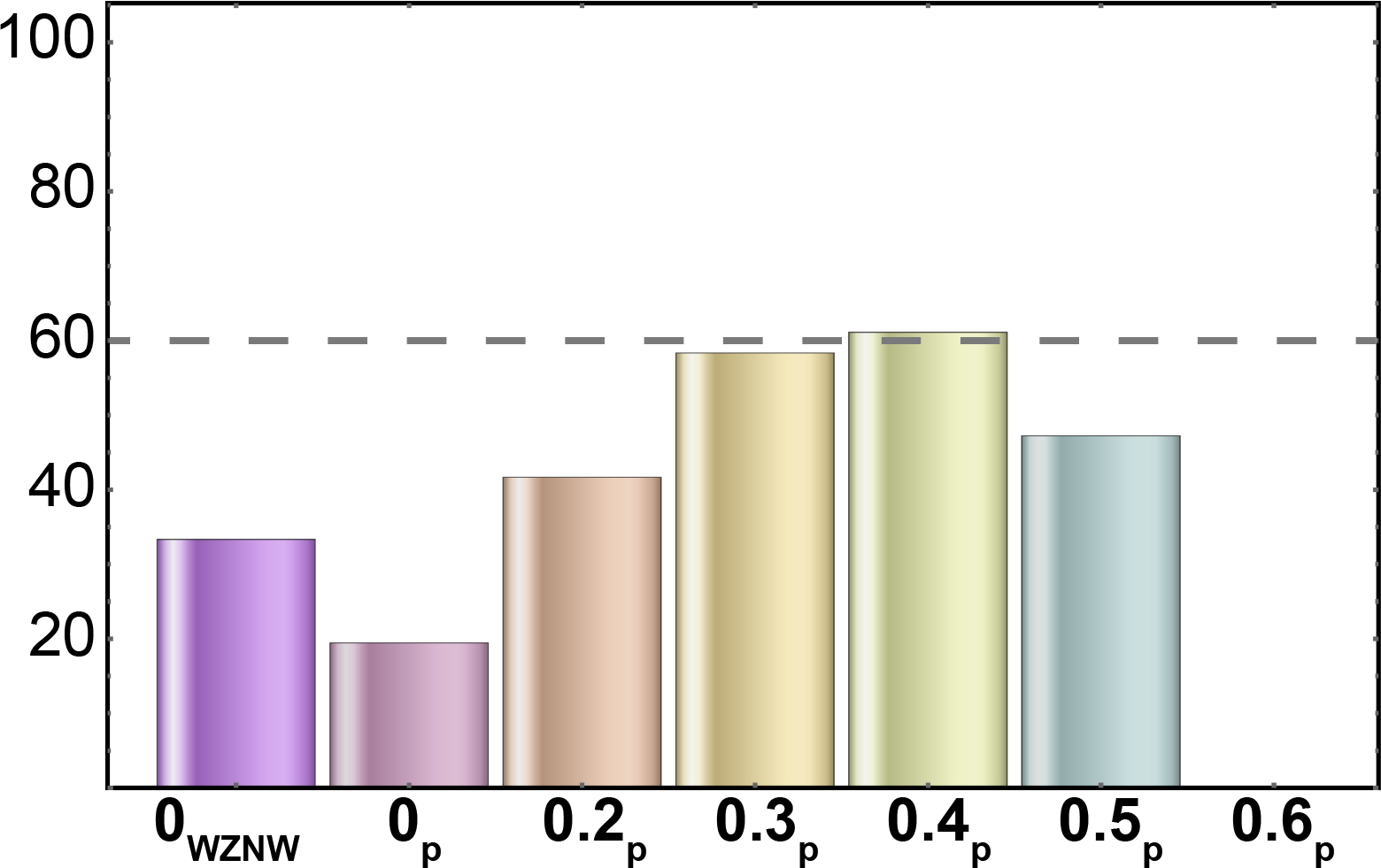}
}
\hspace{14pt}
\subfigure{
\includegraphics[height=0.2\textheight]{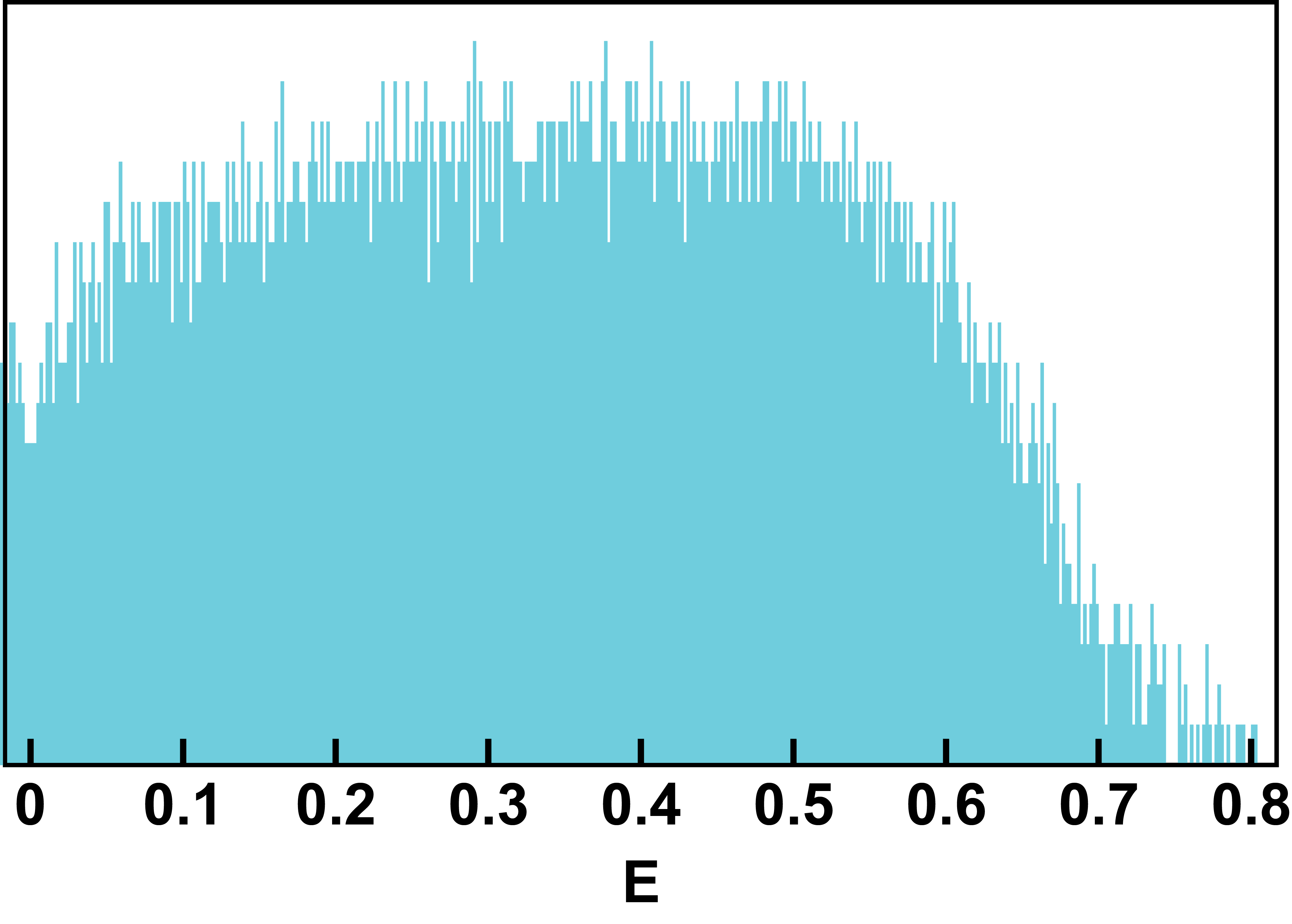}
}
\caption{$k = 1$, $N = 24$. Box sizes $b = 2,6$.}
\end{figure}


\begin{figure}[h!]
\subfigure{
\includegraphics[width=0.4\textwidth]{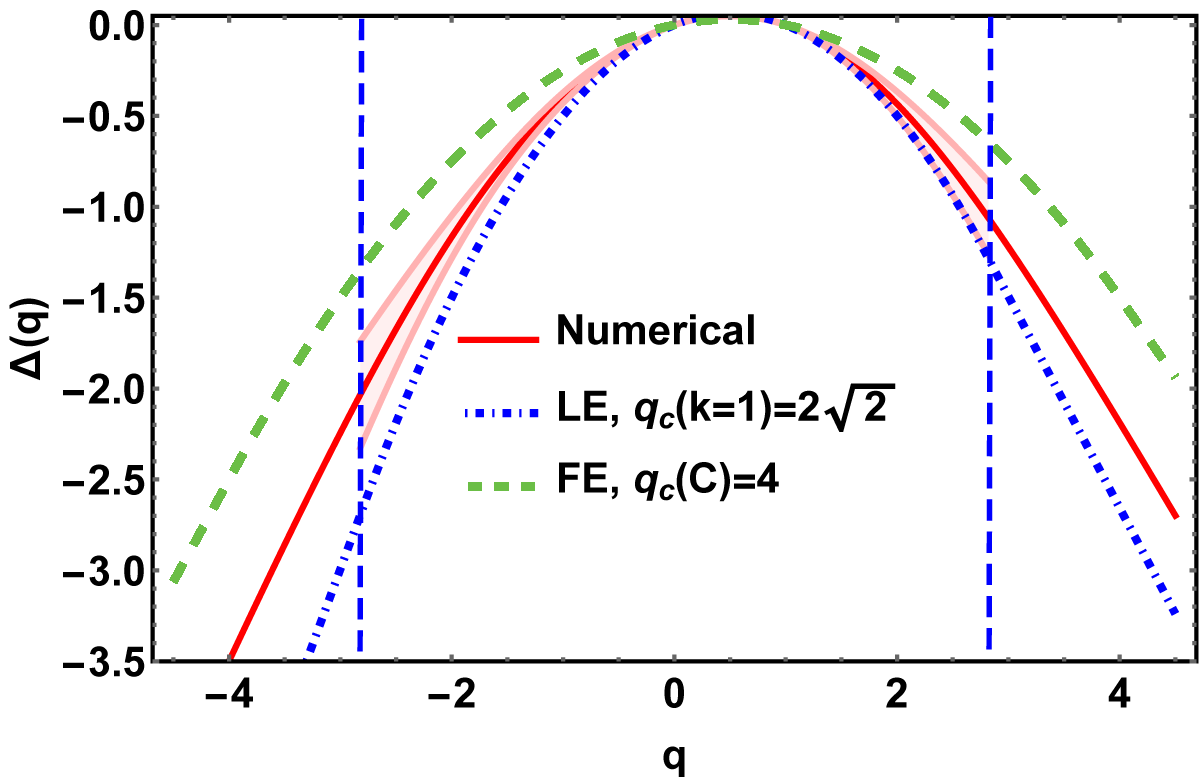}
}
\subfigure{
\includegraphics[width=0.4\textwidth]{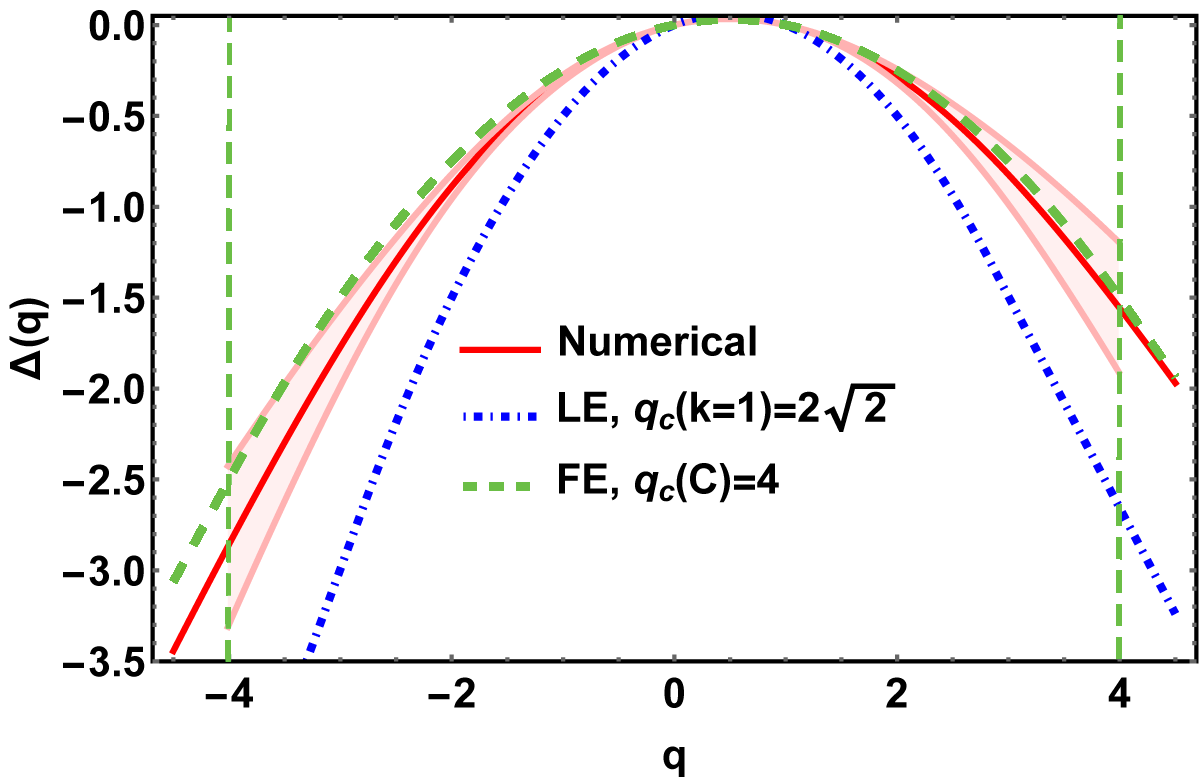}
}
\\
\subfigure{
\includegraphics[height=0.2\textheight]{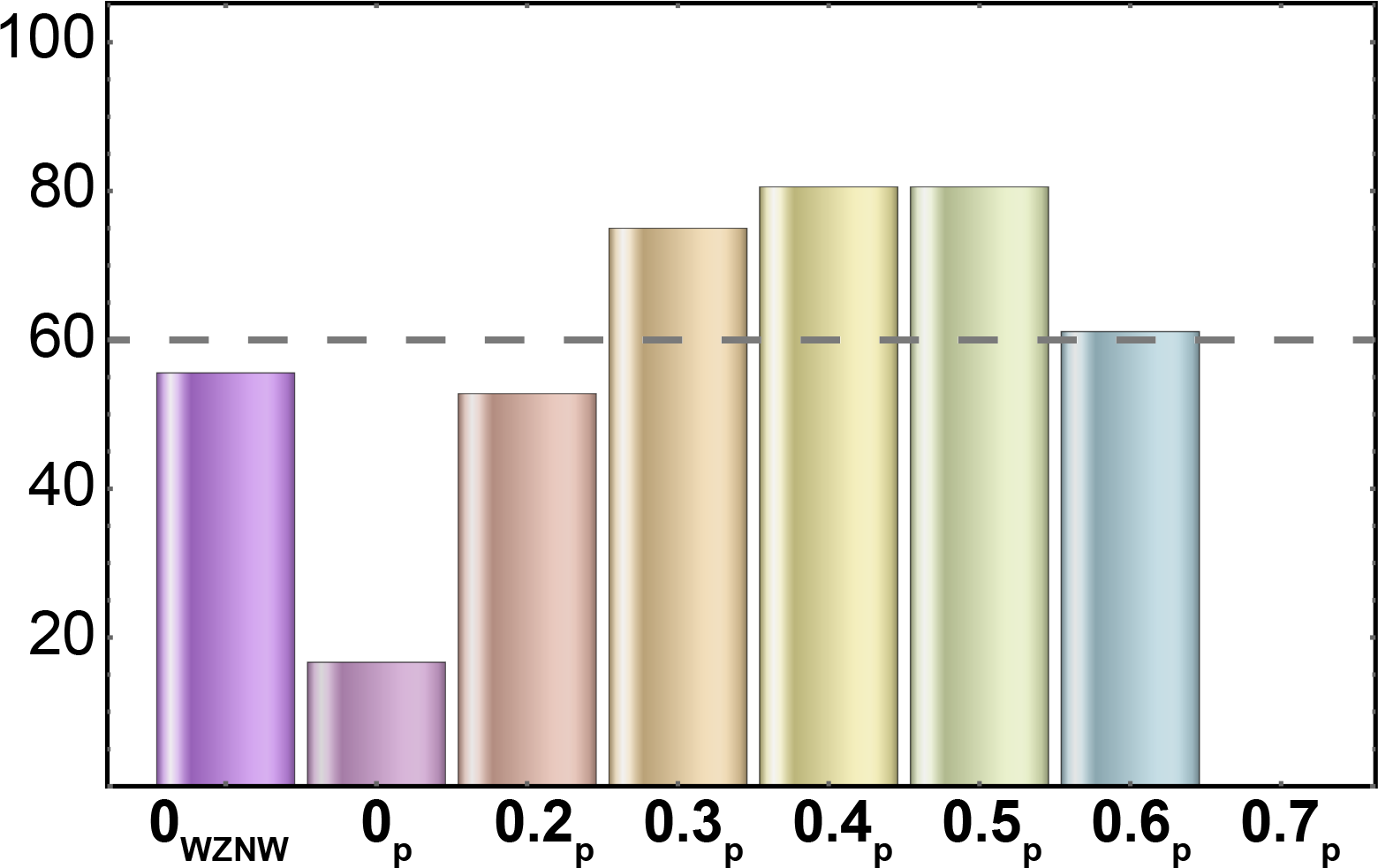}
}
\hspace{14pt}
\subfigure{
\includegraphics[height=0.2\textheight]{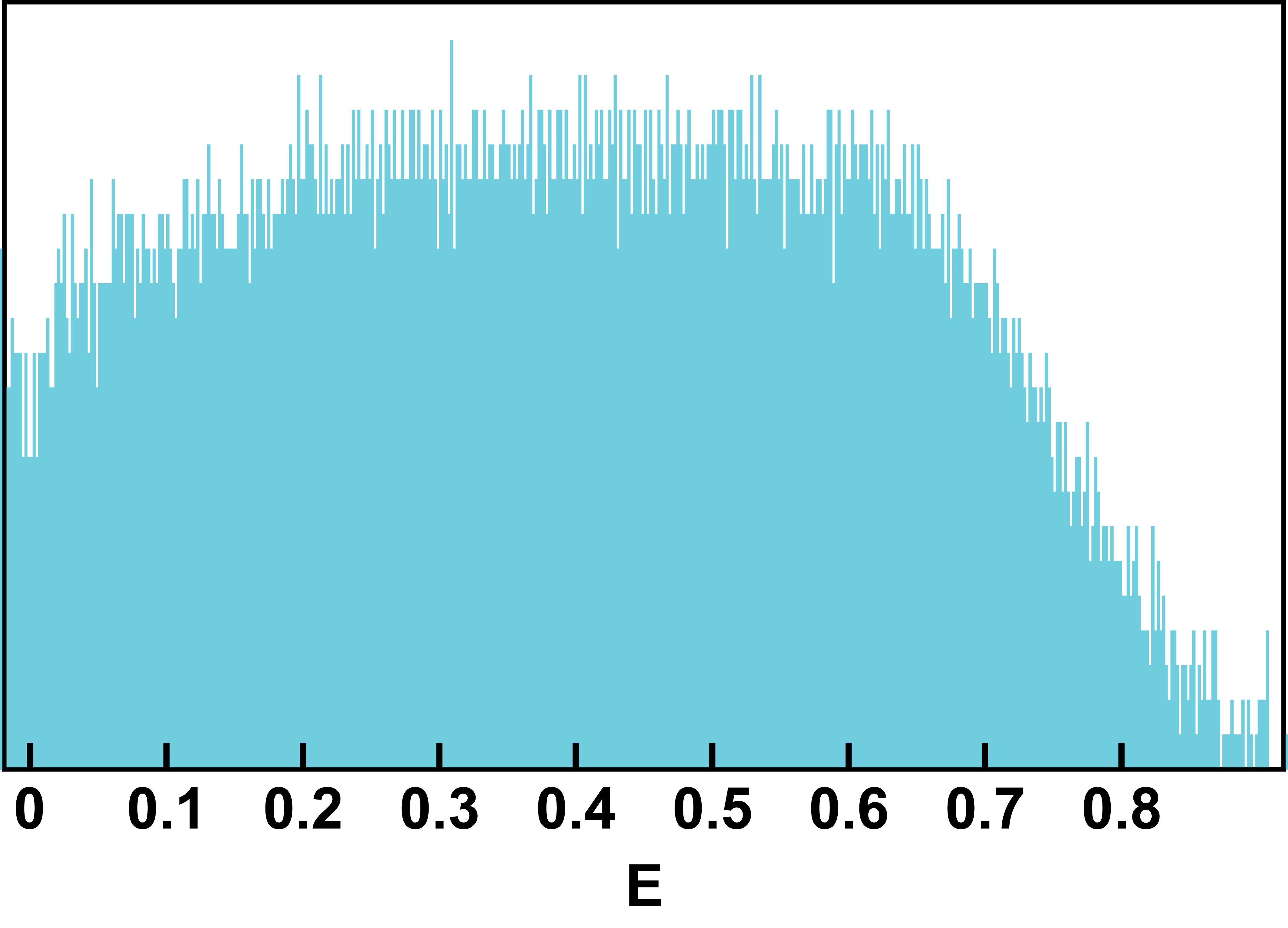}
}
\caption{$k = 1$, $N = 28$. Box sizes $b = 2,8$.}
\end{figure}


\begin{figure}[h!]
\subfigure{
\includegraphics[width=0.4\textwidth]{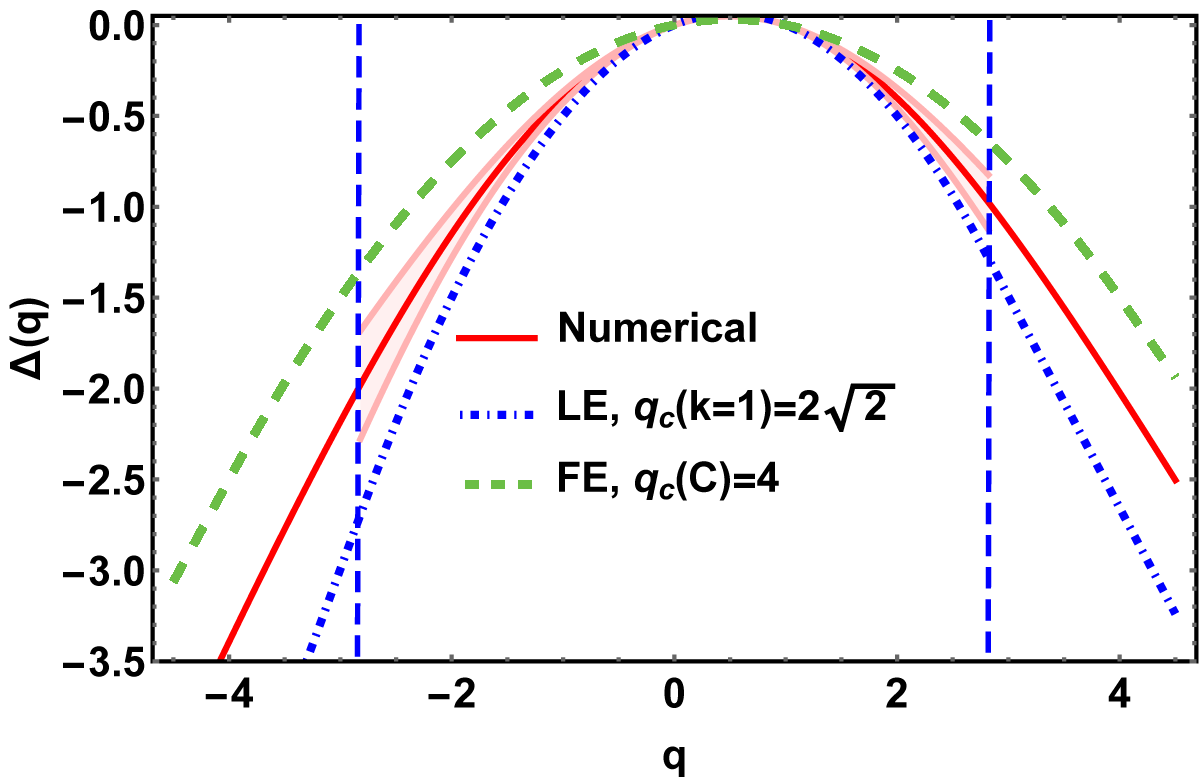}
}
\subfigure{
\includegraphics[width=0.4\textwidth]{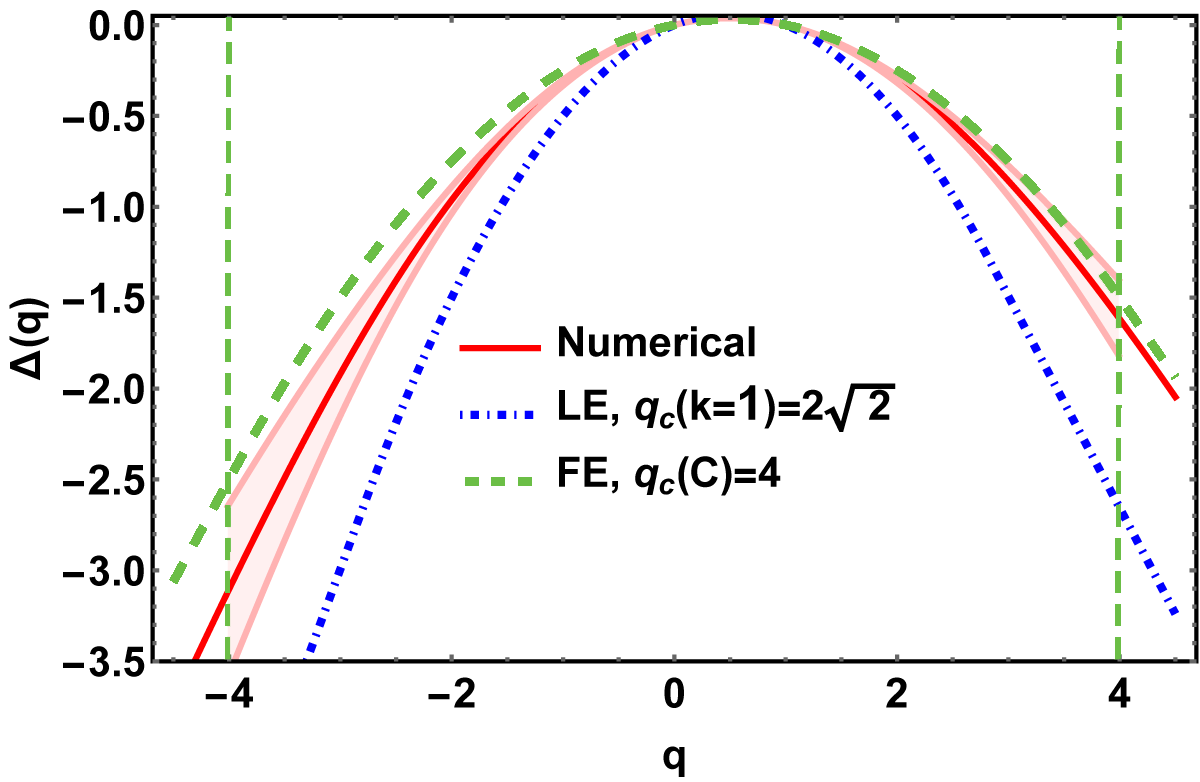}
}
\\
\subfigure{
\includegraphics[height=0.2\textheight]{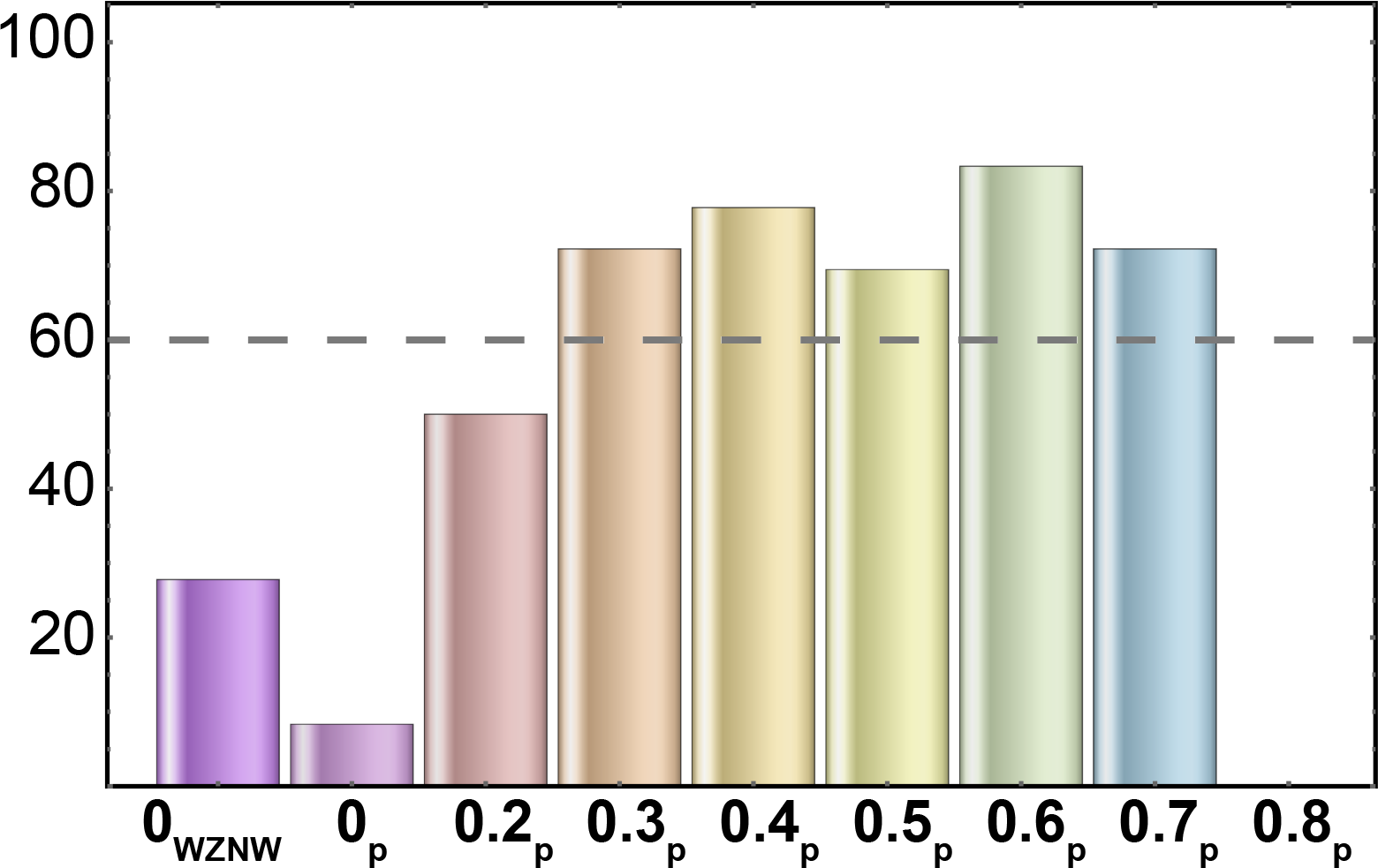}
}
\hspace{14pt}
\subfigure{
\includegraphics[height=0.2\textheight]{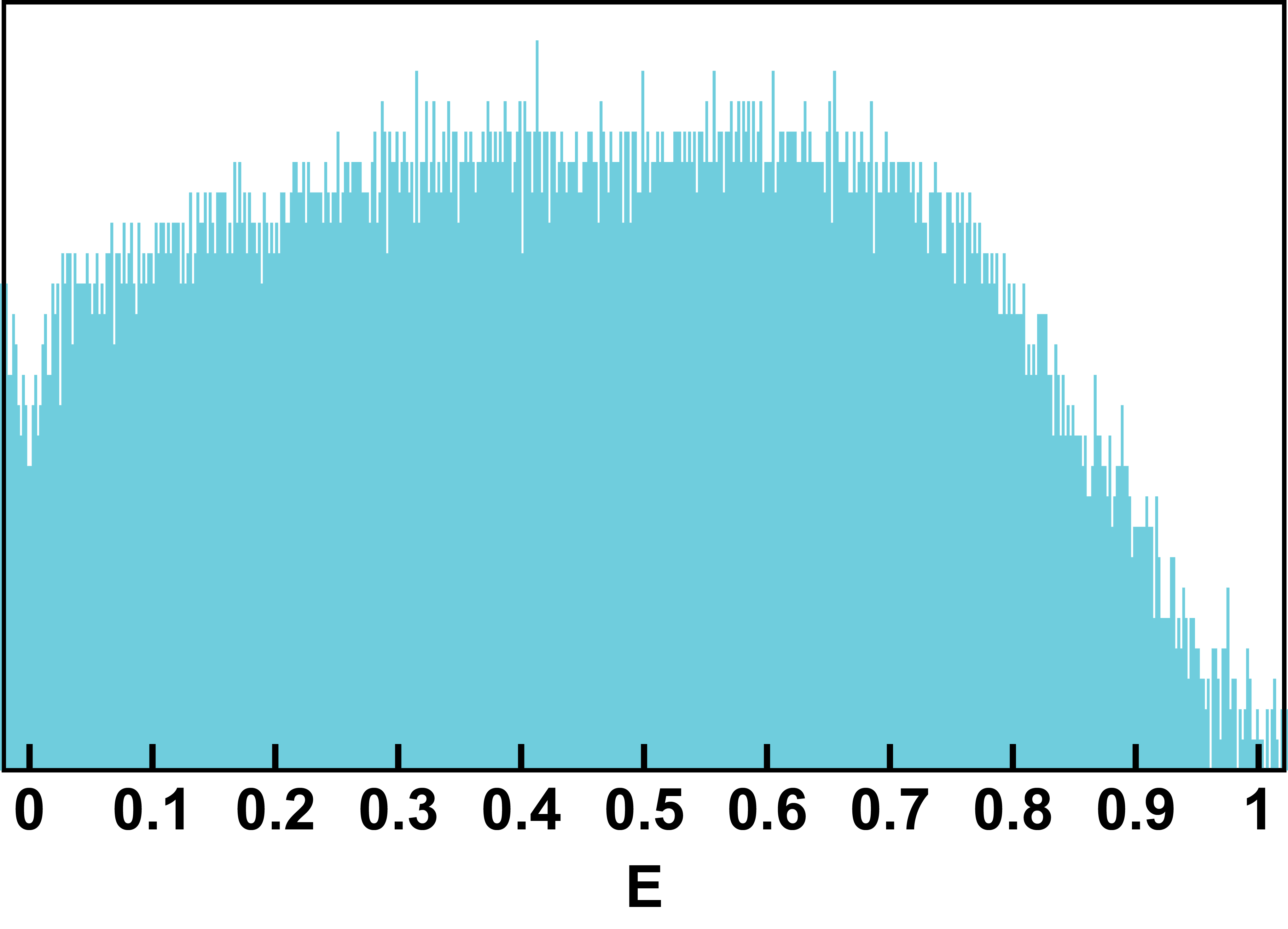}
}
\caption{$k = 1$, $N = 32$. Box sizes $b = 2,8$.}
\end{figure}


\begin{figure}[h!]
\subfigure{
\includegraphics[width=0.4\textwidth]{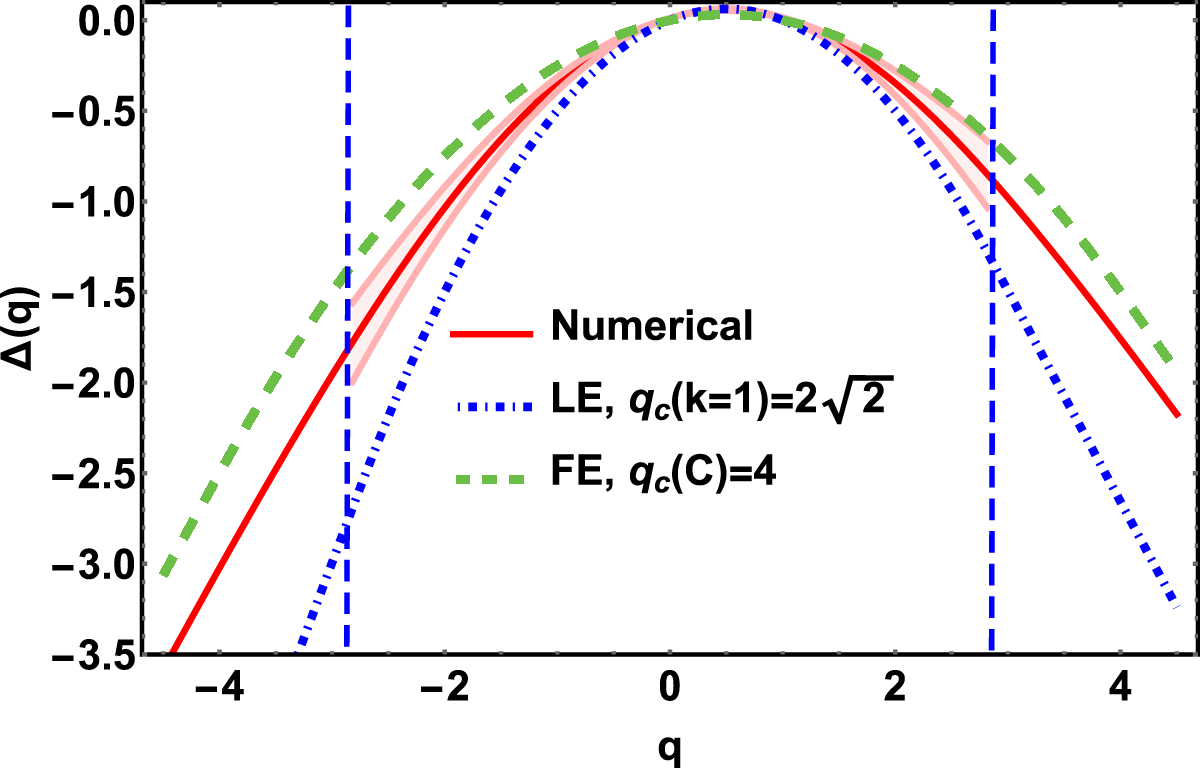}
}
\subfigure{
\includegraphics[width=0.4\textwidth]{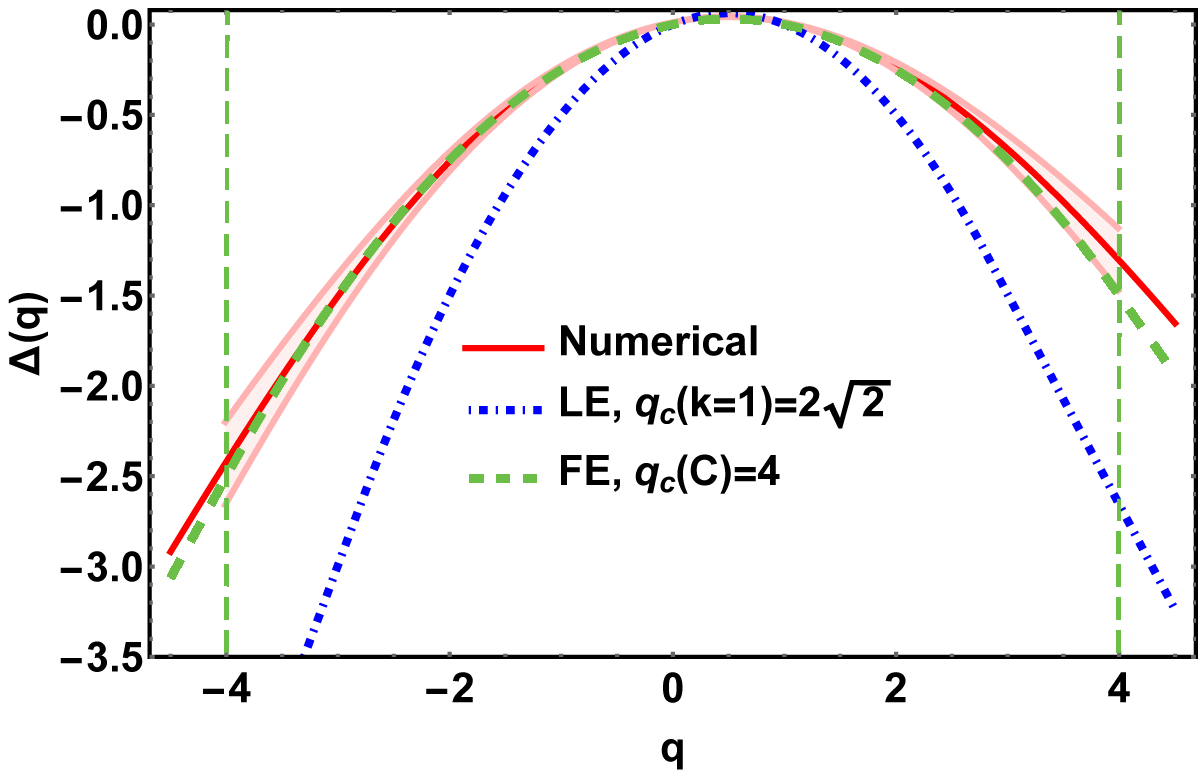}
}
\\
\subfigure{
\includegraphics[height=0.2\textheight]{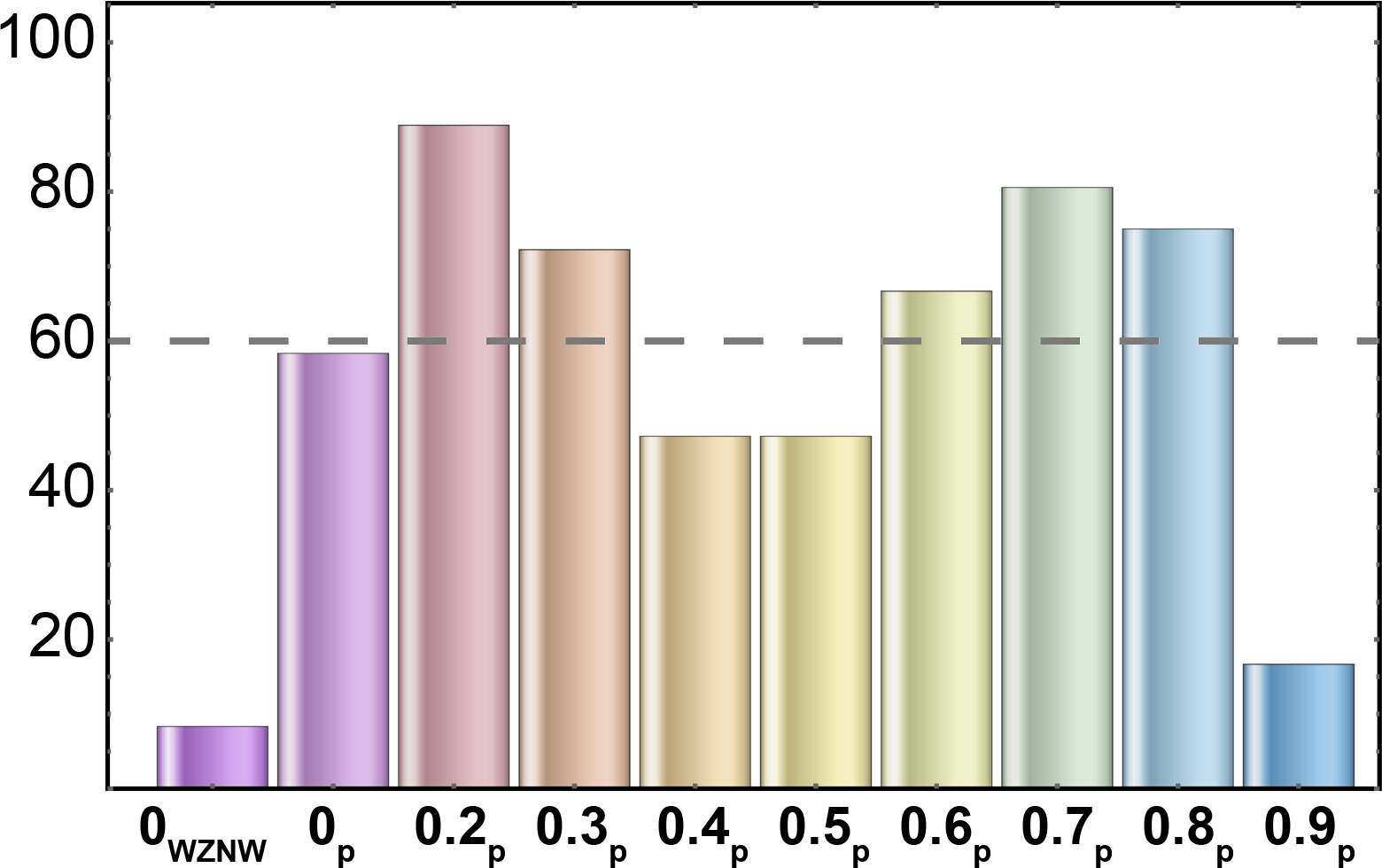}
}
\hspace{14pt}
\subfigure{
\includegraphics[height=0.2\textheight]{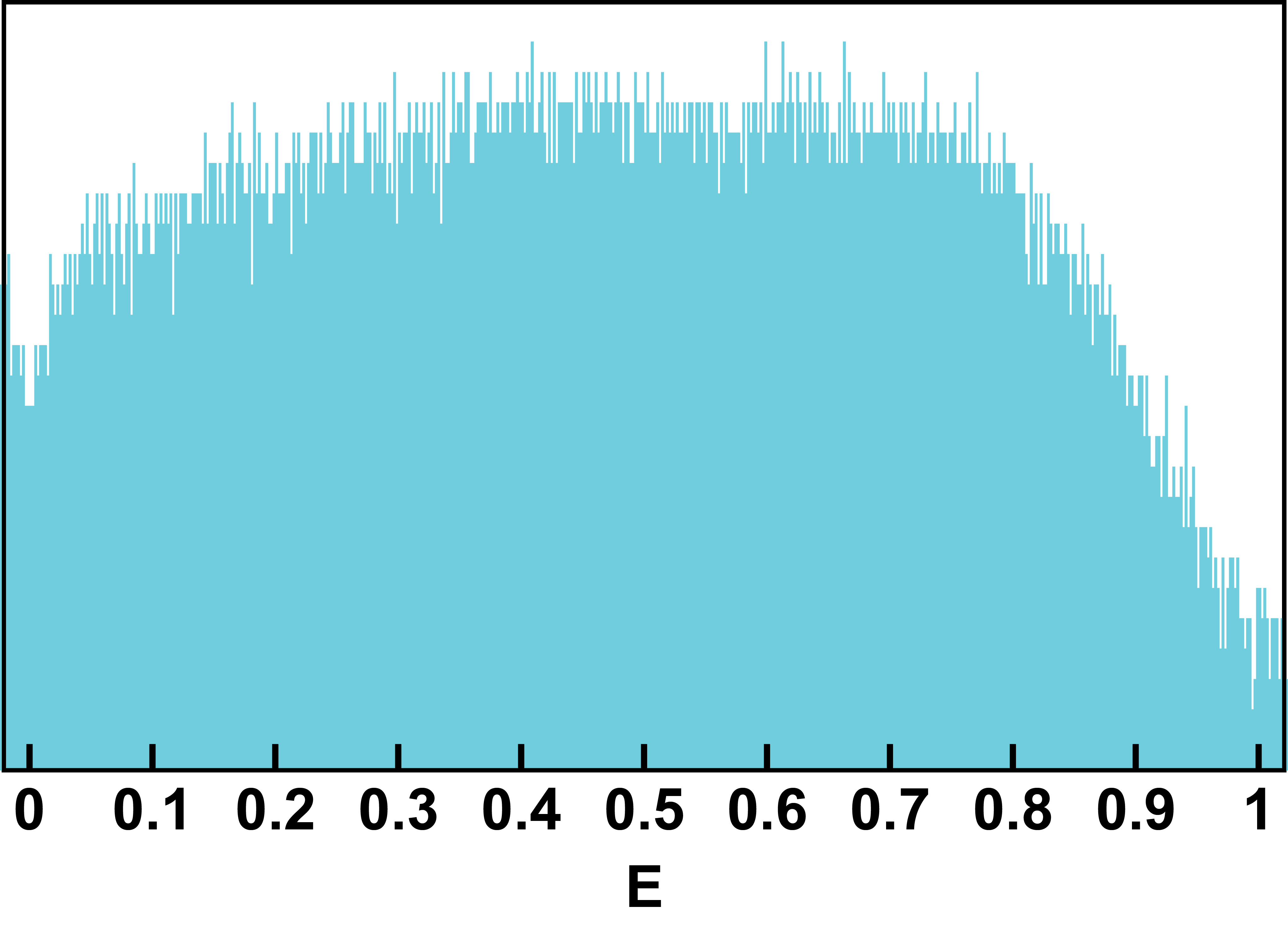}
}
\caption{$k = 1$, $N = 34$. Box sizes $b = 2,23$.}
\end{figure}


\begin{figure}[h!]
\subfigure{
\includegraphics[width=0.4\textwidth]{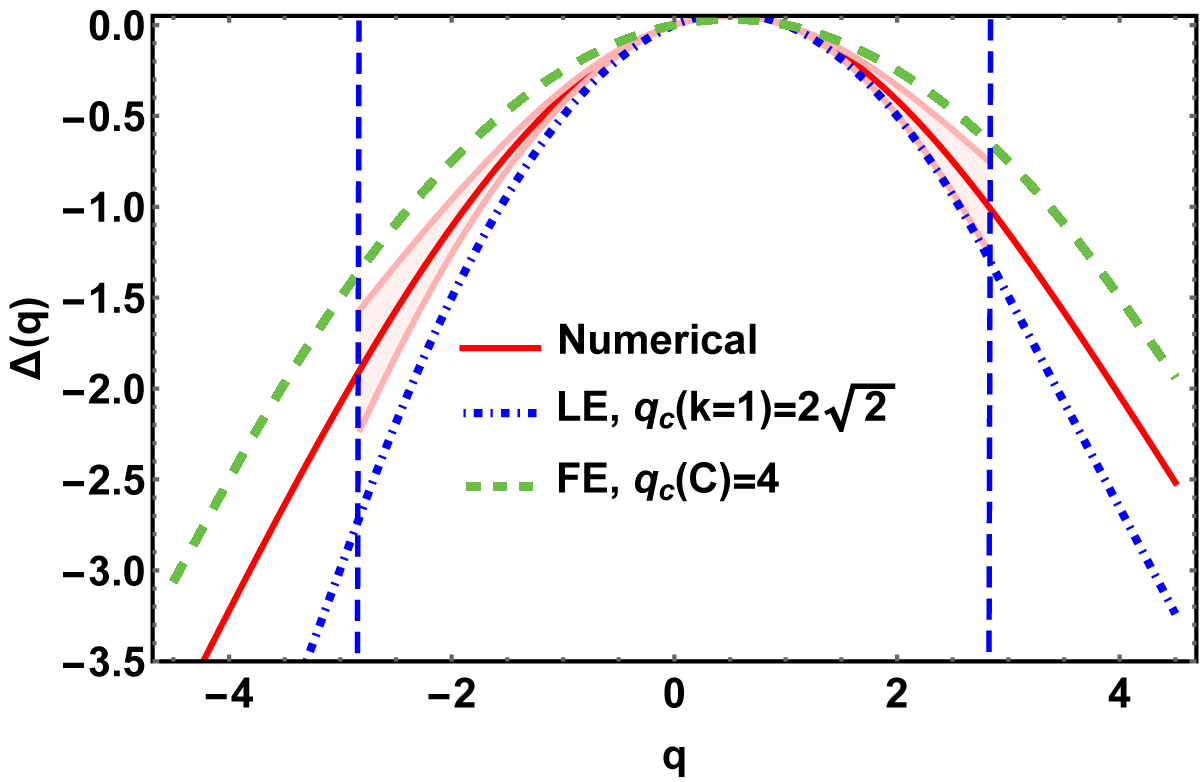}
}
\subfigure{
\includegraphics[width=0.4\textwidth]{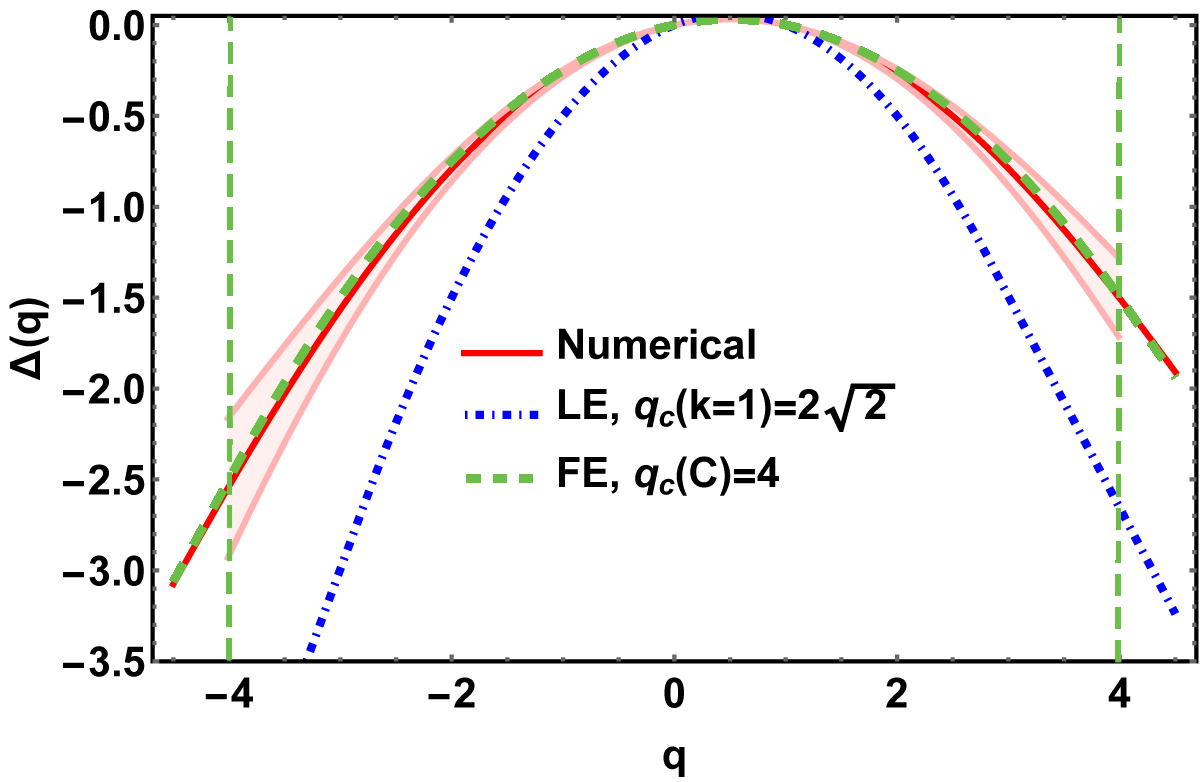}
}
\\
\subfigure{
\includegraphics[height=0.2\textheight]{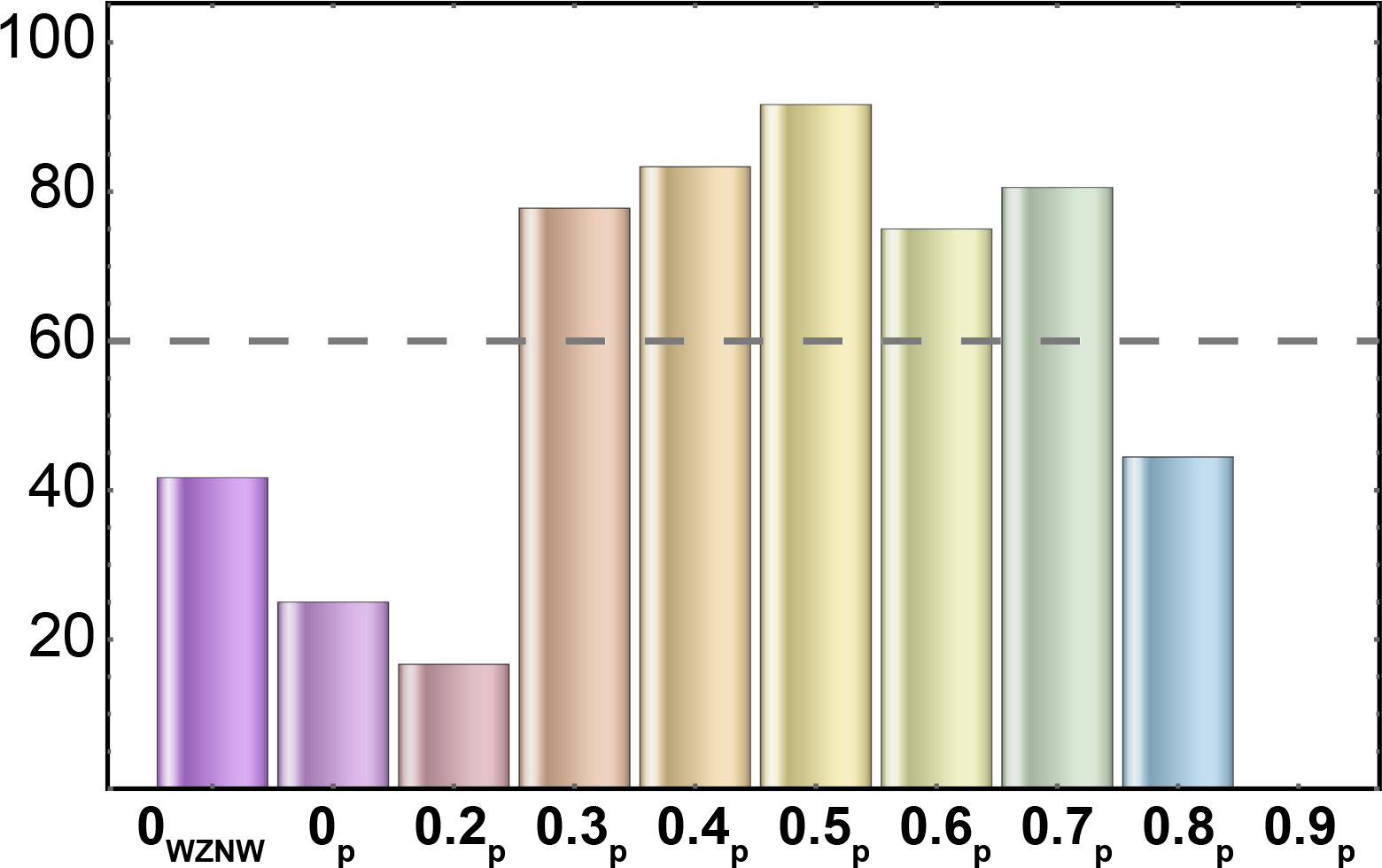}
}
\hspace{14pt}
\subfigure{
\includegraphics[height=0.2\textheight]{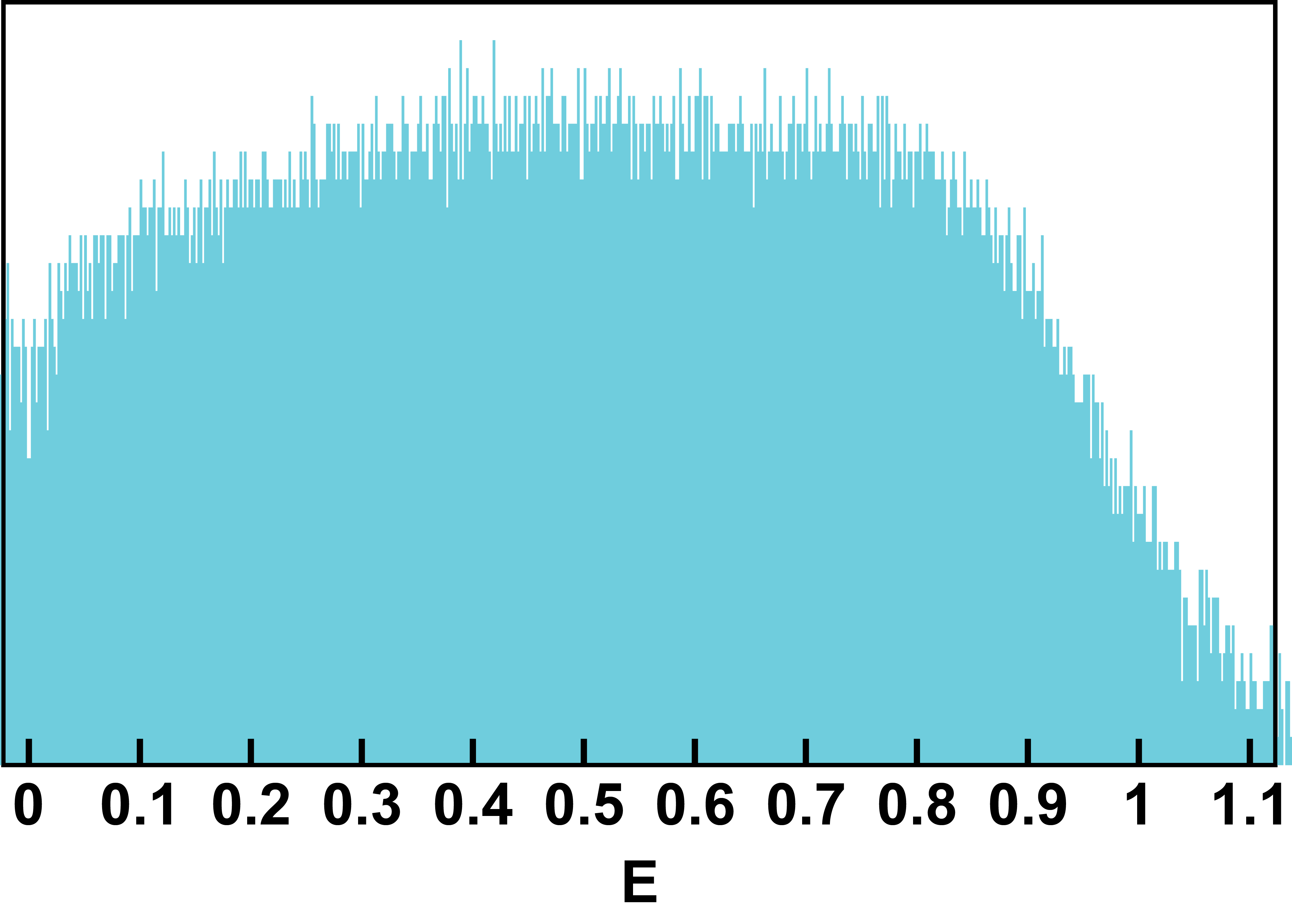}
}
\caption{$k = 1$, $N = 36$. Box sizes $b = 3,6$.}
\end{figure}


\begin{figure}[h!]
\subfigure{
\includegraphics[width=0.4\textwidth]{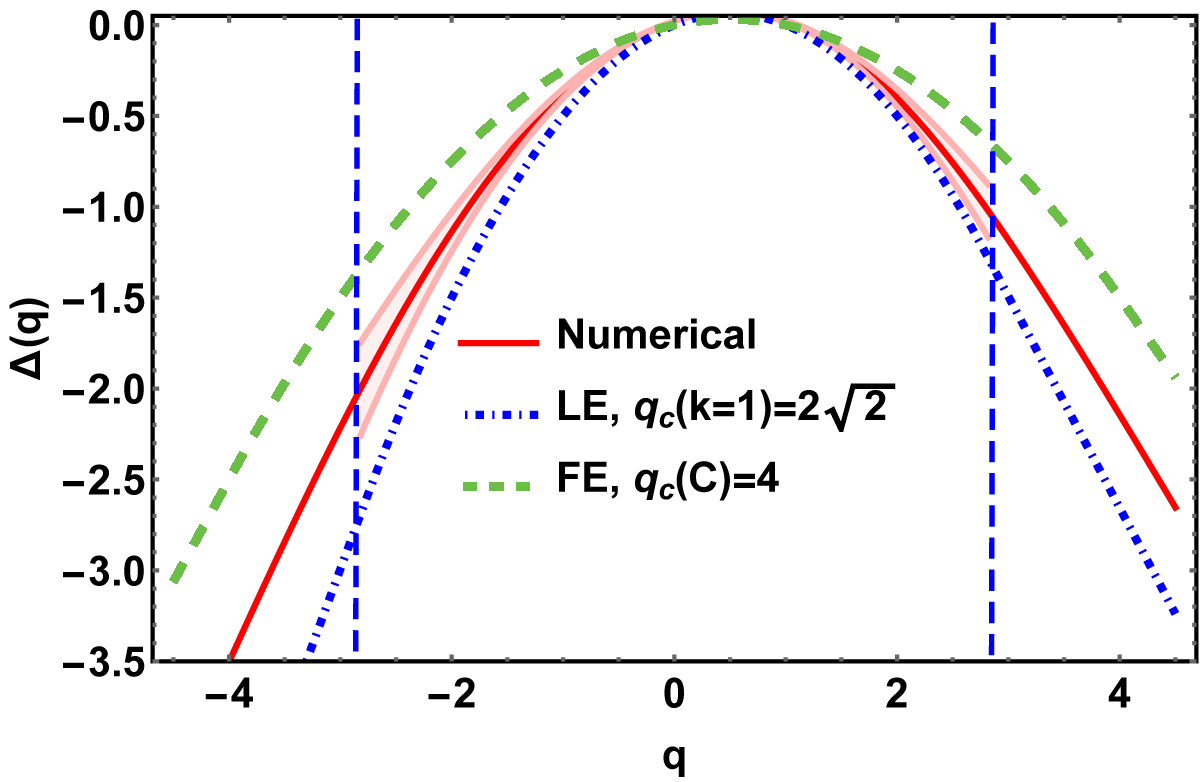}
}
\subfigure{
\includegraphics[width=0.4\textwidth]{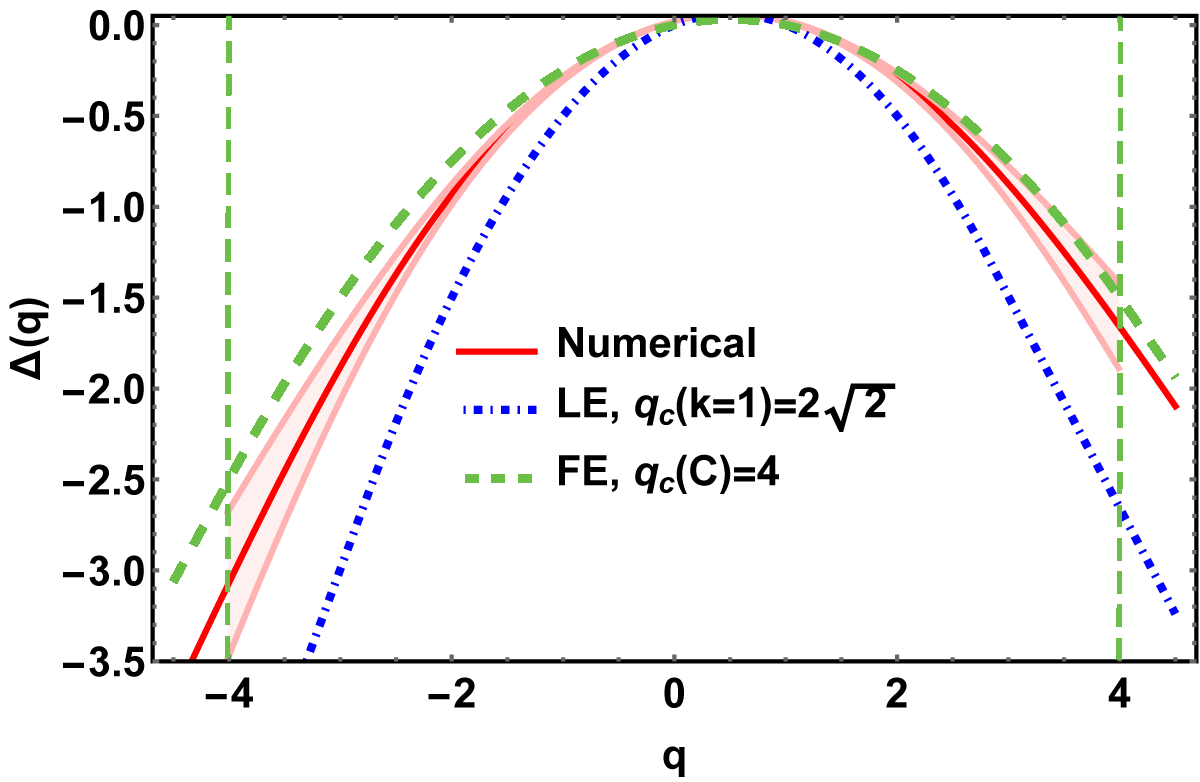}
}
\\
\subfigure{
\includegraphics[height=0.2\textheight]{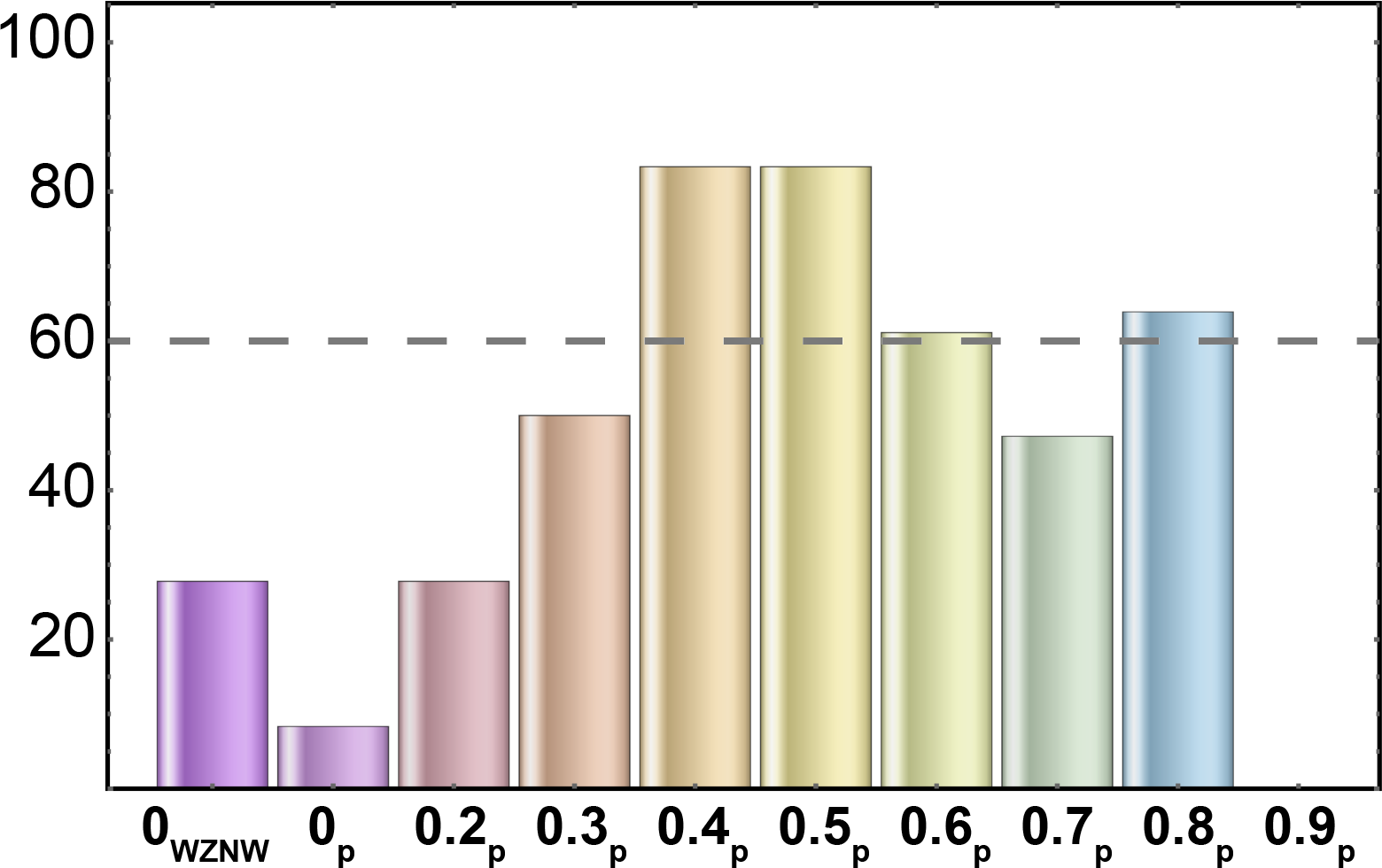}
}
\hspace{14pt}
\subfigure{
\includegraphics[height=0.2\textheight]{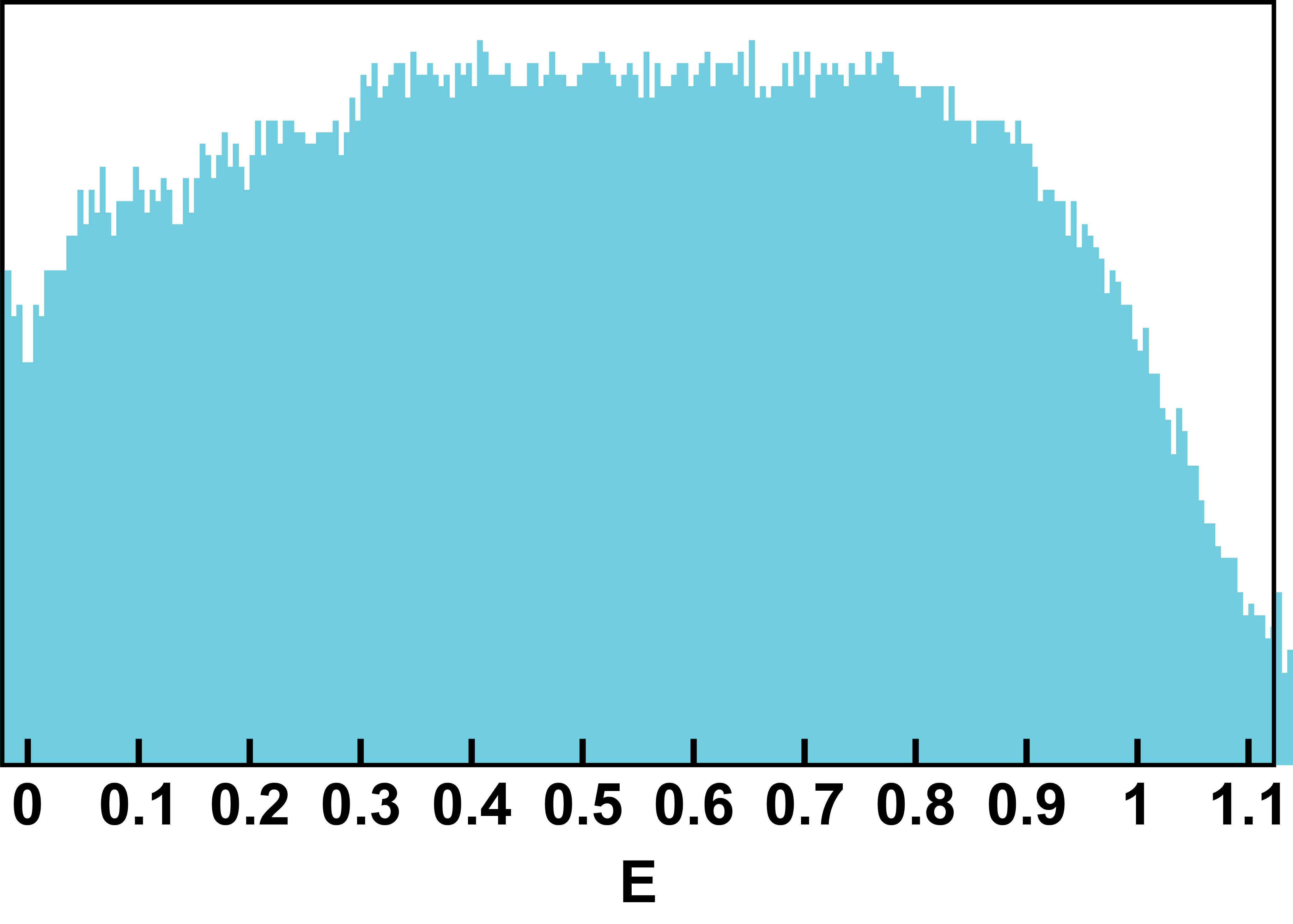}
}
\caption{$k = 1$, $N = 38$. Box sizes $b = 2,7$.}
\end{figure}


\begin{figure}[h!]
\subfigure{
\includegraphics[width=0.4\textwidth]{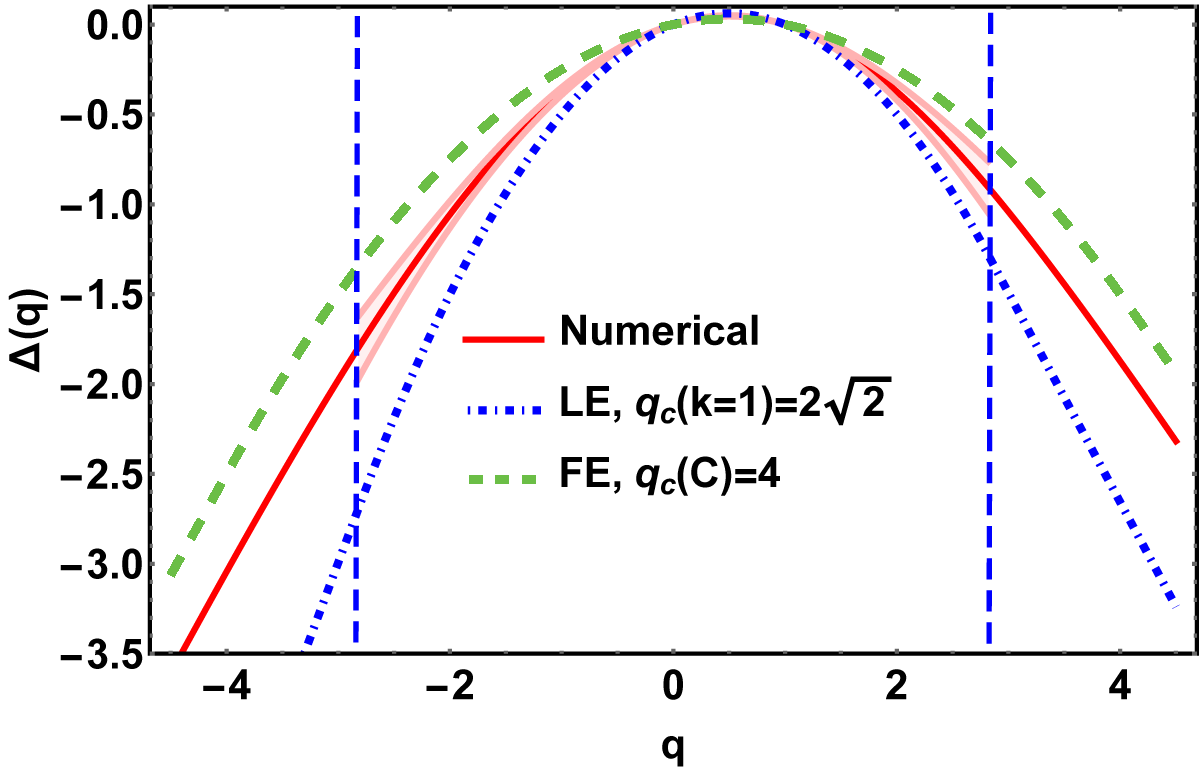}
}
\subfigure{
\includegraphics[width=0.4\textwidth]{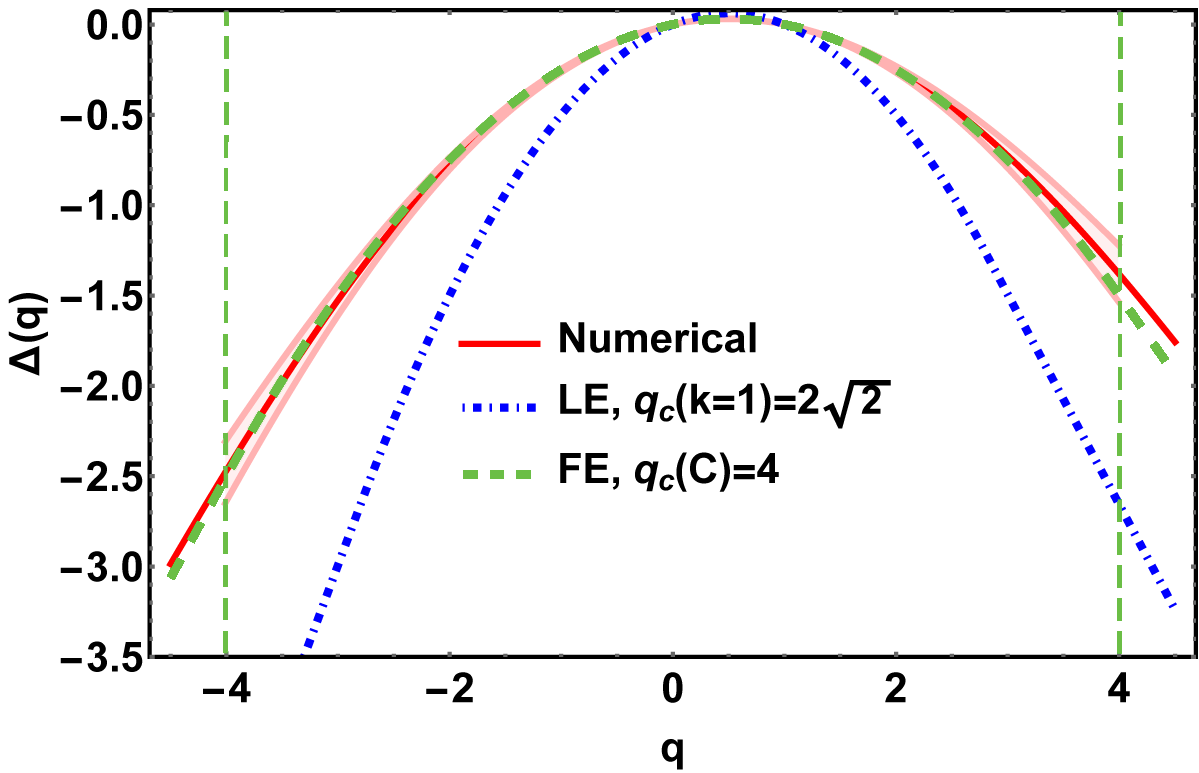}
}
\\
\subfigure{
\includegraphics[height=0.2\textheight]{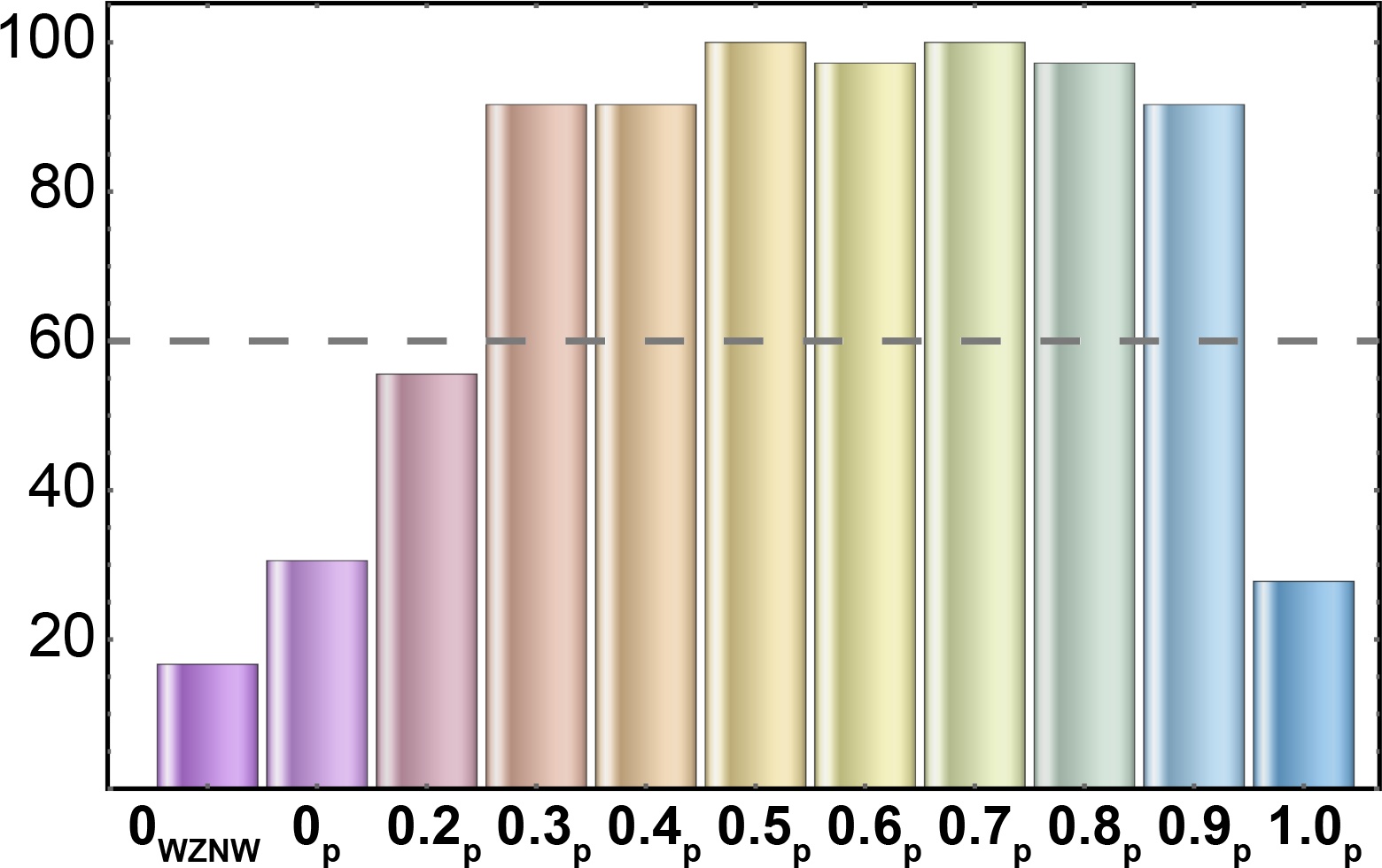}
}
\hspace{14pt}
\subfigure{
\includegraphics[height=0.2\textheight]{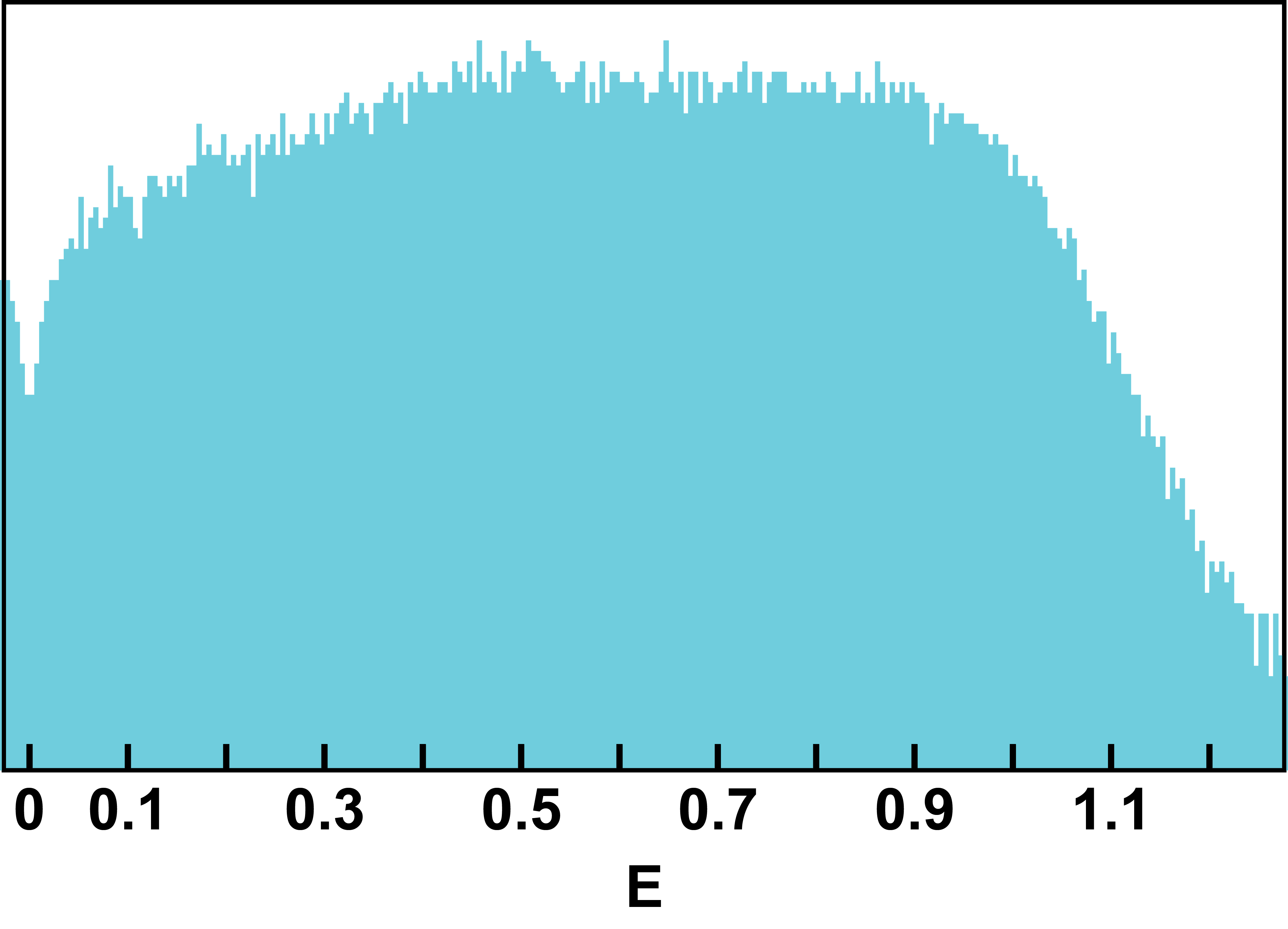}
}
\caption{$k = 1$, $N = 42$. Box sizes $b = 2,14$.}
\end{figure}


\begin{figure}[h!]
\subfigure{
\includegraphics[width=0.4\textwidth]{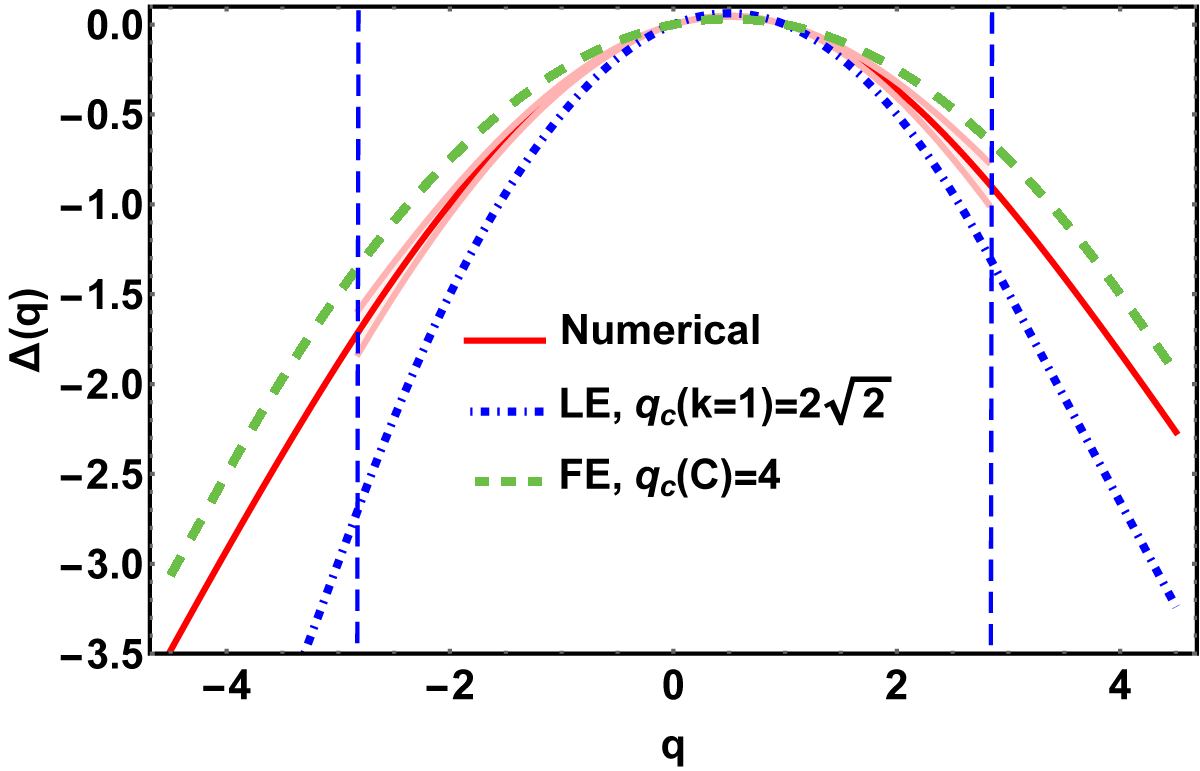}
}
\subfigure{
\includegraphics[width=0.4\textwidth]{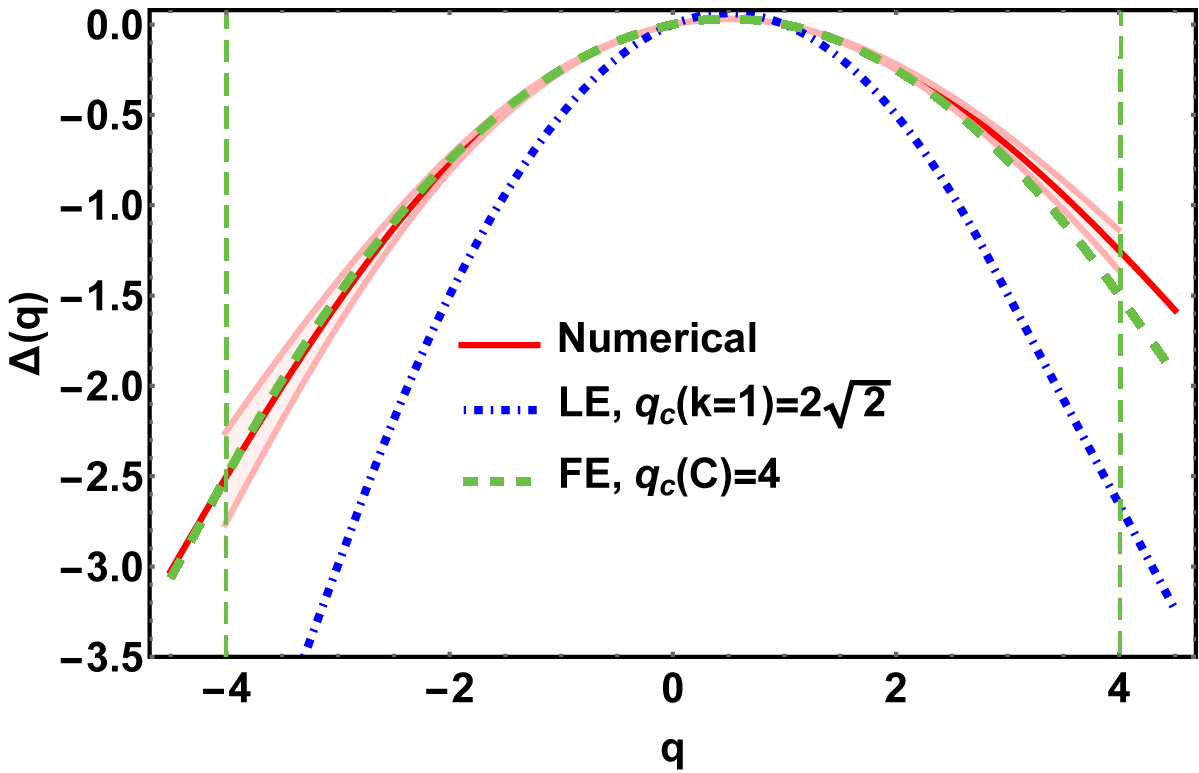}
}
\\
\subfigure{
\includegraphics[height=0.2\textheight]{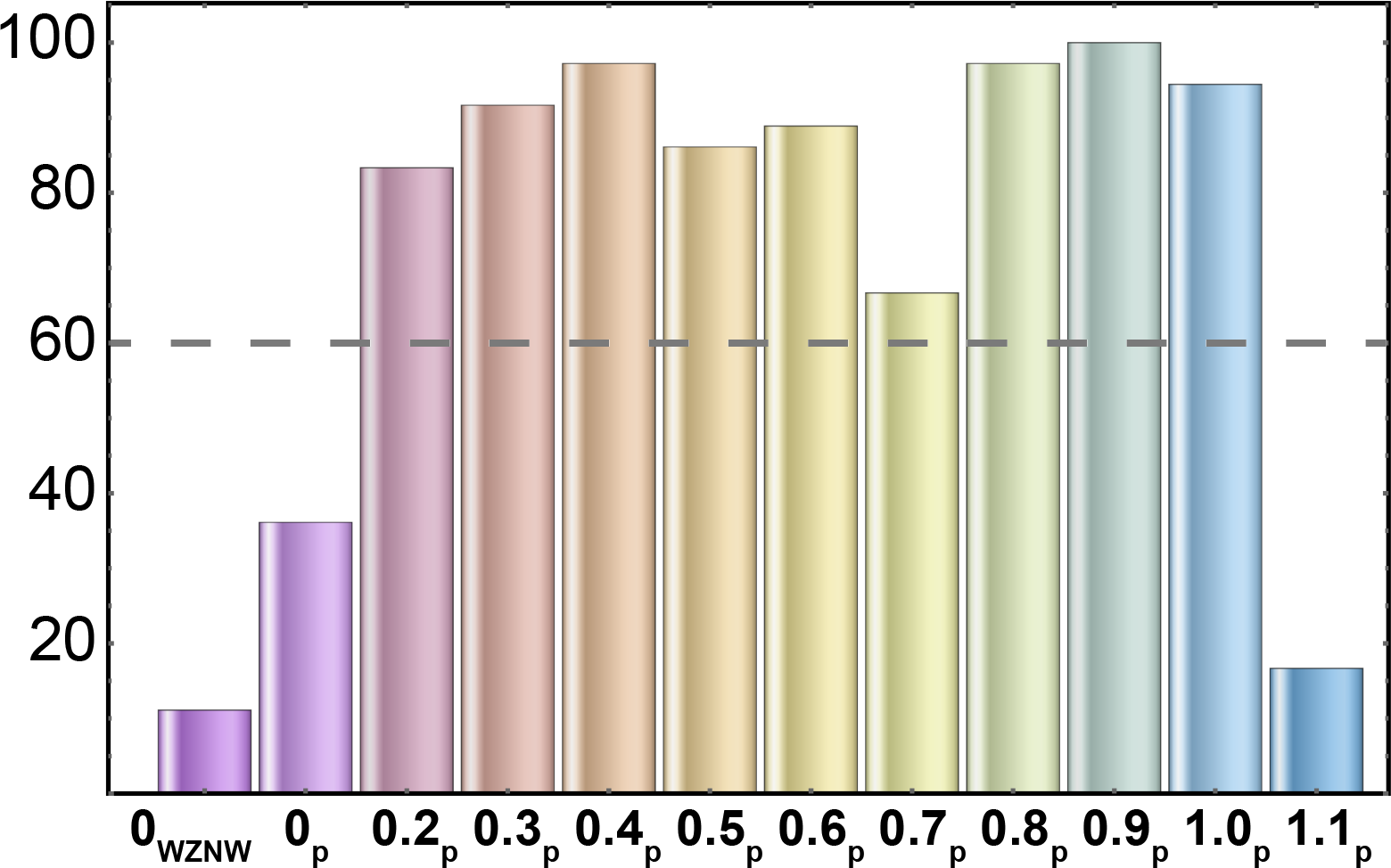}
}
\hspace{14pt}
\subfigure{
\includegraphics[height=0.2\textheight]{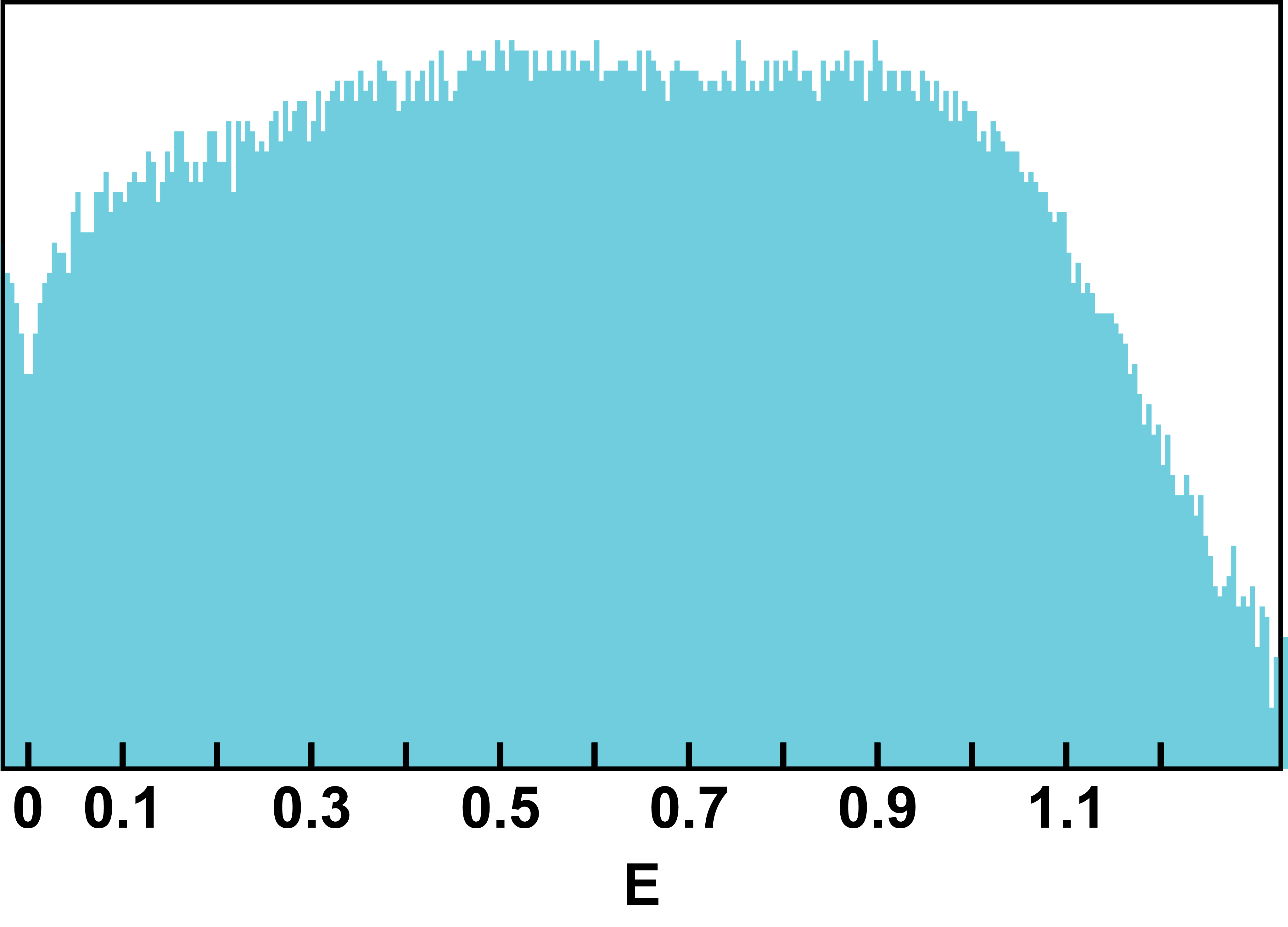}
}
\caption{$k = 1$, $N = 44$. Box sizes $b = 2,22$.}
\end{figure}


\begin{figure}[h!]
\subfigure{
\includegraphics[width=0.4\textwidth]{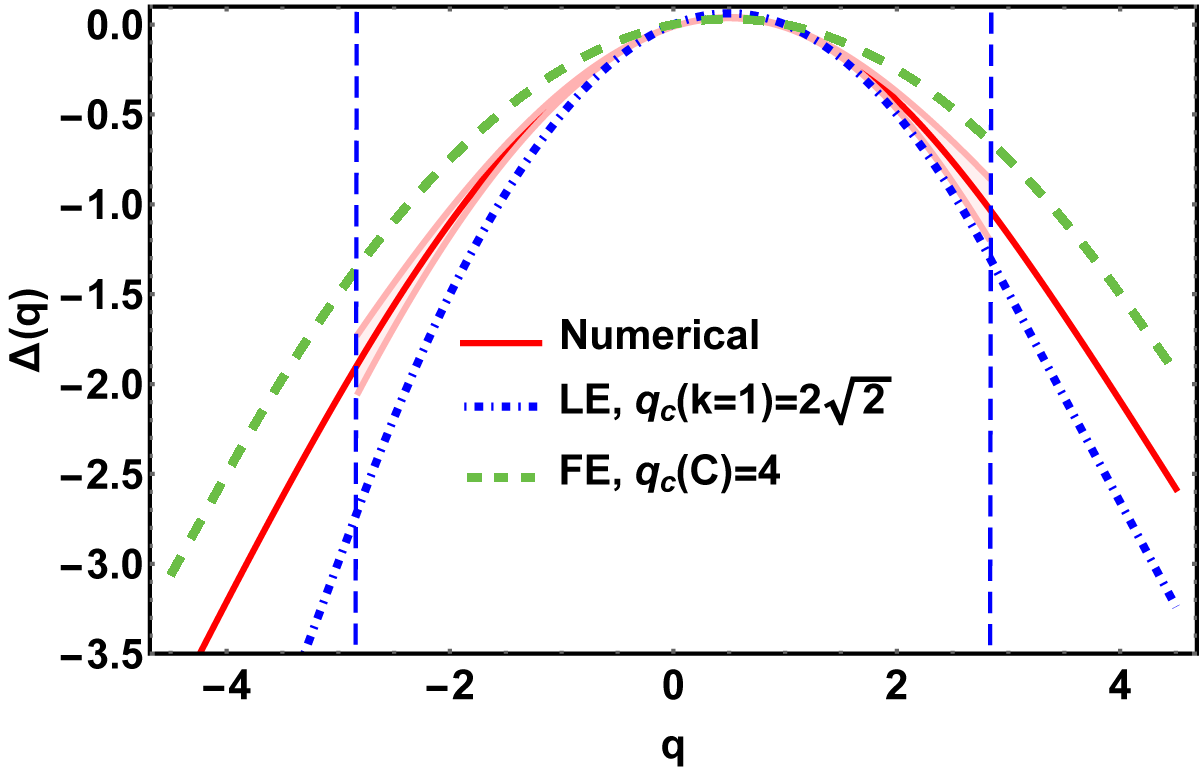}
}
\subfigure{
\includegraphics[width=0.4\textwidth]{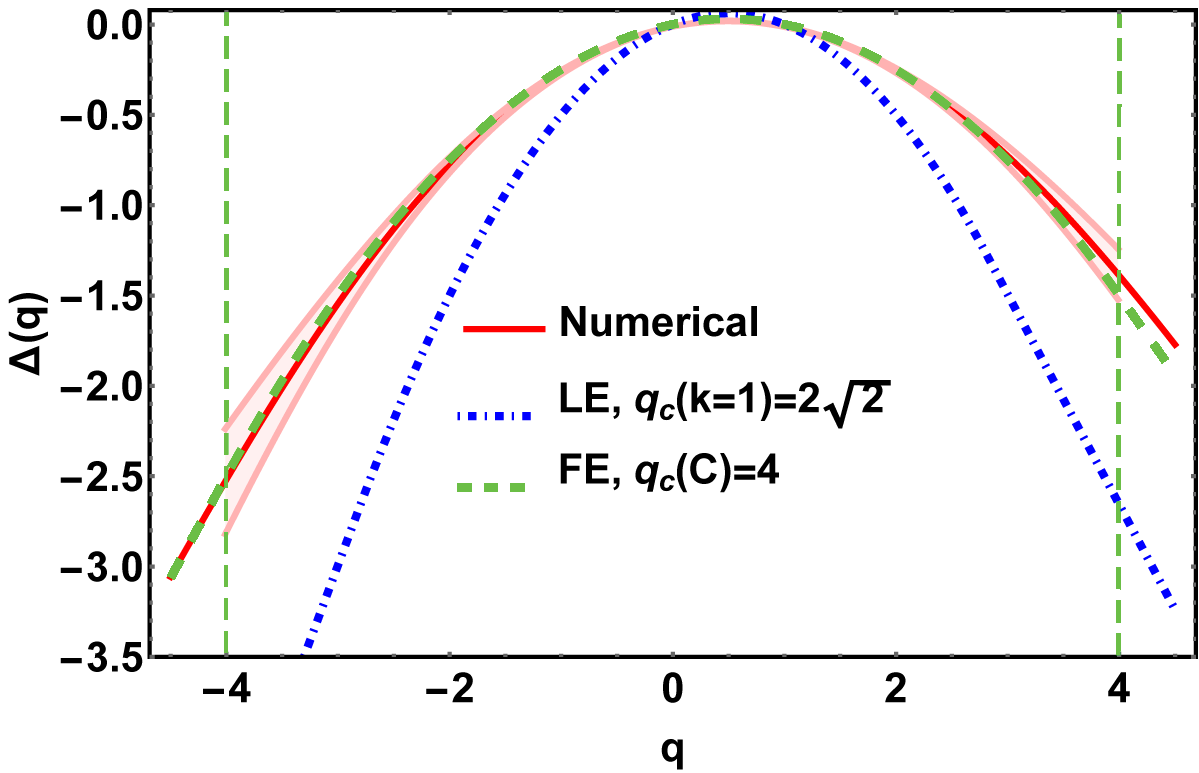}
}
\\
\subfigure{
\includegraphics[height=0.2\textheight]{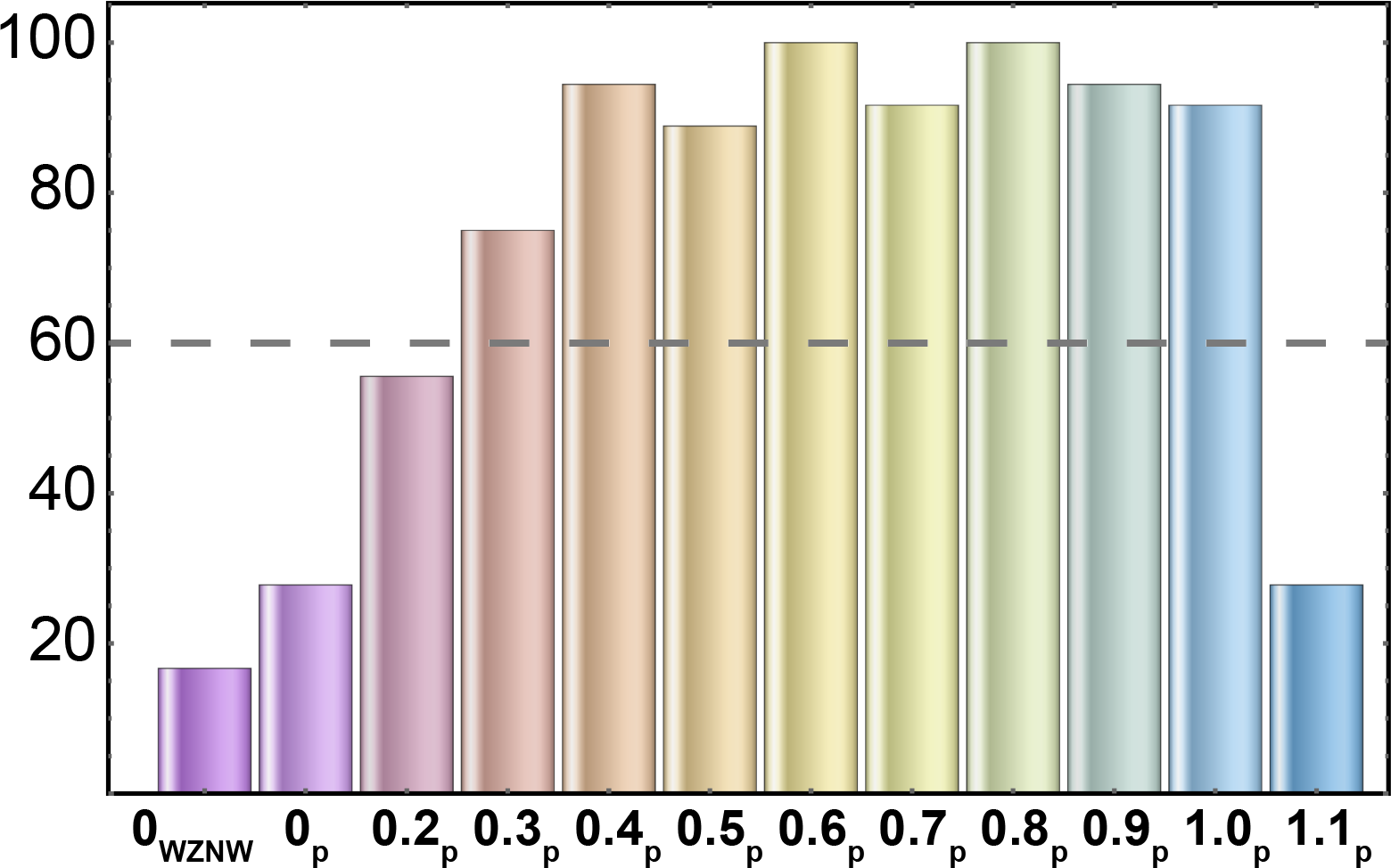}
}
\hspace{14pt}
\subfigure{
\includegraphics[height=0.2\textheight]{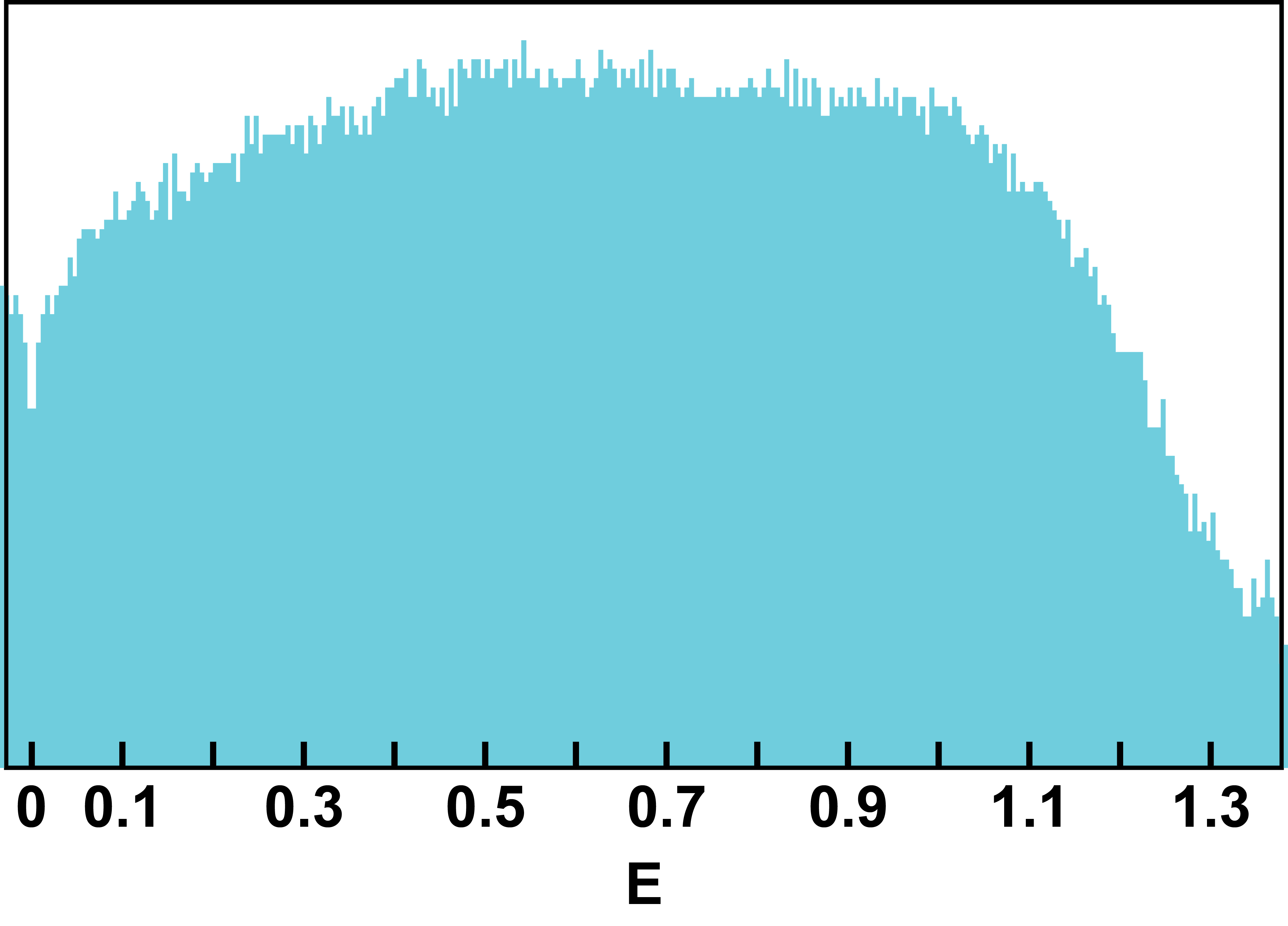}
}
\caption{$k = 1$, $N = 46$. Box sizes $b = 2,13$.}
\end{figure}


\begin{figure}[h!]
\subfigure{
\includegraphics[width=0.4\textwidth]{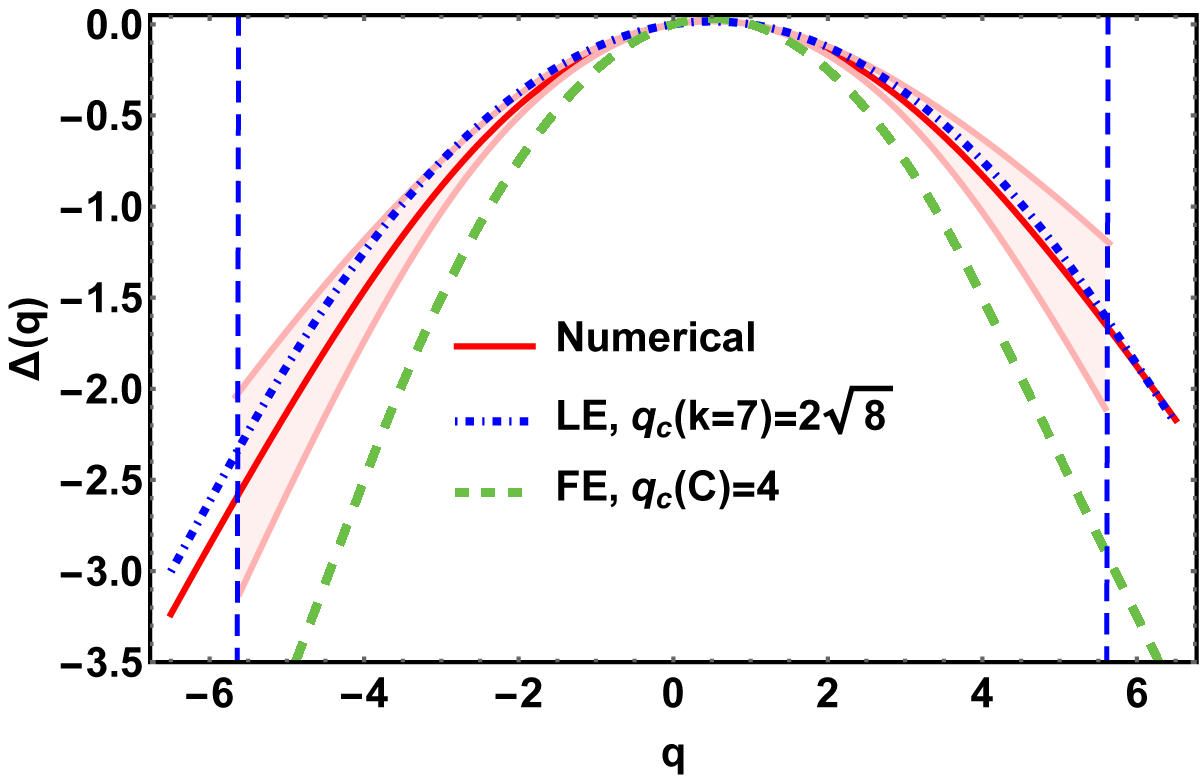}
}
\subfigure{
\includegraphics[width=0.4\textwidth]{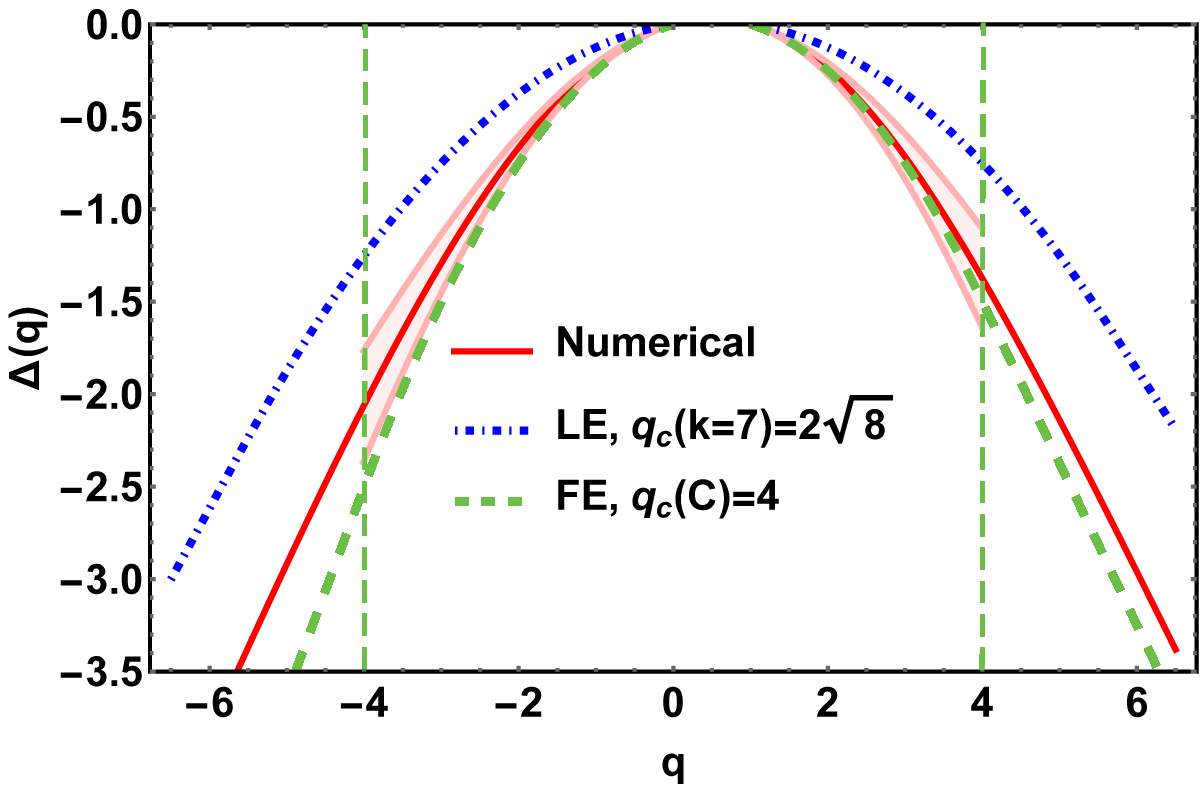}
}
\\
\subfigure{
\includegraphics[height=0.2\textheight]{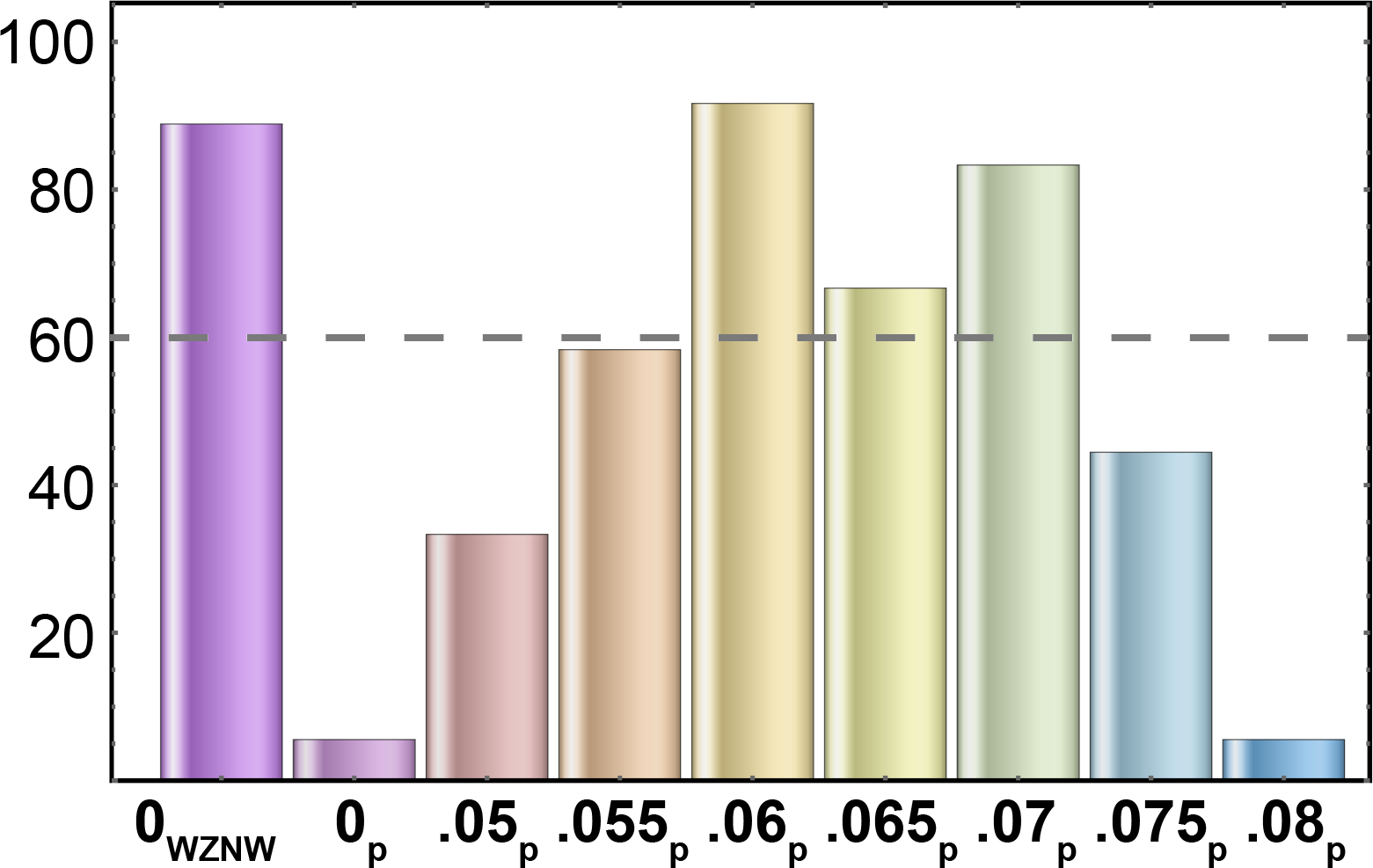}
}
\hspace{14pt}
\subfigure{
\includegraphics[height=0.2\textheight]{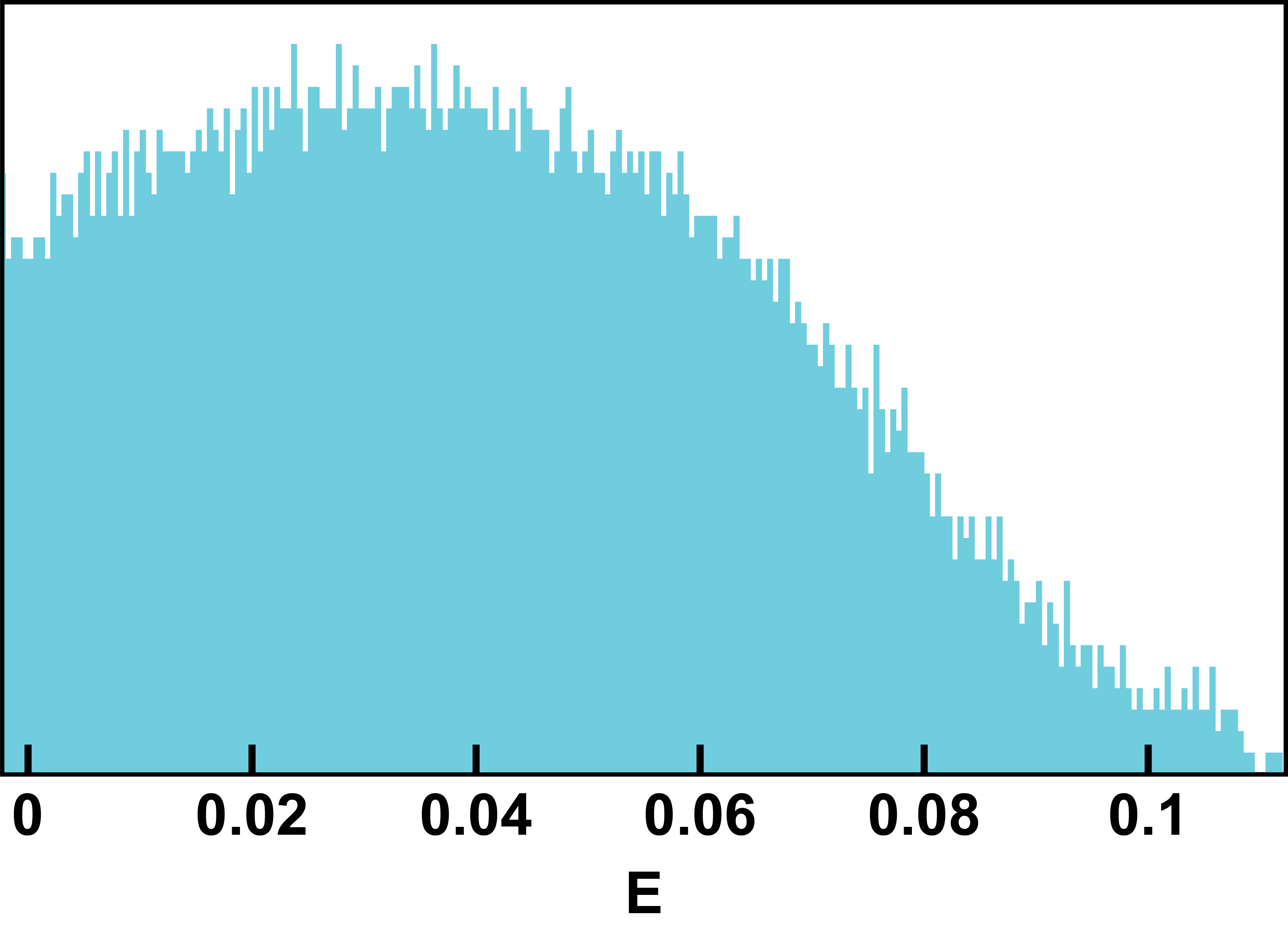}
}
\caption{$k = 7$, $N = 24$. Box sizes $b = 3,12$.}
\end{figure}


\begin{figure}[h!]
\subfigure{
\includegraphics[width=0.4\textwidth]{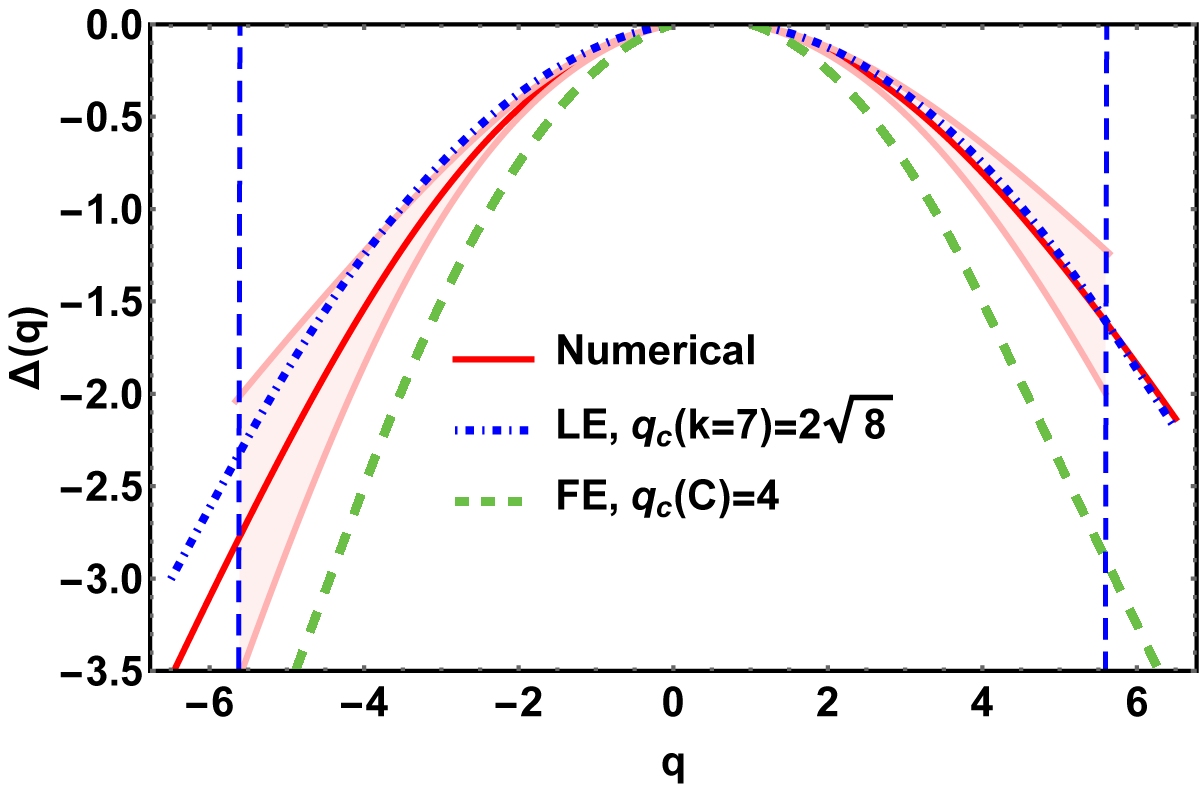}
}
\subfigure{
\includegraphics[width=0.4\textwidth]{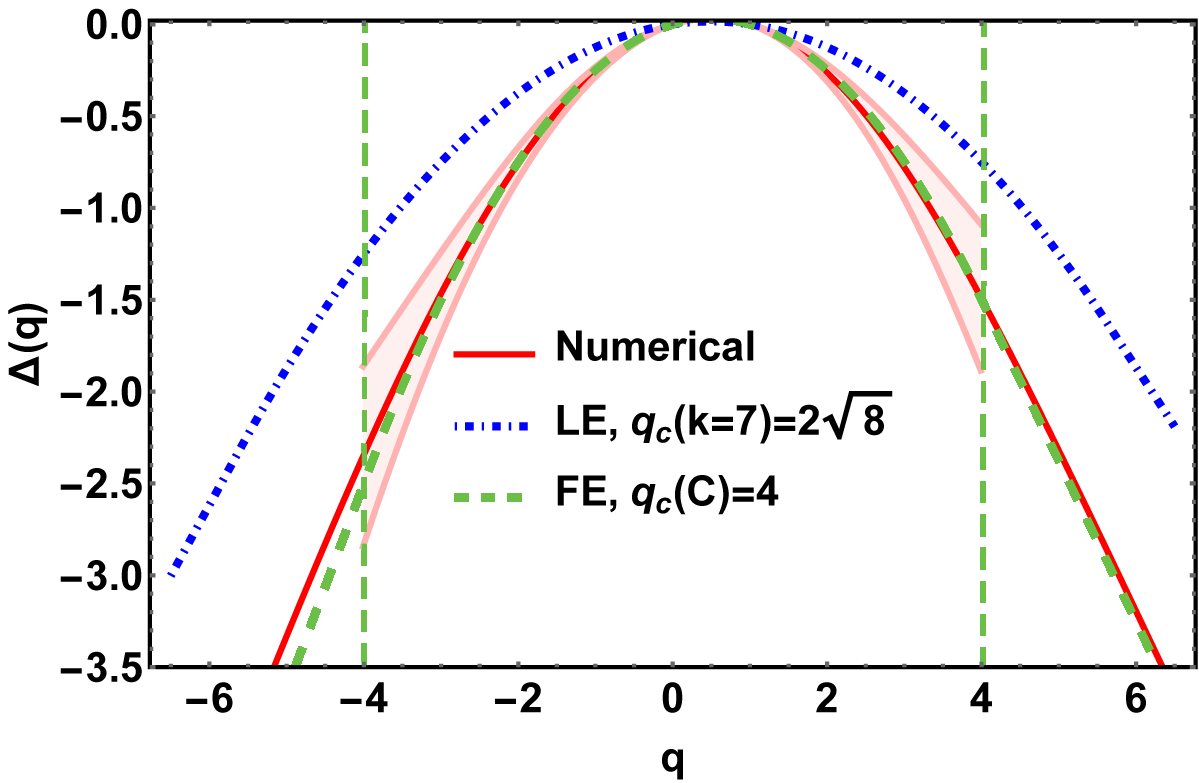}
}
\\
\subfigure{
\includegraphics[height=0.2\textheight]{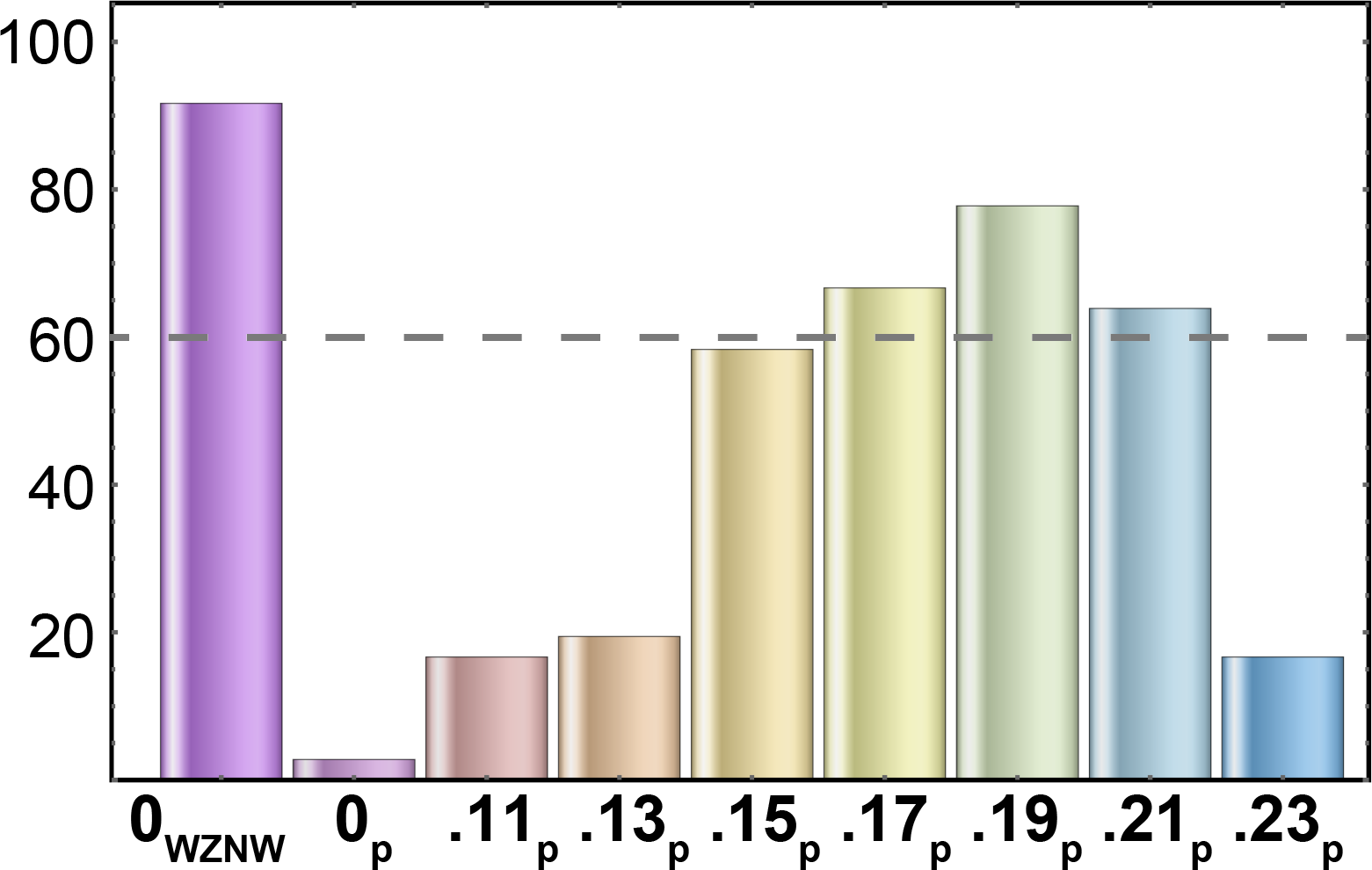}
}
\hspace{14pt}
\subfigure{
\includegraphics[height=0.2\textheight]{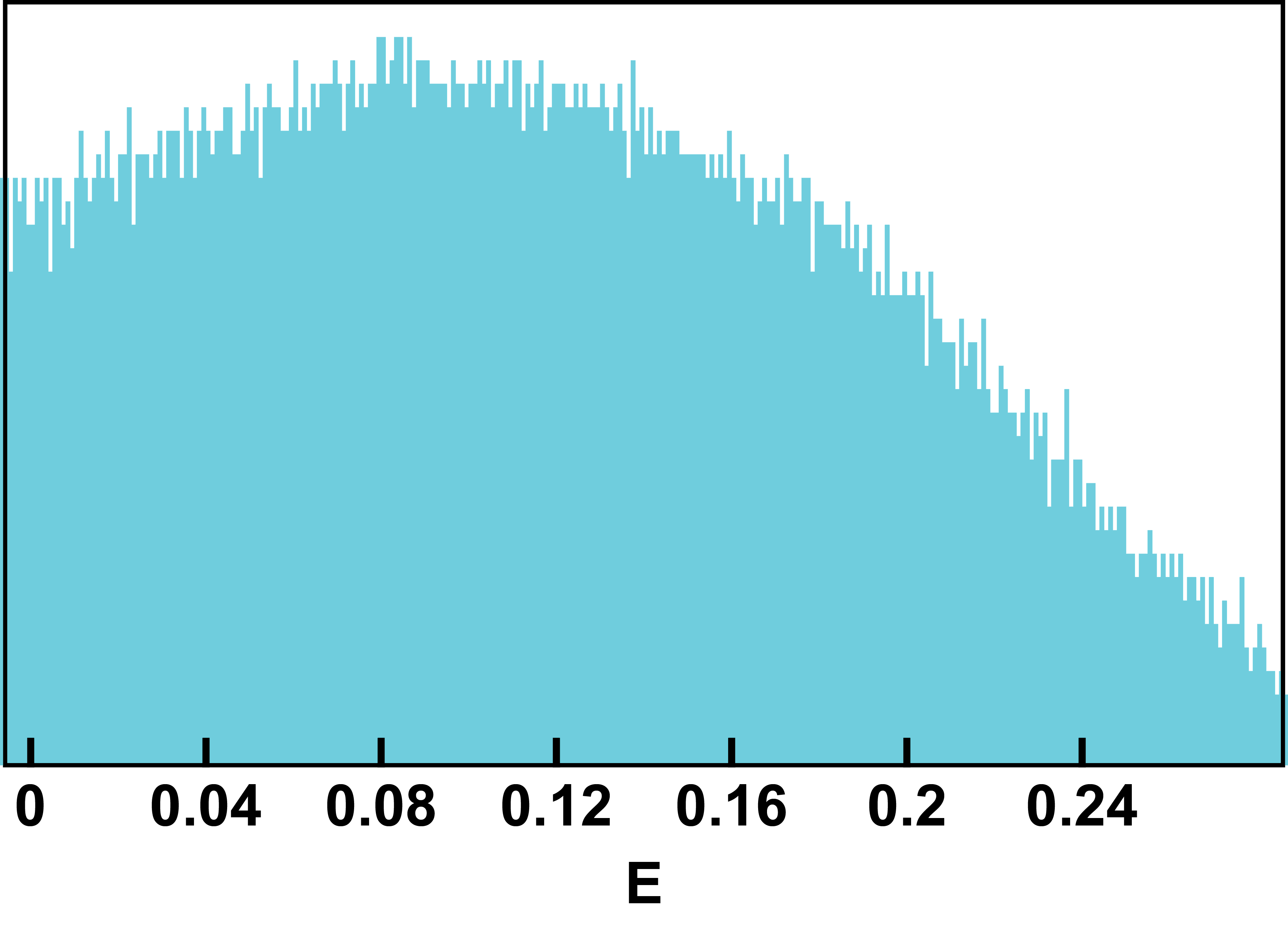}
}
\caption{$k = 7$, $N = 28$. Box sizes $b = 4,8$.}
\end{figure}


\begin{figure}[h!]
\subfigure{
\includegraphics[width=0.4\textwidth]{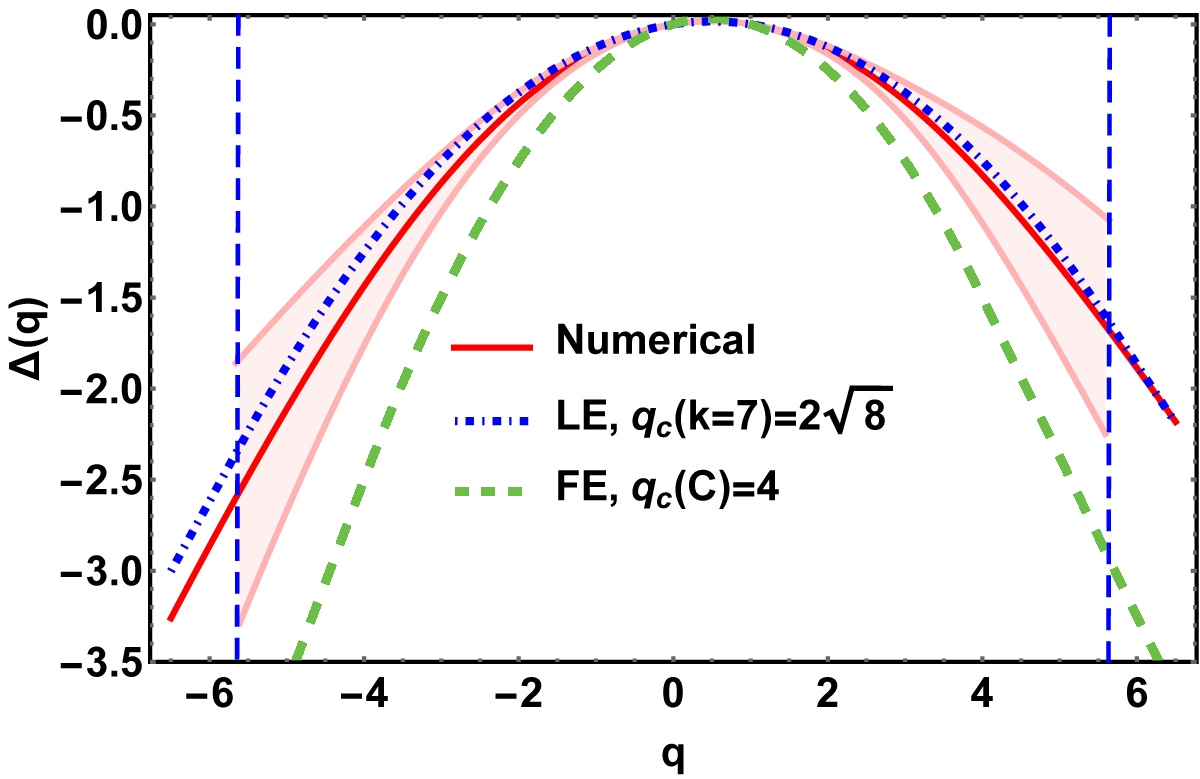}
}
\subfigure{
\includegraphics[width=0.4\textwidth]{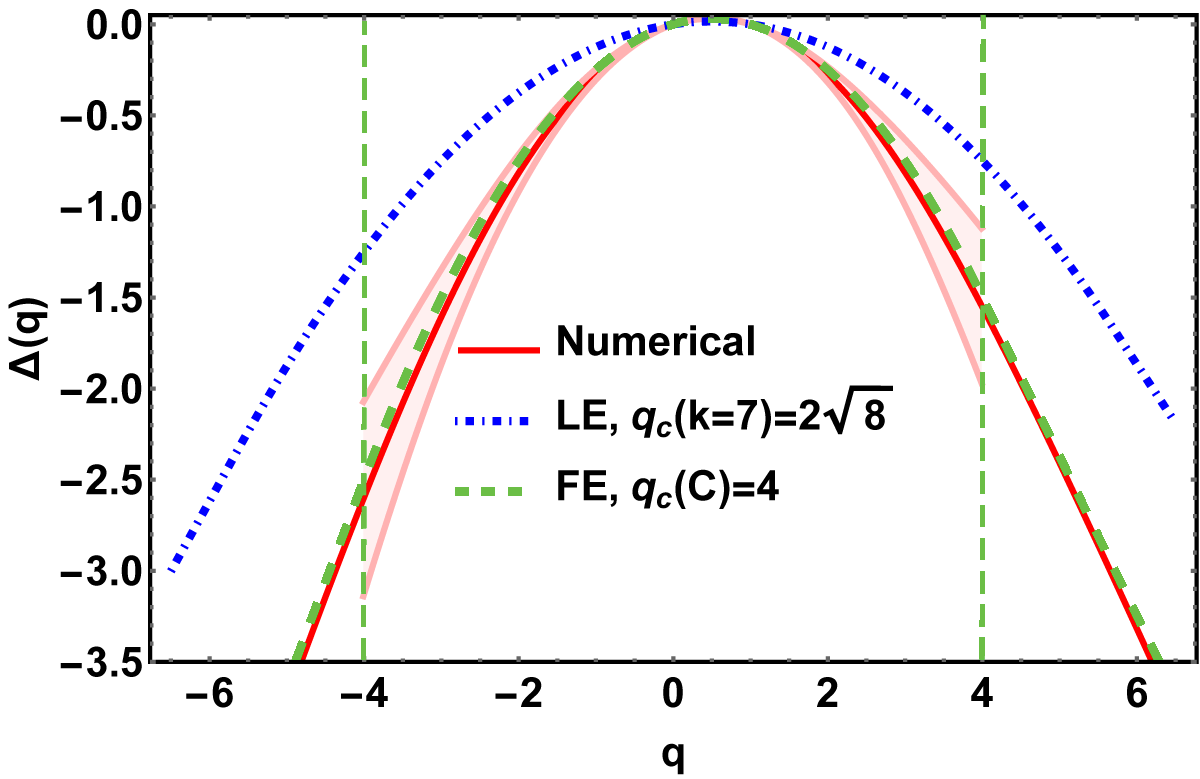}
}
\\
\subfigure{
\includegraphics[height=0.2\textheight]{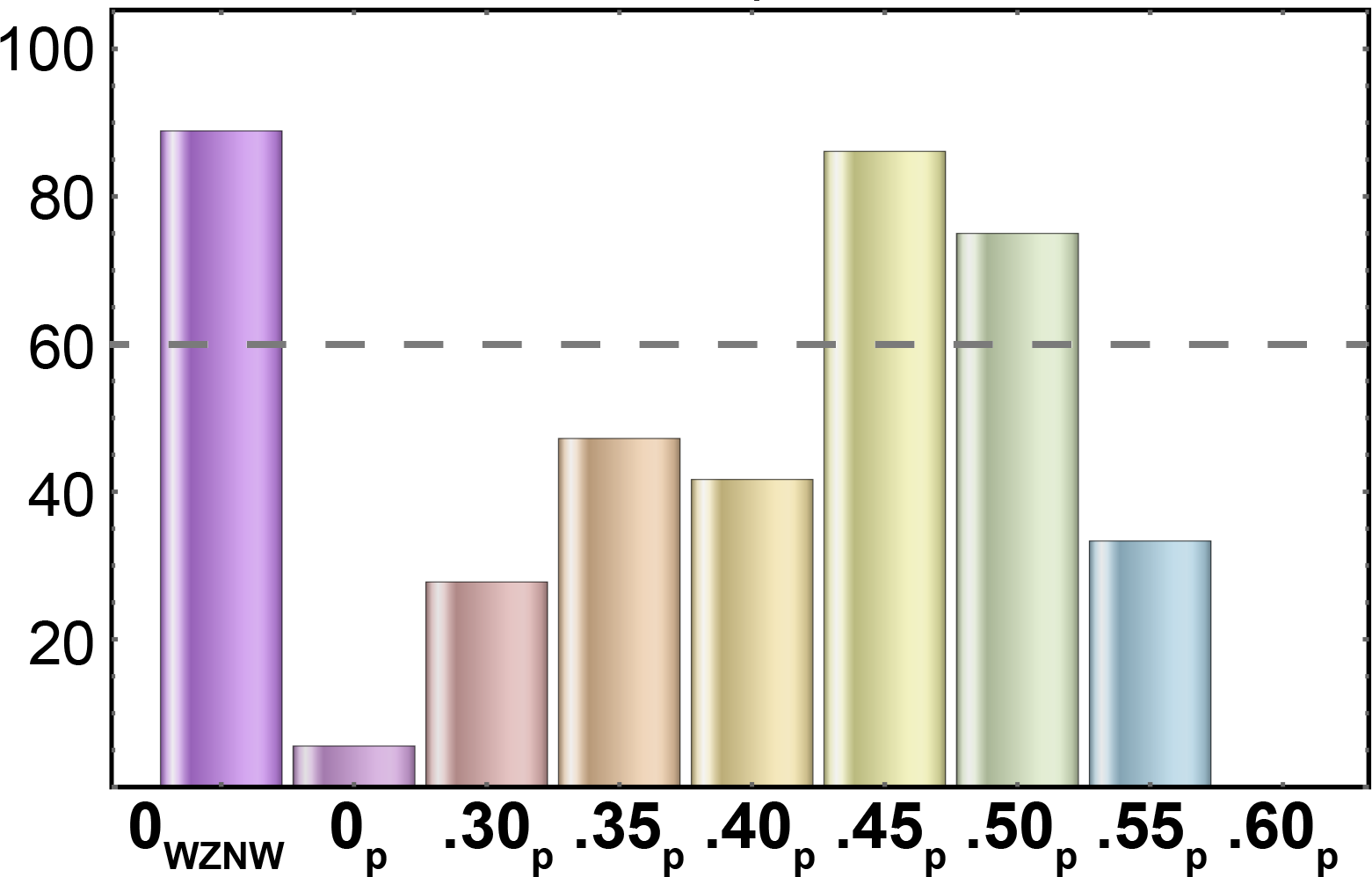}
}
\hspace{14pt}
\subfigure{
\includegraphics[height=0.2\textheight]{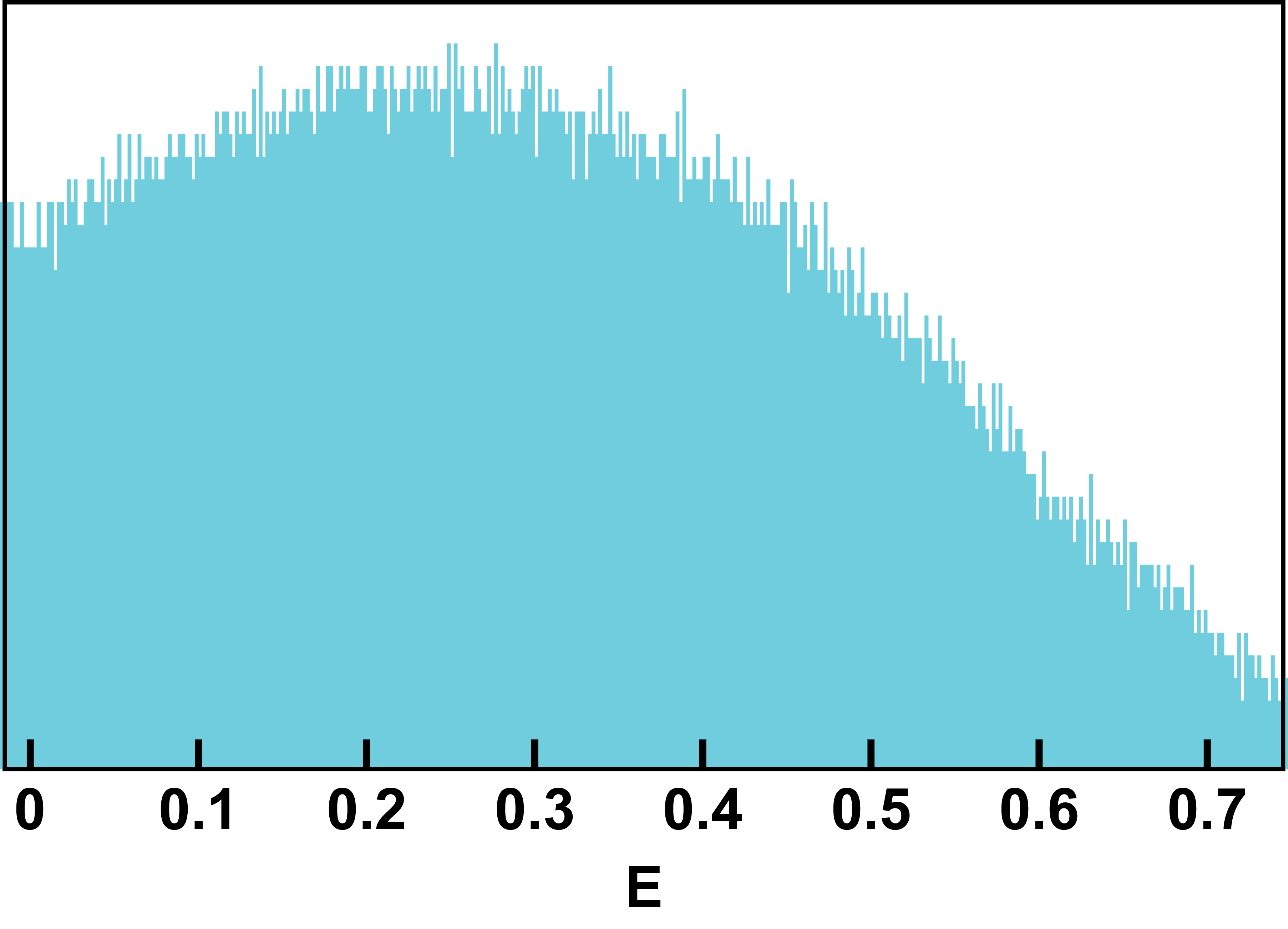}
}
\caption{$k = 7$, $N = 32$. Box sizes $b = 4,8$.}
\end{figure}


\begin{figure}[h!]
\subfigure{
\includegraphics[width=0.4\textwidth]{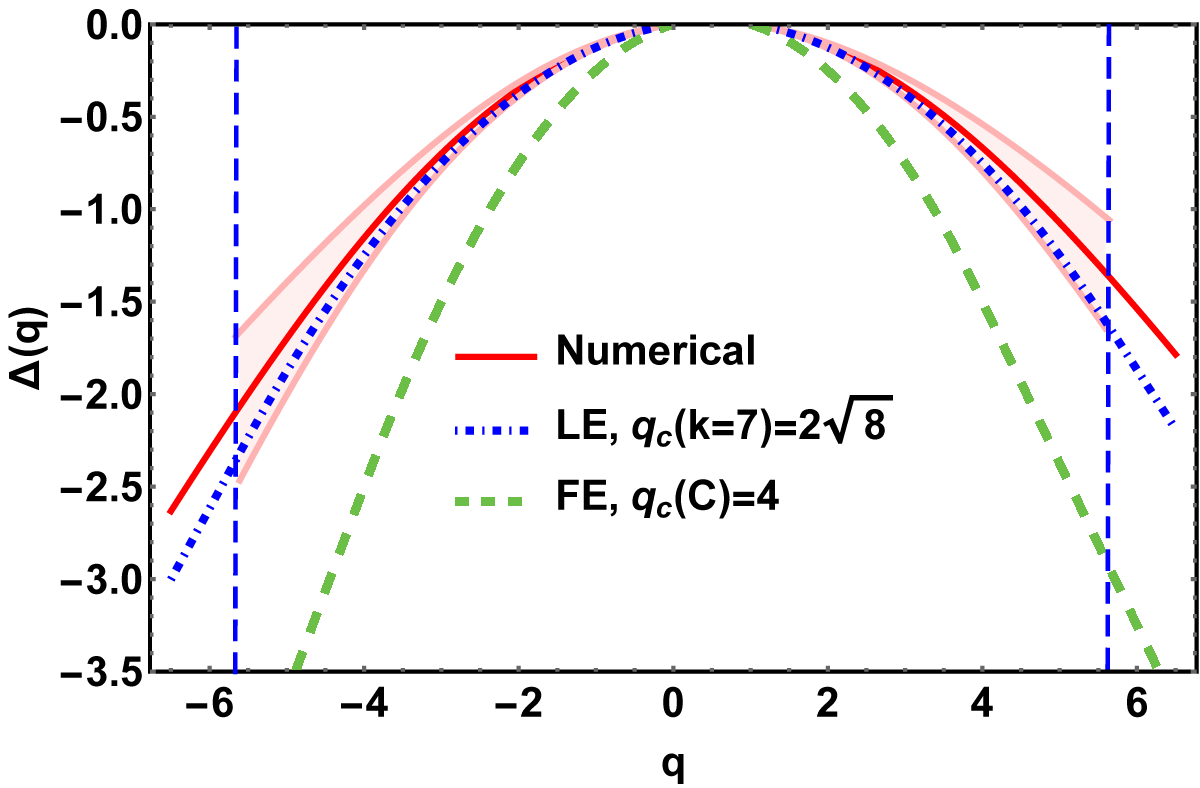}
}
\subfigure{
\includegraphics[width=0.4\textwidth]{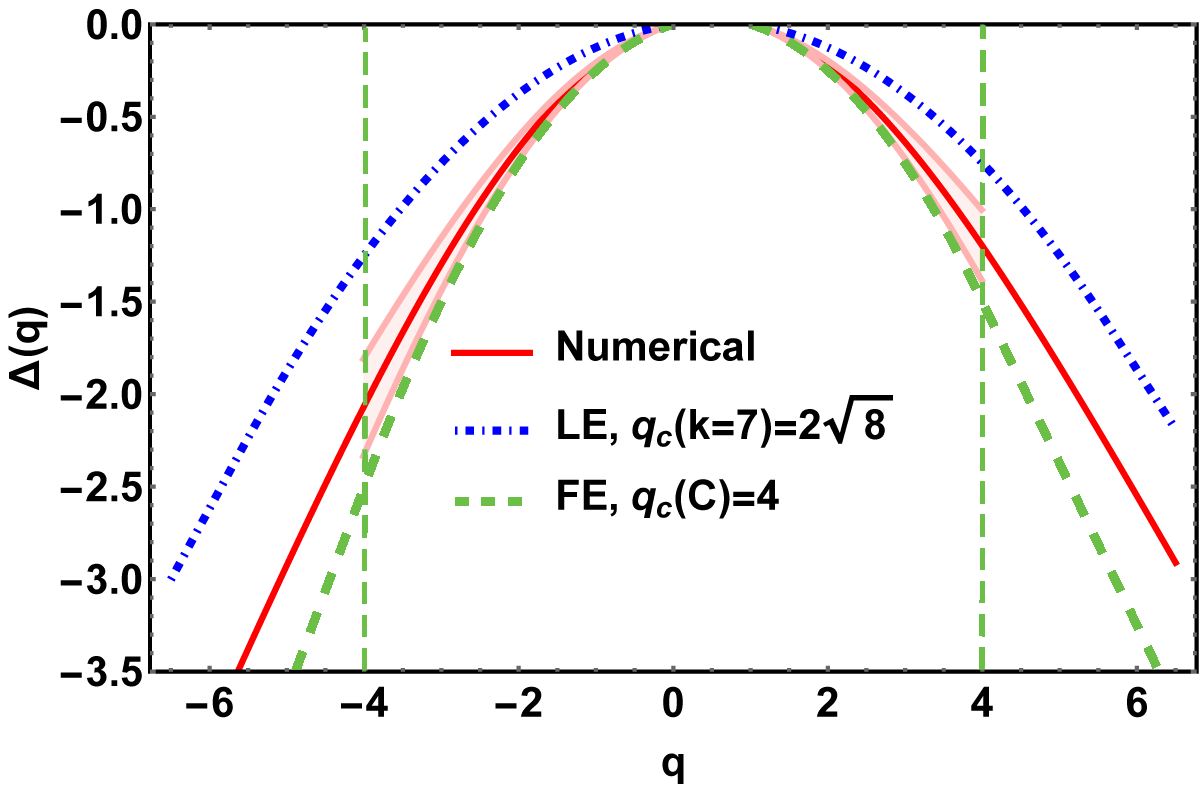}
}
\\
\subfigure{
\includegraphics[height=0.2\textheight]{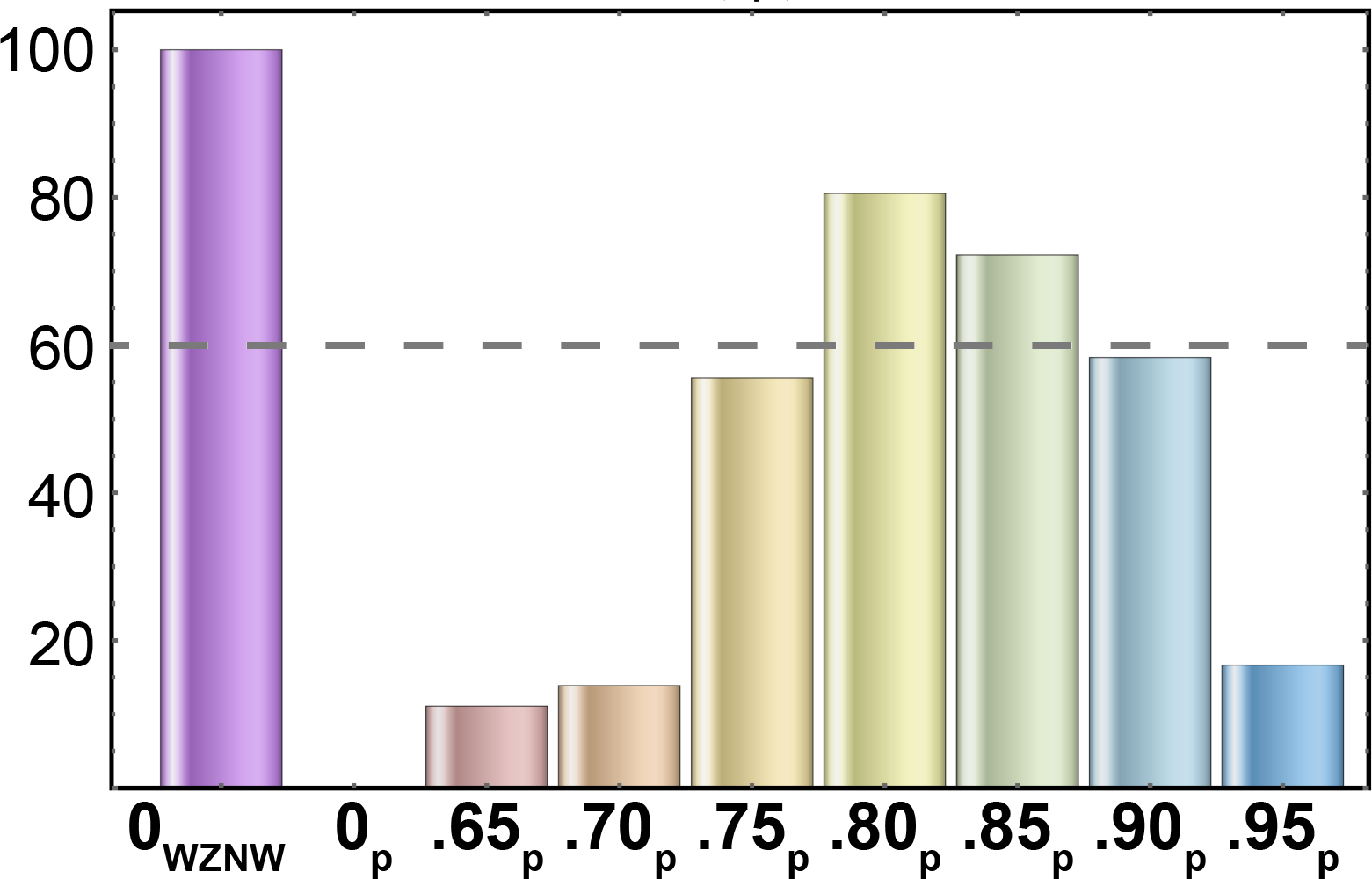}
}
\hspace{14pt}
\subfigure{
\includegraphics[height=0.2\textheight]{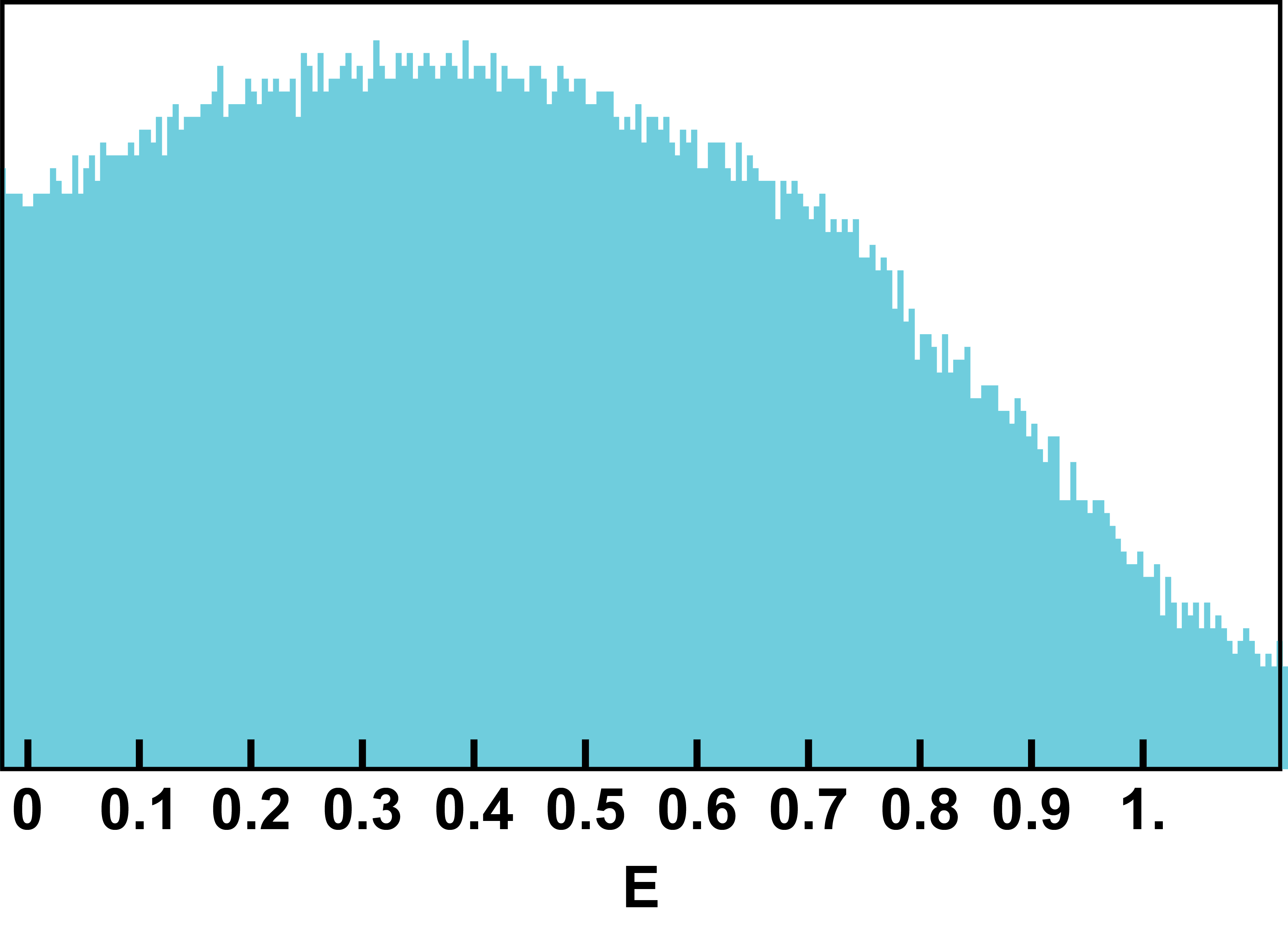}
}
\caption{$k = 7$, $N = 34$. Box sizes $b = 4,17$.}
\end{figure}


\begin{figure}[h!]
\subfigure{
\includegraphics[width=0.4\textwidth]{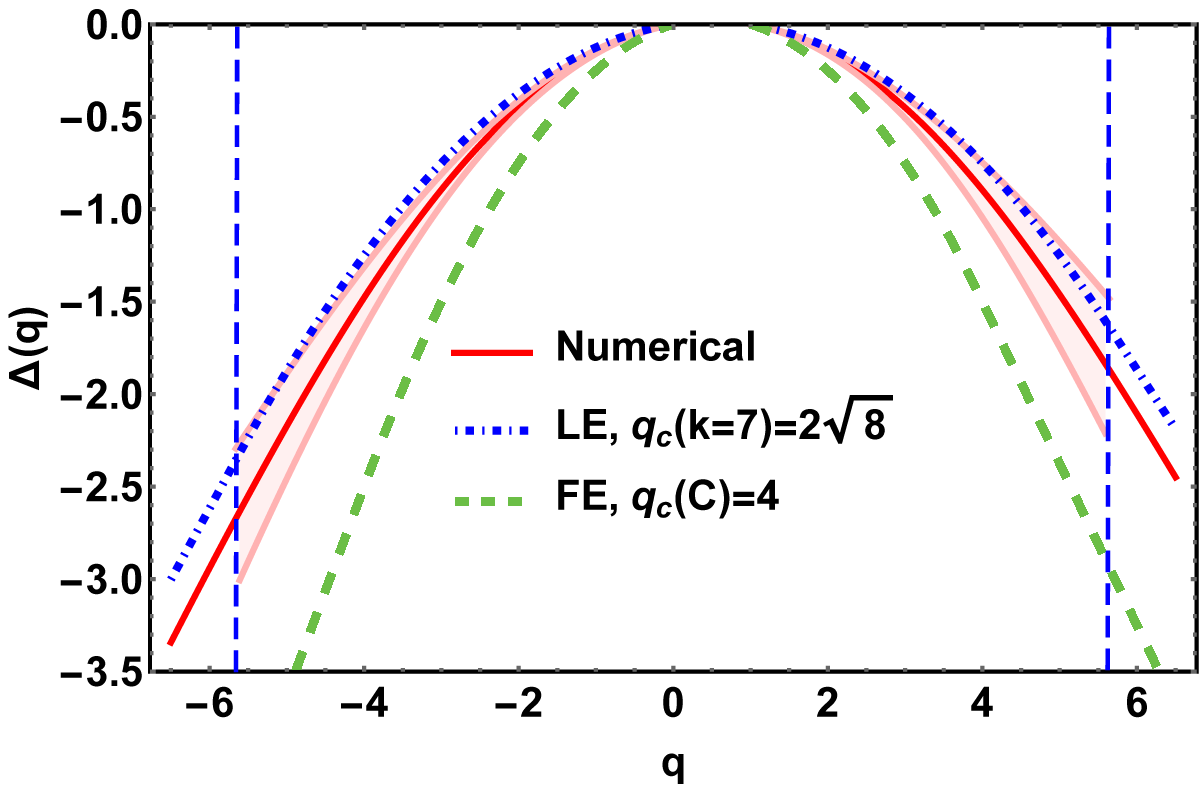}
}
\subfigure{
\includegraphics[width=0.4\textwidth]{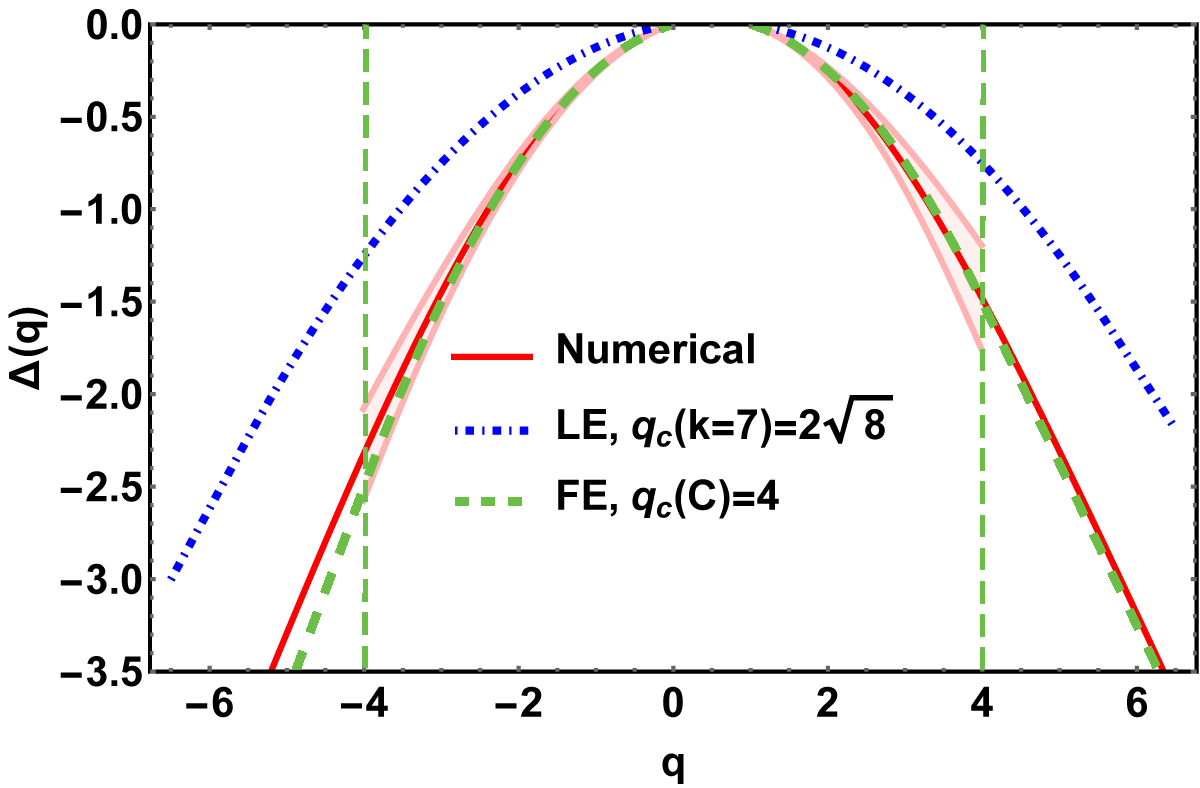}
}
\\
\subfigure{
\includegraphics[height=0.2\textheight]{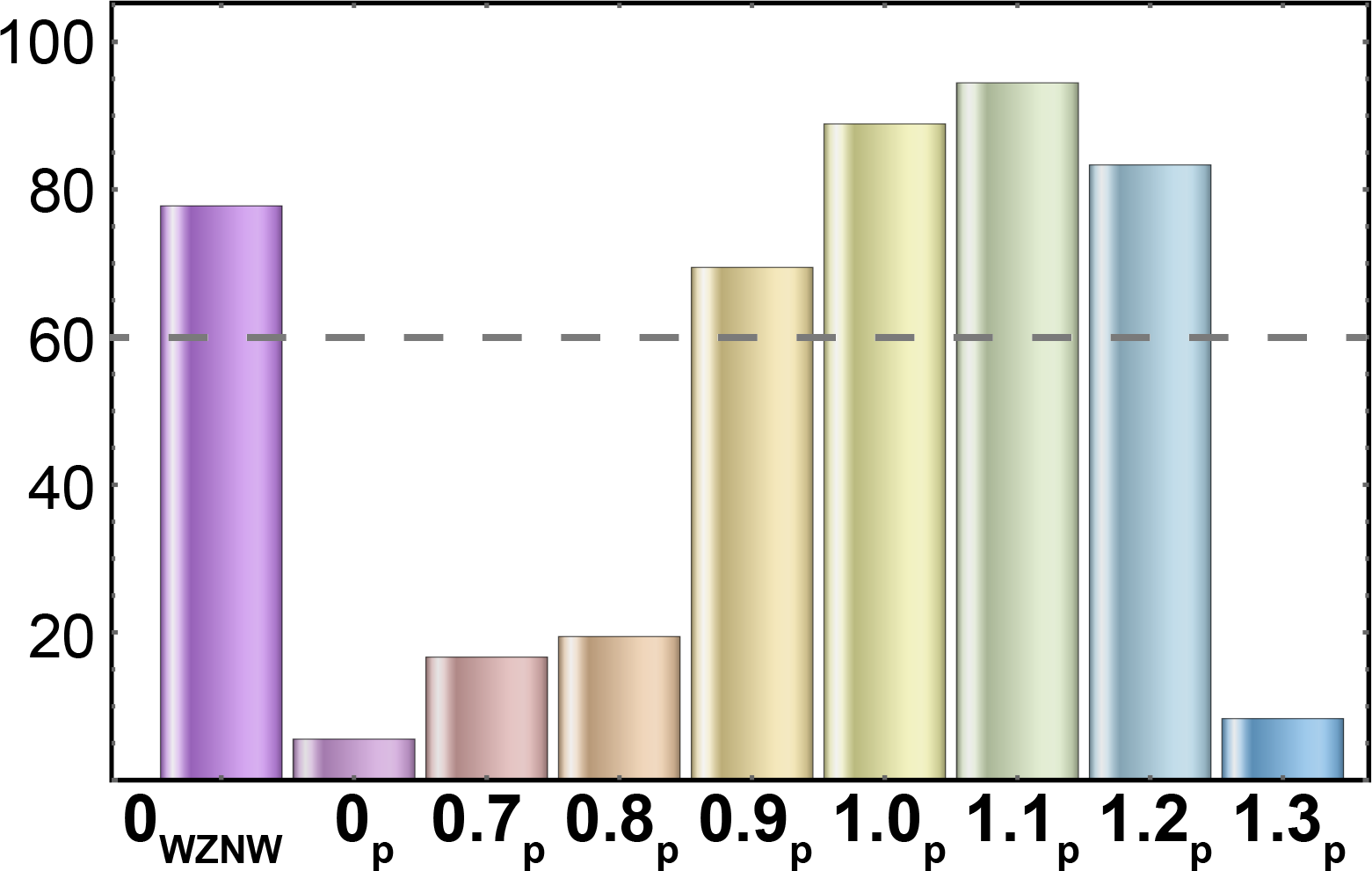}
}
\hspace{14pt}
\subfigure{
\includegraphics[height=0.2\textheight]{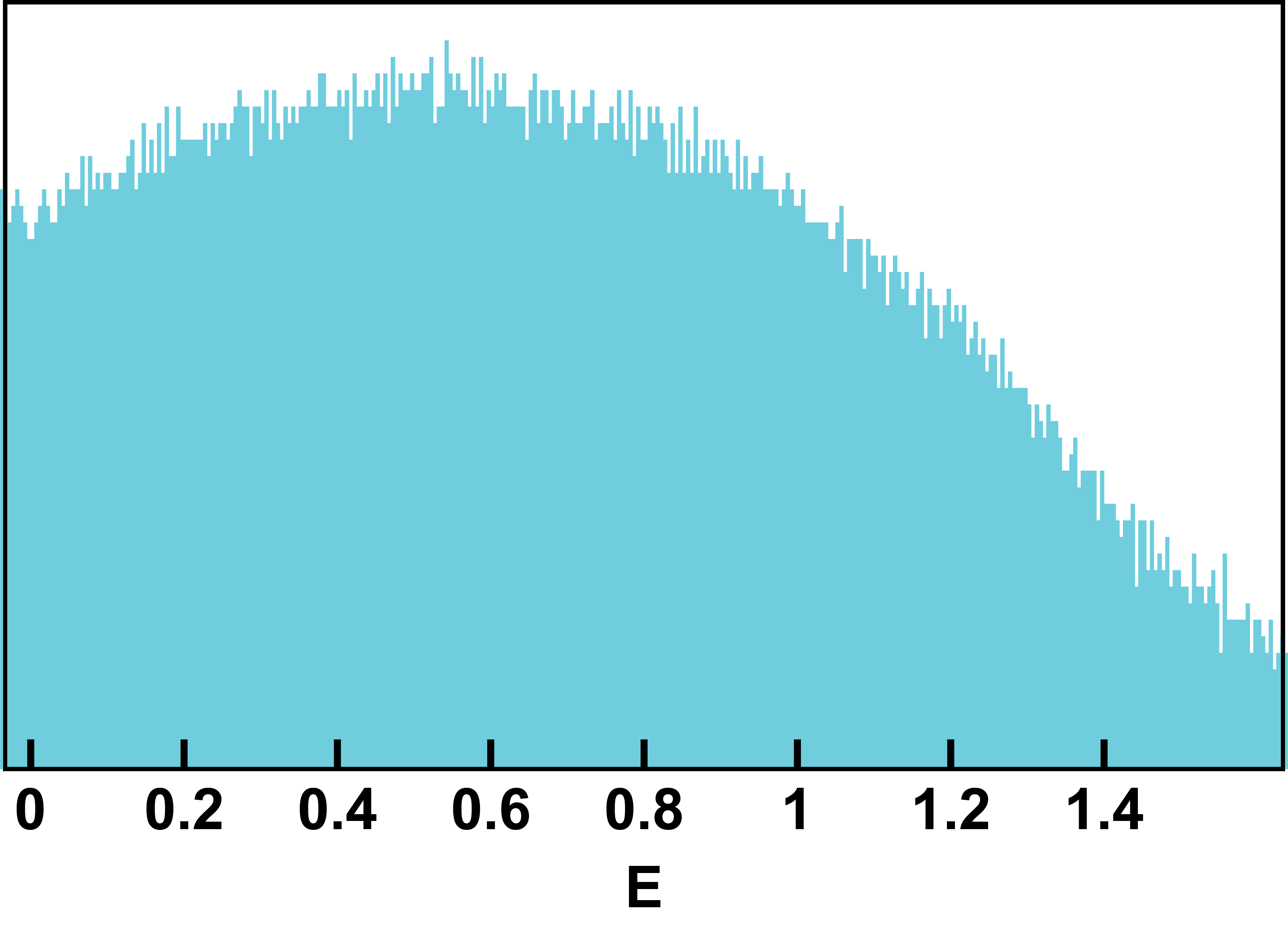}
}
\caption{$k = 7$, $N = 36$. Box sizes $b = 3,12$.}
\end{figure}


\begin{figure}[h!]
\subfigure{
\includegraphics[width=0.4\textwidth]{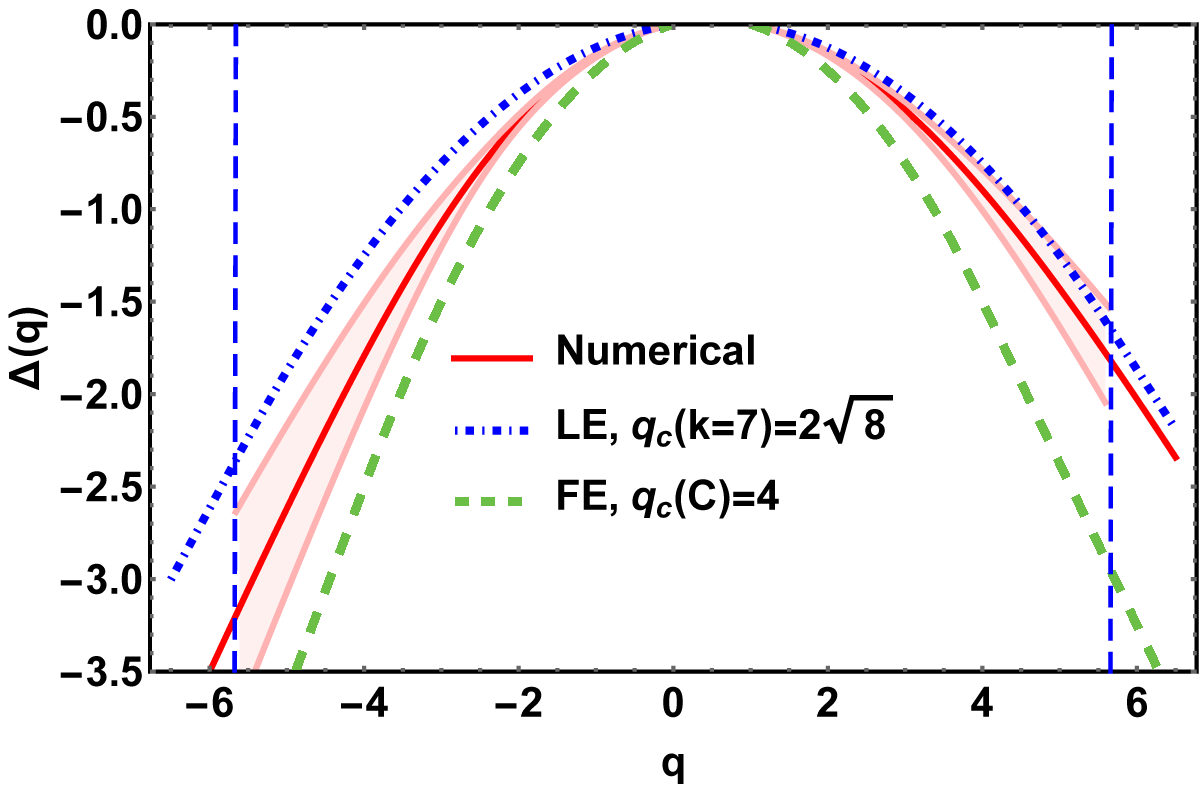}
}
\subfigure{
\includegraphics[width=0.4\textwidth]{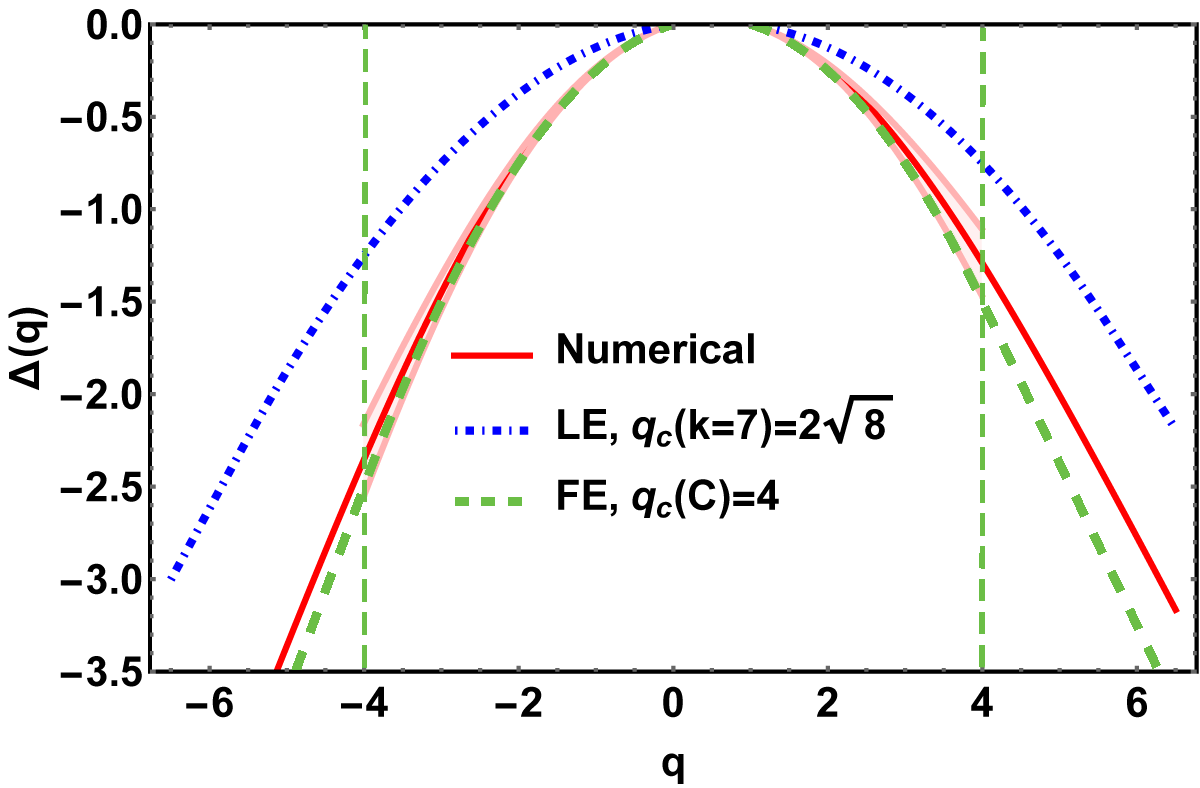}
}
\\
\subfigure{
\includegraphics[height=0.2\textheight]{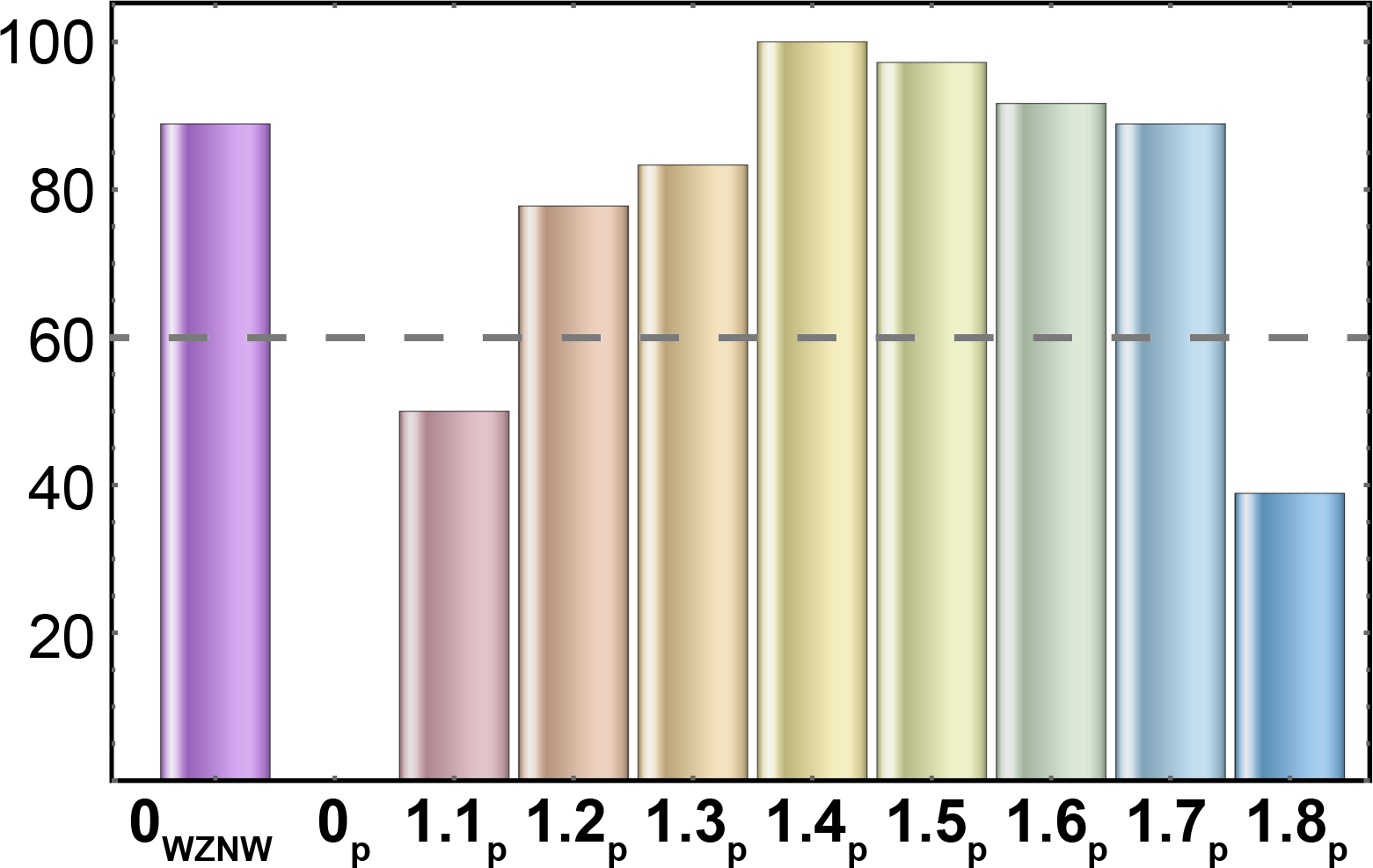}
}
\hspace{14pt}
\subfigure{
\includegraphics[height=0.2\textheight]{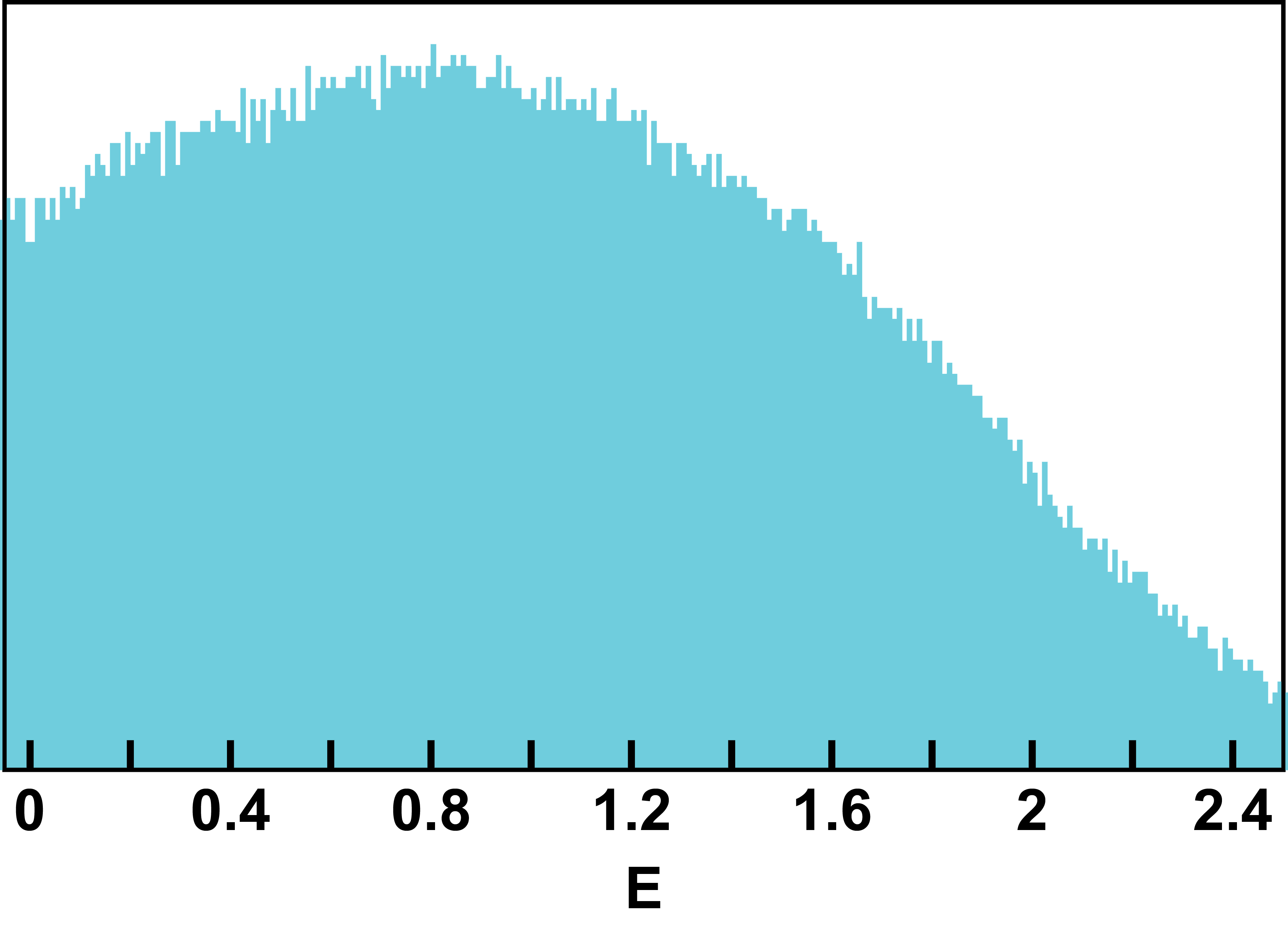}
}
\caption{$k = 7$, $N = 38$. Box sizes $b = 2,19$.}
\end{figure}


\begin{figure}[h!]
\subfigure{
\includegraphics[width=0.4\textwidth]{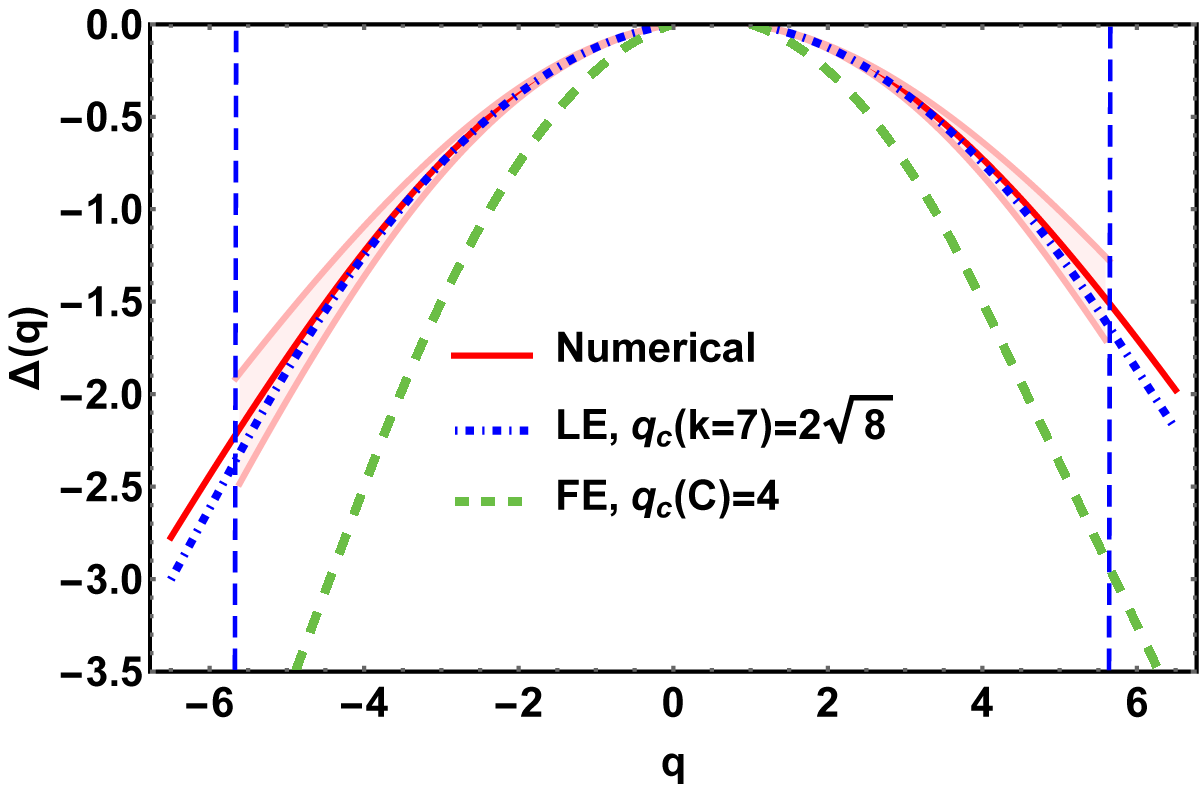}
}
\subfigure{
\includegraphics[width=0.4\textwidth]{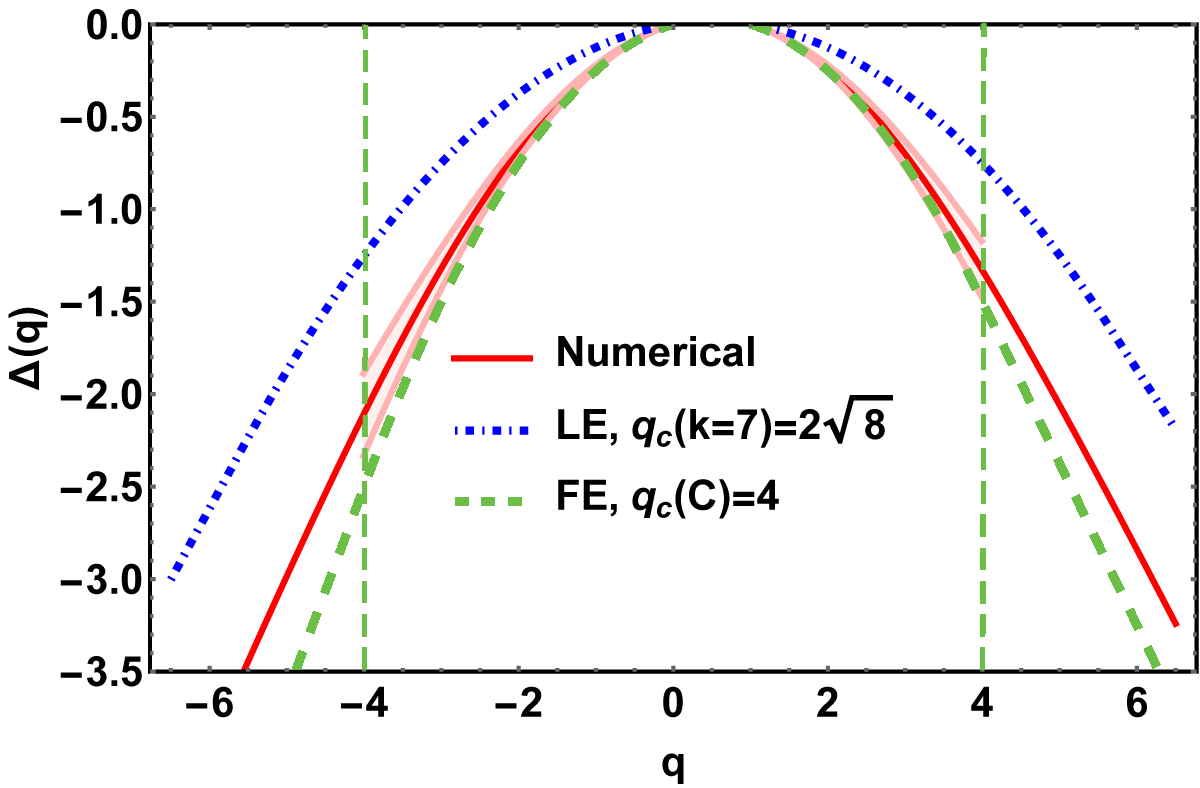}
}
\\
\subfigure{
\includegraphics[height=0.2\textheight]{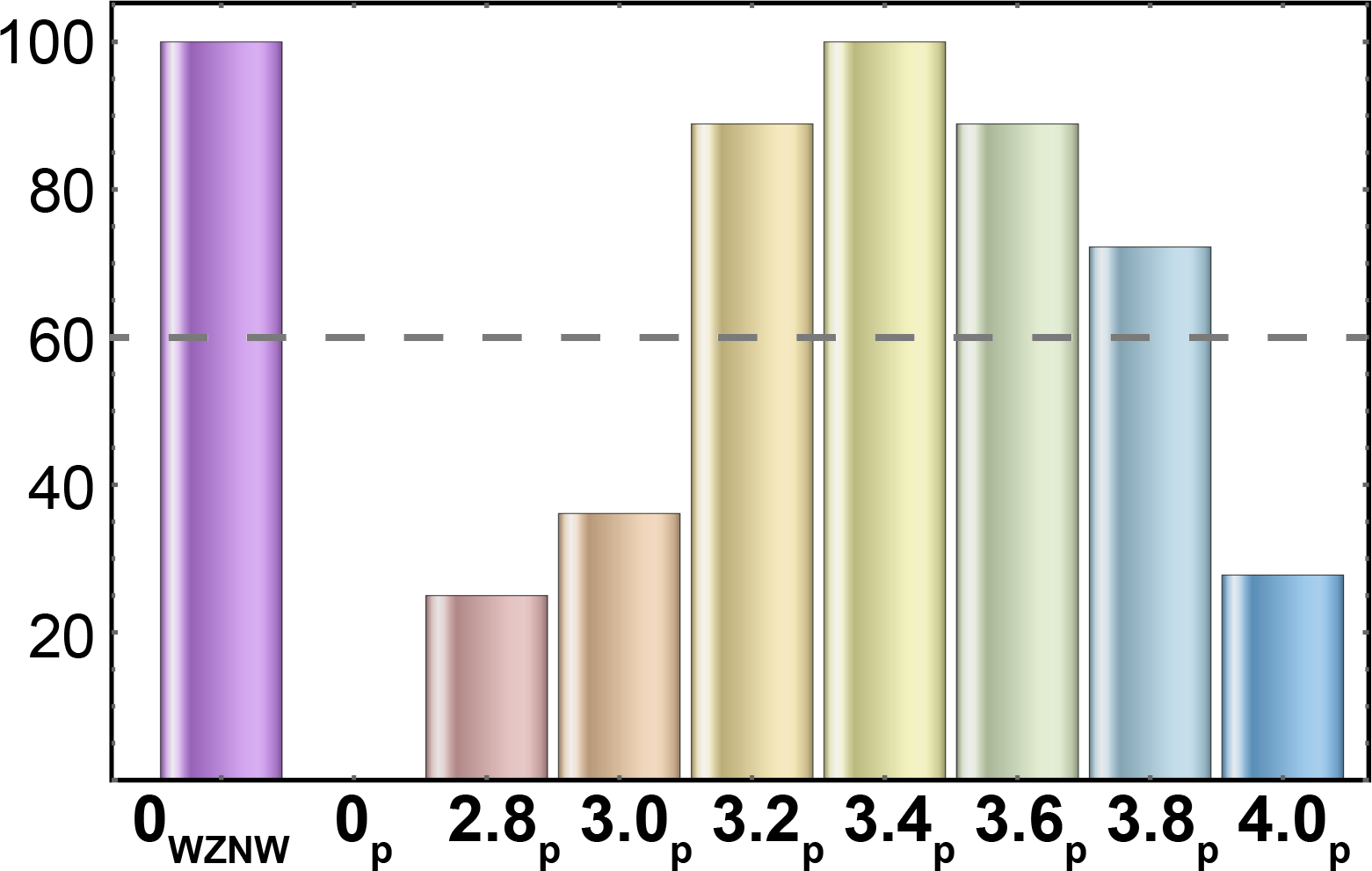}
}
\hspace{14pt}
\subfigure{
\includegraphics[height=0.2\textheight]{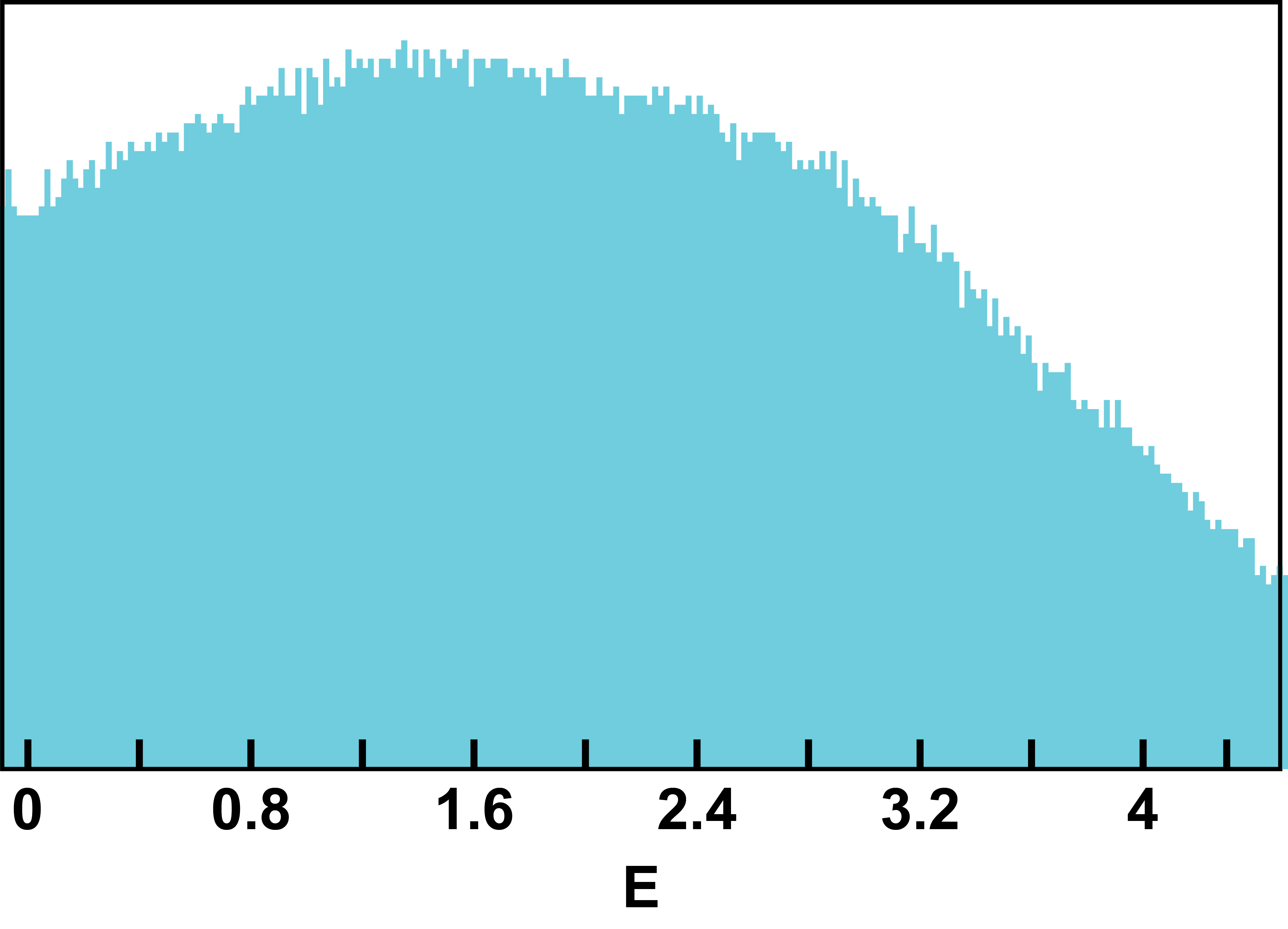}
}
\caption{$k = 7$, $N = 42$. Box sizes $b = 3,21$.}
\end{figure}


\begin{figure}[h!]
\subfigure{
\includegraphics[width=0.4\textwidth]{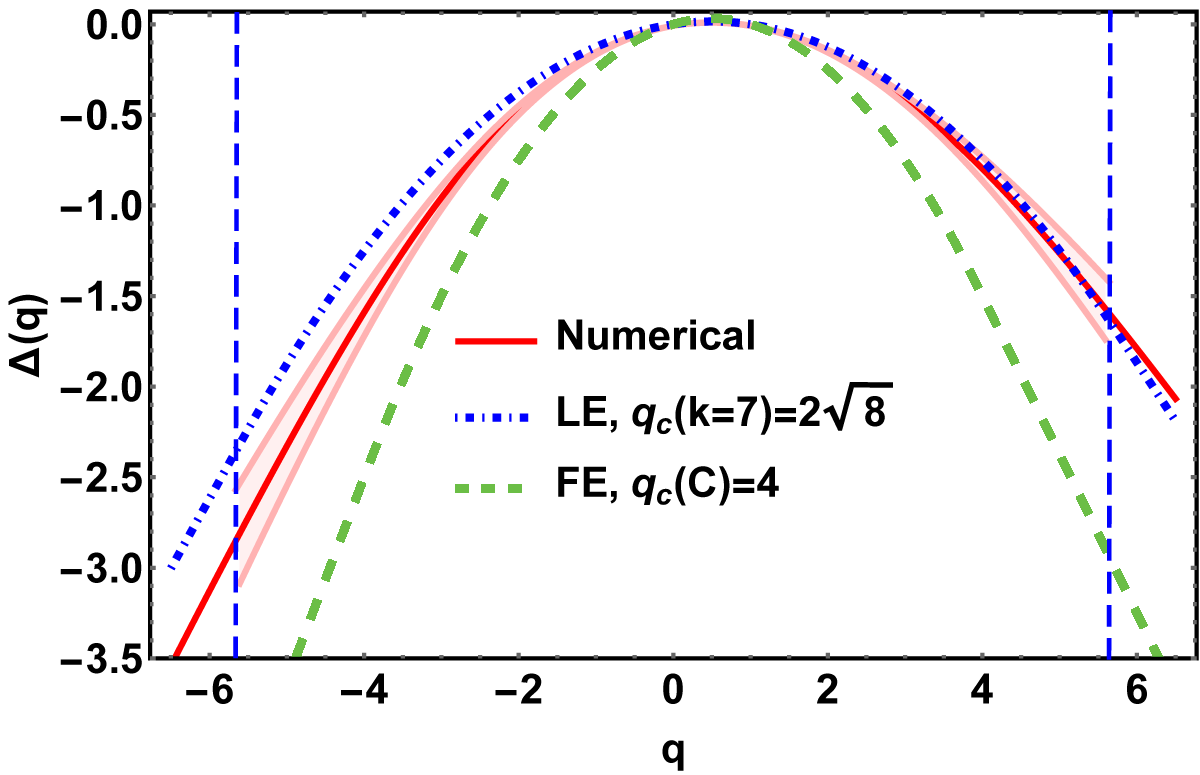}
}
\subfigure{
\includegraphics[width=0.4\textwidth]{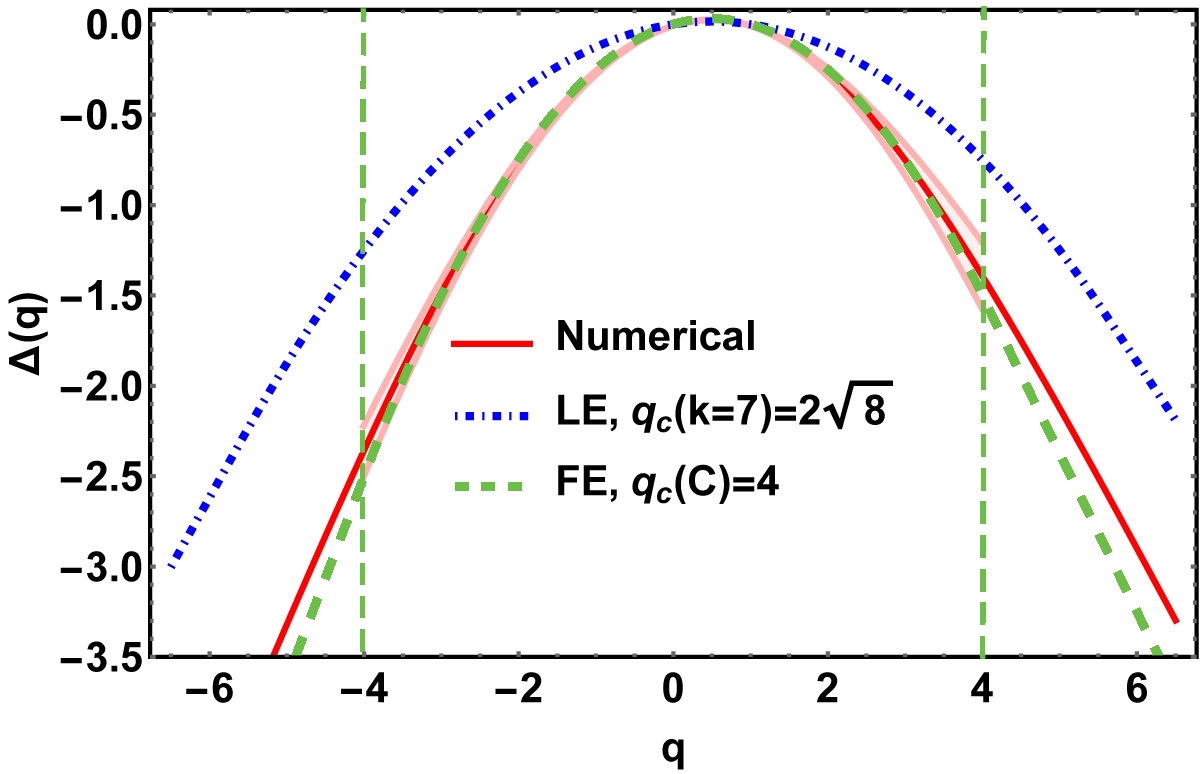}
}
\\
\subfigure{
\includegraphics[height=0.2\textheight]{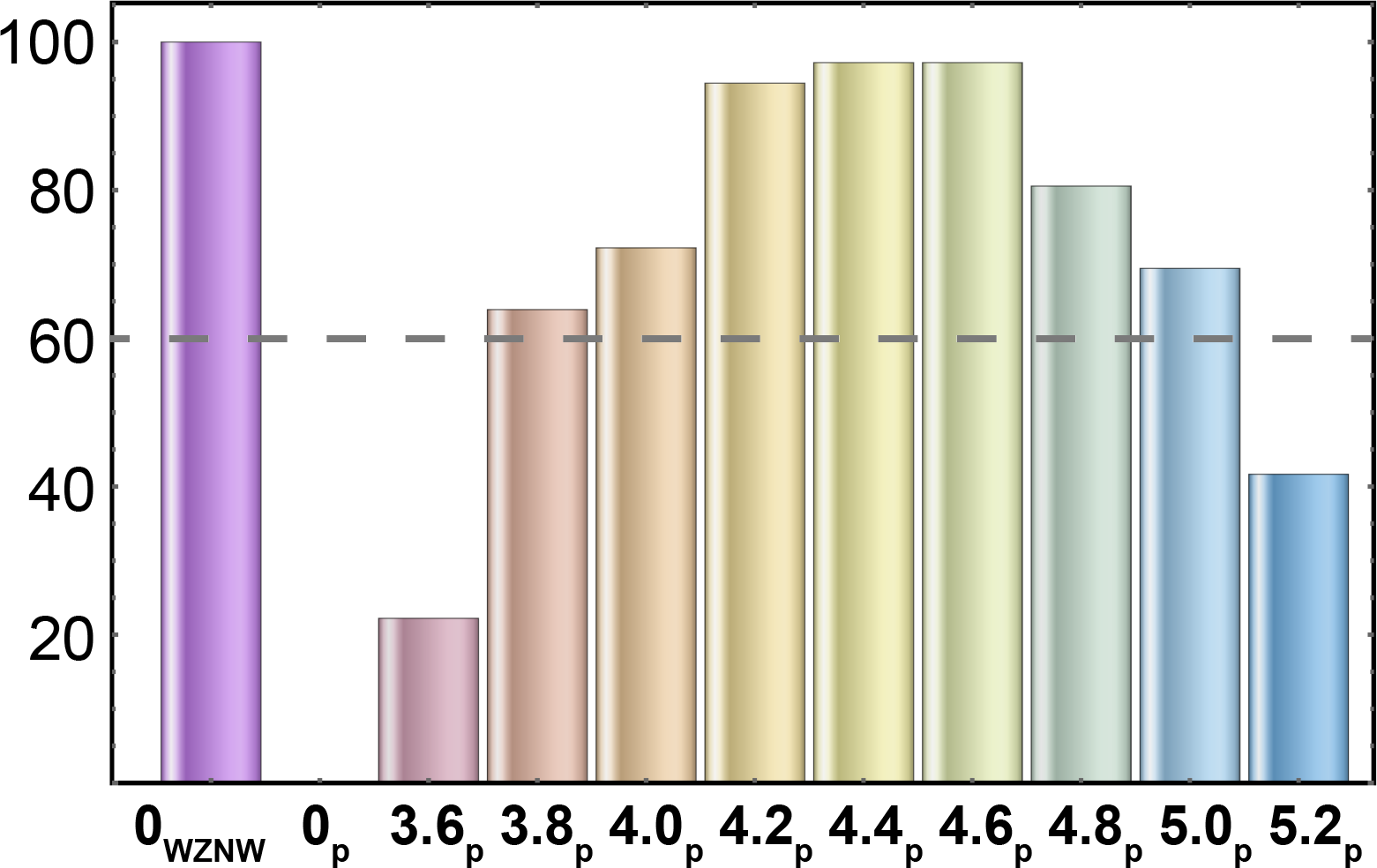}
}
\hspace{14pt}
\subfigure{
\includegraphics[height=0.2\textheight]{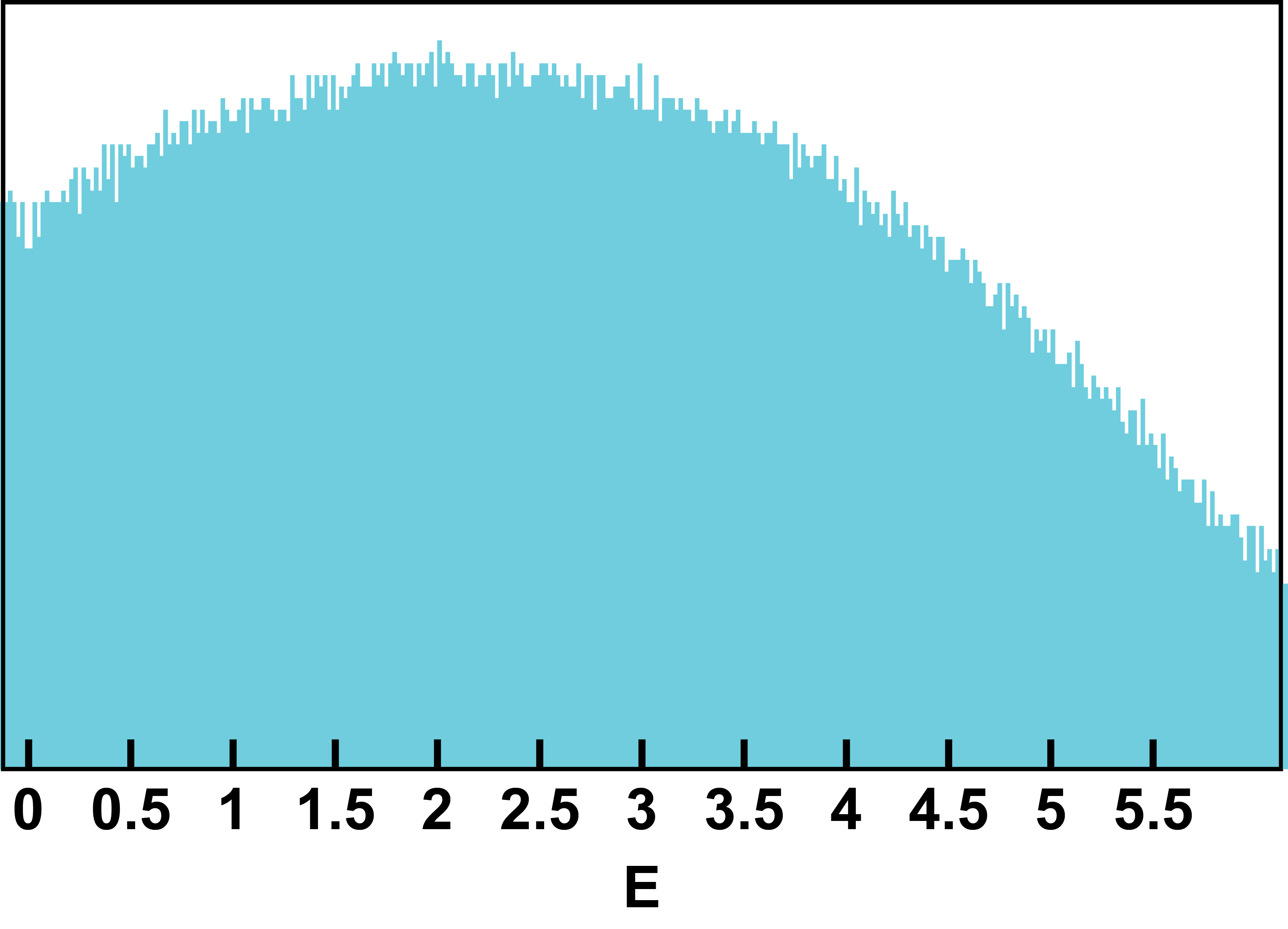}
}
\caption{$k = 7$, $N = 44$. Box sizes $b = 2,29$.}
\end{figure}


\begin{figure}[h!]
\subfigure{
\includegraphics[width=0.4\textwidth]{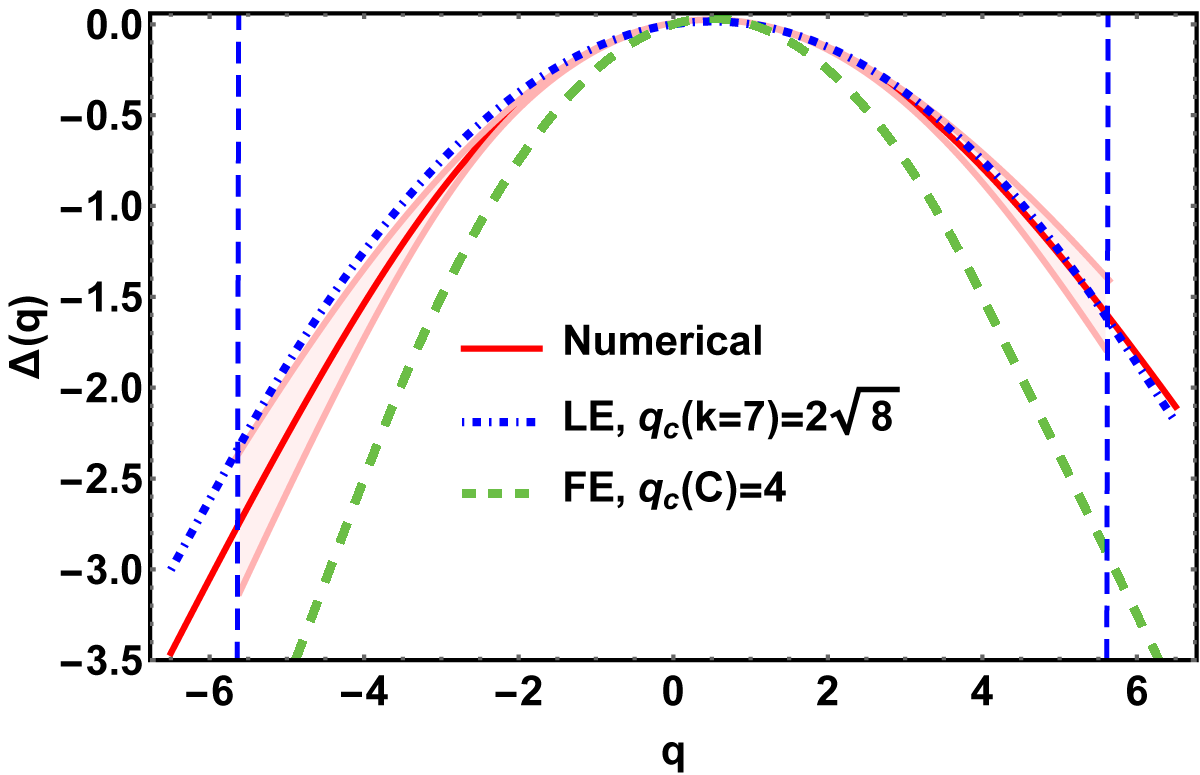}
}
\subfigure{
\includegraphics[width=0.4\textwidth]{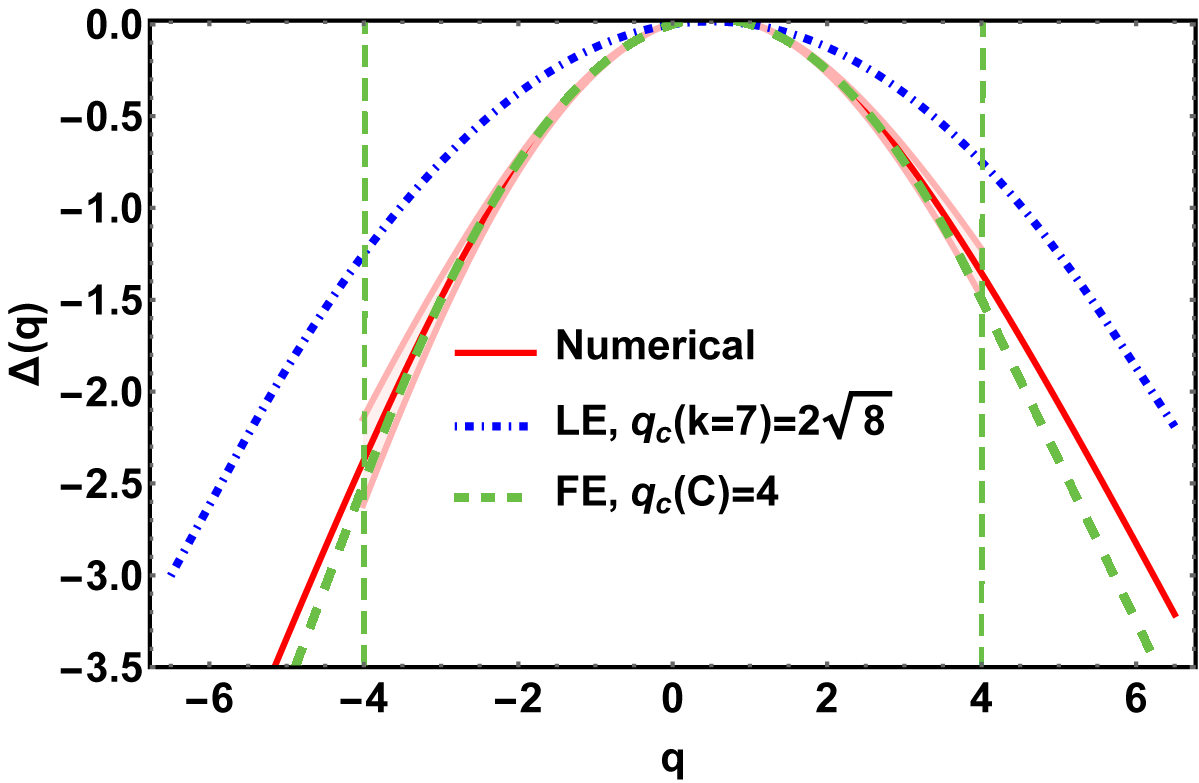}
}
\\
\subfigure{
\includegraphics[height=0.2\textheight]{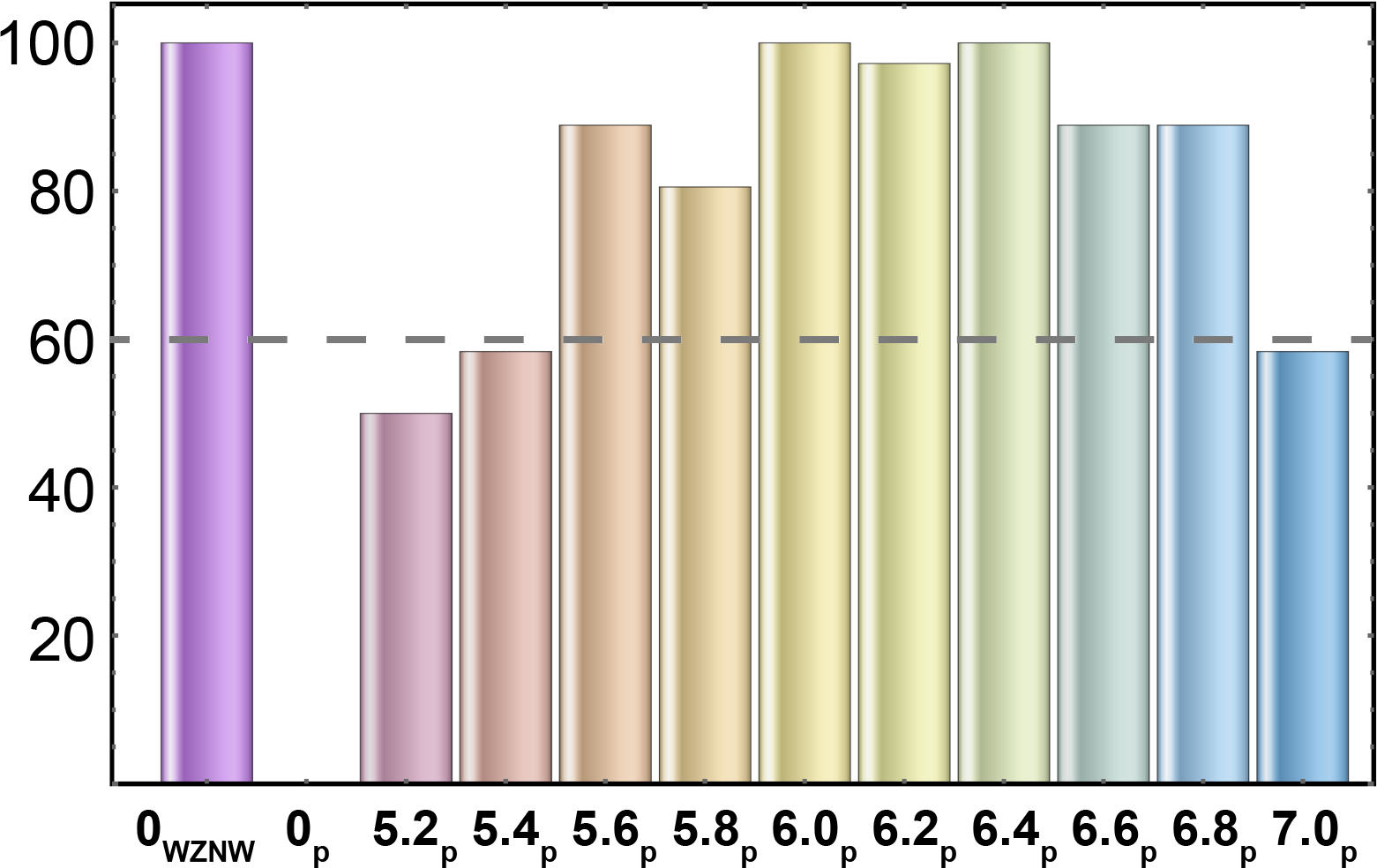}
}
\hspace{14pt}
\subfigure{
\includegraphics[height=0.2\textheight]{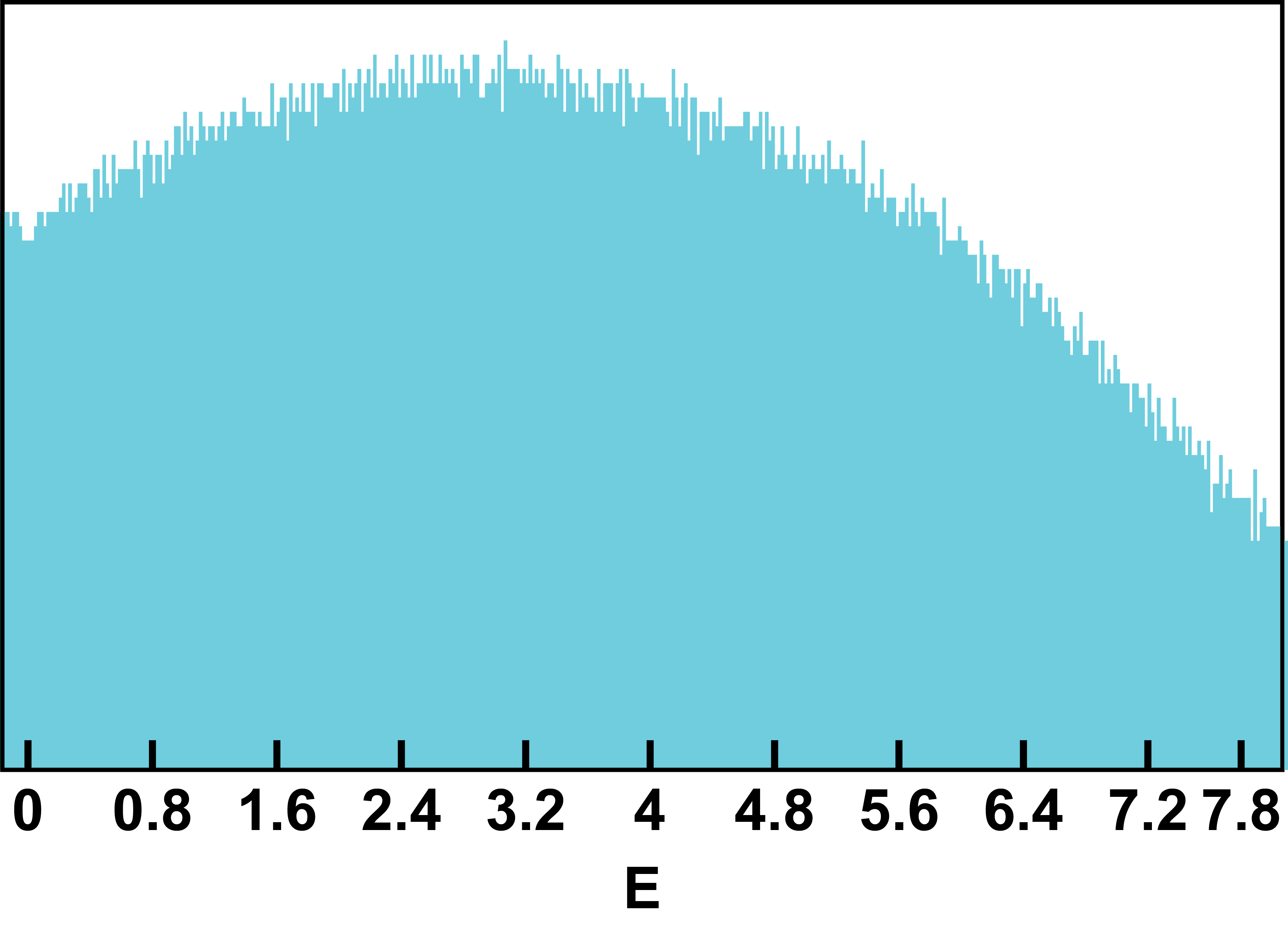}
}
\caption{$k = 7$, $N = 46$. Box sizes $b = 2,31$.}
\end{figure}



\begin{thebibliography}{100}
\bibitem{S--Footnote}
	Note that quasiparticles in the BdG description of a superconductor
	cannot be ``doped'' to a nonzero chemical potential, because the zero of 
	energy is always determined by the correlation gap (which occurs at the Fermi energy in the BCS regime). 
	This is true both for quasiparticles in a non-topological, gapless (e.g.\ $d$-wave) superconductor, and for 
	surface state Dirac or Majorana fermions in a TSC.  
	For the sigma model designed to calculate the physical spin conductivity, the nonzero energy wave functions formally 
	enter the zero temperature response by retaining $\omega \neq 0$ in Eqs.~(\ref{S--CI}) and (\ref{S--C}). 
	For class CI, this is not a ``soft'' insertion, since it is guaranteed that the finite-energy wave functions belong to a different class.
\end{thebibliography}
\end{document}